\documentclass[11pt]{article}
\usepackage{amsfonts,amssymb,amsmath,amsthm,amscd,latexsym}
\usepackage[latin1]{inputenc}

\def\Ad{\mathrm{Ad}}
\def\ad{\mathrm{ad}}
\newcommand{\inv}[0]{{-1}}

\newcommand{\cif}[0]{\mathcal{C}^\infty}

\def\ba{{\mbox{\boldmath $a$}}}

\def\bx{{\mbox{\boldmath $x$}}}
\def\Bx{{\mbox{\boldmath $X$}}}

\def\by{{\mbox{\boldmath $y$}}}
\def\By{{\mbox{\boldmath $Y$}}}

\def\bq{{\mbox{\boldmath $q$}}}

\def\bk{{\mbox{\boldmath $k$}}}
\def\bp{{\mbox{\boldmath $p$}}}

\def\bq{{\mbox{\boldmath $q$}}}

\def\bn{{\mbox{\boldmath $n$}}}
\def\bm{{\mbox{\boldmath $m$}}}
\def\bl{{\mbox{\boldmath $l$}}}

\newcommand{\gothg}{\mathfrak g }

\newcommand{\gothh}{\mathfrak h }

\newcommand{\ZZ}{\mathbb{Z}}

\newcommand{\RR}{\mathbb{R}}
\newcommand{\CC}{\mathbb{C}}
\newcommand{\MM}{\mathbb{M}}

\newcommand{\prgr}{G\ltimes \mathfrak{g}^*}

\newcommand{\dca}[0]{{\overline{a}}}
\newcommand{\dcb}[0]{{\overline{b}}}
\newcommand{\dcx}[0]{{\overline{x}}}

\newcommand{\ai}[0]{{A_i}}
\newcommand{\bi}[0]{{B_i}}

\newcommand{\aj}[0]{{A_j}}
\newcommand{\bjj}[0]{{B_j}}

\newcommand{\hyp}[0]{\mathbb{H}^2}

\newtheorem{theorem}{Theorem}[section]
\newtheorem{lemma}[theorem]{Lemma}
\newtheorem{corollary}[theorem]{Corollary}

\begin{document}
\parskip 6pt
\parindent 0pt

\begin{center}
\baselineskip 24 pt {\Large \bf Geometrical (2+1)-gravity and the
Chern-Simons formulation: Grafting, Dehn twists, Wilson loop
observables and the cosmological constant}

\baselineskip 18 pt

\vspace{1cm} {C.~Meusburger}\footnote{\tt  cmeusburger@perimeterinstitute.ca}\\
Perimeter Institute for Theoretical Physics\\
31 Caroline Street North,
Waterloo, Ontario N2L 2Y5, Canada\\

\vspace{0.5cm}

{26 July 2006}

\end{center}

\begin{abstract}

We relate the geometrical and the Chern-Simons description of
(2+1)-dimensional gravity for spacetimes of topology
$\mathbb{R}\times S_g$, where $S_g$ is an oriented two-surface of
genus $g>1$, for Lorentzian signature and general cosmological
constant and the Euclidean case with negative cosmological
constant.  We show how the variables parametrising the phase space
in the Chern-Simons formalism are obtained from the geometrical
description and how the geometrical construction of
(2+1)-spacetimes via grafting along closed, simple geodesics gives
rise to transformations on the phase space. We demonstrate that
these transformations are generated via the Poisson bracket by one
of the two canonical Wilson loop observables associated to the
geodesic, while the other acts as the Hamiltonian for
infinitesimal Dehn twists. For spacetimes with Lorentzian
signature, we discuss the role of the cosmological constant as a
deformation parameter in the geometrical and the Chern-Simons
formulation of the theory. In particular, we show that the Lie
algebras of the Chern-Simons gauge groups can be identified with
the (2+1)-dimensional Lorentz algebra over a commutative ring,
characterised
 by a formal parameter
$\Theta_\Lambda$ whose square is minus the cosmological constant.
In this framework, the Wilson loop observables that generate
grafting and Dehn twists are obtained as the real and the
$\Theta_\Lambda$-component of a
 Wilson loop observable with values in the ring, and the grafting
transformations can be viewed as infinitesimal Dehn twists with
the parameter $\Theta_\Lambda$.

\end{abstract}


\section{Introduction}

The quantisation of Einstein's theory of gravity is often viewed
as the problem of constructing a quantum theory of geometry. In
particular, a physically meaningful quantum theory of gravity
should allow one to recover spacetime geometry from the gauge
theory-like formulations used in most quantisation approaches.
While the quantisation of gravity in (3+1) dimensions is far from
complete, the (2+1)-dimensional version of the theory has been
used successfully as a testing ground for various quantisation
formalisms \cite{Carlipbook, carliprev}. As in the
(3+1)-dimensional case, most of these formalisms are based on
gauge theoretical descriptions of the theory. To apply these
results to concrete physics questions, it would be therefore be
necessary to recover their geometrical interpretation. Yet the
relation between the phase space variables used in these
approaches and spacetime
 geometry is not fully clarified even in
the classical theory.

The simplifications in (2+1)-dimensional gravity  compared to the
(3+1)-dimensional case are due to the absence of local
gravitational degrees of freedom and the finite-dimensionality of
its phase space. In the geometrical formulation of the theory,
this manifests itself in the fact that vacuum solutions of
Einstein's equations are flat or of constant curvature. They are
therefore locally isometric to certain model spacetimes, into
which any simply
 connected region of the spacetime can
be embedded. The physical degrees of freedom are purely
topological and encoded in transition functions, which take values
in the isometry group of the model spacetime and relate the
embedding of different spacetime regions. From a gauge theoretical
perspective, the absence of local gravitational degrees of freedom
in  (2+1)-dimensional gravity results in its formulation as a
Chern-Simons gauge theory
 with the isometry group of the associated model
spacetime as the gauge group \cite{AT, Witten1}. The Einstein
equations then take the form of a flatness condition on the gauge
field, and their solutions can be  locally trivialised,
i.~e.~written as pure gauge. The physical degrees of freedom are
then encoded in a set of elements of the gauge group which relate
the trivialisations on different regions of the spacetime
manifold.

The advantage of the Chern-Simons formulation of (2+1)-dimensional
gravity is that it allows one to apply gauge theoretical concepts
and methods to achieve an explicit parametrisation of the phase
space that serves as a starting point for quantisation. As gauge
fields solving the equations of motions are flat, physical states
can be characterised in terms of the holonomies along closed
curves in the spacetime manifold. Conjugation invariant functions
of such holonomies then define a complete set of gauge invariant
Wilson loop observables, which were first investigated in the
context of (2+1)-dimensional gravity in \cite{RN,
RN1,RN2,RN3,RN4,ahrss,martin}. Moreover, by parametrising the
phase space in terms  of the holonomies along a set of generators
of the fundamental group, one obtains an efficient description of
the Poisson structure \cite{FR,AMII}. These descriptions were used
in \cite{we1} to investigate the classical phase space of theory
and are the basis of Alekseev, Grosse and Schomerus combinatorial
quantisation formalism \cite{AGSI, AGSII} and the related
approaches in \cite{BNR, we2}.  The drawback of  the Chern-Simons
formulation is that it complicates the physical interpretation of
the theory by obscuring the underlying spacetime geometry. Except
for particularly simple spacetimes such as static spacetimes and
the torus universe, it is in general difficult to reconstruct
spacetime geometry from the gauge theoretical variables that
parametrise the phase space. In a geometrical framework, the
relation between holonomies and geometry was first investigated by
Mess \cite{mess}, who gives a shows how the holonomies determine
the geometry of the spacetime. More recent results on this problem
were obtained by Benedetti and Guadagnini \cite{bg} and by
Benedetti and Bonsante \cite{bb,bb2}, who focus on the
construction of (2+1)-dimensional spacetimes via grafting and
relate the resulting spacetimes for different values of the
cosmological constant.  However, despite these results, the
relation between spacetime geometry and the description of the
phase space of (2+1)-dimensional gravity in the Chern-Simons
formalism is still not fully clarified. While the results in
\cite{mess,bg,bb,bb2} establish a relation between holonomies and
geometry in the geometrical formulation of the theory, they do not
relate these variables to the quantities encoding the physical
degrees of freedom in the Chern-Simons formalism. In particular,
it is not clear how the embedding of spacetime regions into model
spacetimes and the associated transition functions are related to
the corresponding concepts in Chern-Simons theory, the
trivialisation of the gauge field and the gauge group elements
linking the trivialisations on different regions. Moreover, a full
understanding of the relation between spacetime geometry and the
Chern-Simons formulation  should clarify the role of phase space
and Poisson structure. This includes a geometrical interpretation
of the phase space transformations generated by the Wilson loop
observables  as well as the question how constructions that change
the geometry of a spacetime such as grafting and Dehn twists
manifest themselves on the phase space of the theory.

These questions concerning the relation between geometrical and
Chern-Simons formulation in (2+1)-dimensional gravity are the
subject of the present paper, in which we consider vacuum
spacetimes of topology $\RR\times S_g$, where $S_g$ is an
orientable two-surface of general genus $g>1$. Our results are
valid for spacetimes of Lorentzian signature and with  general
cosmological constant and for the Euclidean case with negative
cosmological constant. They can be summarised as follows.

{\bf 1. Embedding and trivialisation:} We relate the embedding of
spacetime regions into model manifolds in the geometrical
formulation and the trivialisation of the gauge field in the
Chern-Simons formalism and derive explicit formulas linking the
variables which encode the physical degrees of freedom in the two
approaches.

{\bf 2. Grafting transformations on phase space:} We show how the
geometrical construction of (2+1)-spacetimes by grafting along
closed, simple geodesics gives rise to a transformation on the
phase space in the Chern-Simons formulation and derive explicit
expressions for the action of this transformation on the
holonomies along a set of generators of the fundamental group.

{\bf 3. The transformations generated by Wilson loop observables:}
We investigate the two basic Wilson loop observables associated to
a closed, simple curve on $S_g$ and to the two linearly
independent $\Ad$-invariant, symmetric bilinear forms on the Lie
algebra of the gauge group. We derive explicit expressions for the
 phase space transformations these observables generate via the
 Poisson bracket and show that one of these observables acts as a
 Hamiltonian for the grafting transformations, while the other
 generates infinitesimal Dehn twists.

{\bf 4. Relation between grafting and Dehn twists:} We demonstrate
that the phase space transformations representing grafting and
Dehn twists are closely related for all values of the cosmological
constant and that this relation is reflected in a general symmetry
relation for the corresponding Wilson loop observables. We show
that grafting can be viewed as a Dehn twist with a formal
parameter $\Theta_\Lambda$ whose square is identified with minus
the cosmological constant.

{\bf 5. The cosmological constant as a deformation parameter:} We
establish a unified description for spacetimes of Lorentzian
signature in which the cosmological constant plays the role of a
deformation parameter. In the geometrical description, its square
root appears as a parameter relating the embedding into the
different model spacetimes and the action of the associated
isometry groups. In the Chern-Simons formulation, it plays the
role of a deformation parameter in the gauge group and the
associated Lie algebra. More precisely, we demonstrate that the
Lie algebra of the gauge group can be viewed as the
(2+1)-dimensional Lorentz algebra over a commutative ring with a
 multiplication law that depends on the cosmological constant.

Results similar to 1. to 4. were obtained in an earlier paper
\cite{ich} for the case of Lorentzian (2+1)-spacetimes  with
vanishing cosmological constant. Although the general approach in
\cite{ich} is similar, the reasoning and many proofs in \cite{ich}
make use of specific simplifications resulting from the properties
of Minkowski space and the (2+1)-dimensional Poincar\'e group. The
inclusion of these spacetimes in the present paper allows one to
see how these results arise from a general pattern present for all
values of the cosmological constant and to investigate the
 role of the cosmological constant as a deformation parameter.

The paper is structured as follows.

In Sect.~\ref{defnotsect} we introduce definitions and notations
for the Lie groups and Lie algebras considered in this paper and
summarise some facts from hyperbolic geometry used  in the
geometrical description of (2+1)-spacetimes.

Sect.~\ref{geomsect} gives an overview of the geometrical
description of (2+1)-dimensional spacetimes of topology $\RR\times
S_g$ for Lorentzian signature and general cosmological constant
 and for the Euclidean case with negative cosmological constant. We start by
introducing the relevant model spacetimes which are
(2+1)-dimensional Minkowski space, anti de Sitter space and de
Sitter space, respectively, for Lorentzian signature and
vanishing, negative and positive cosmological constant  and the
three-dimensional hyperbolic space for the Euclidean case with
negative cosmological constant. We then review the description of
(2+1)-spacetimes of topology  $\RR\times S_g$ which are obtained
as the quotients of regions in the model spacetimes by certain
actions of a cocompact Fuchsian group. After summarising the
description of static universes, we describe the construction of
evolving universes via grafting along closed, simple geodesics
following the presentation in \cite{bb,bb2}.

In Sect.~\ref{CSsect} we review the formulation of
(2+1)-dimensional gravity as a Hamiltonian Chern-Simons gauge
theory, where the gauge group is the isometry group of the
associated model spacetime, the (2+1)-dimensional Poincar\'e group
$PSU(1,1)\ltimes\RR^2\cong PSL(2,\RR)\ltimes\RR^3$ for Lorentzian
signature and vanishing cosmological constant, the group
$PSU(1,1)\times PSU(1,1)\cong PSL(2,\RR)\times PSL(2,\RR)$ for
Lorentzian signature and negative cosmological constant and
$SL(2,\CC)/\mathbb{Z}_2$ for Lorentzian signature and positive
cosmological constant and for the Euclidean case. We discuss how
the local trivialisation of the gauge field gives rise to a
parametrisation of the phase space in terms of the holonomies
along a set of generators of the fundamental group $\pi_1(S_g)$
and introduce Fock and Rosly's description of the Poisson
structure \cite{FR}.

Sect.~\ref{trivsect}  relates the geometry of (2+1)-spacetimes to
their description in the Chern-Simons formalism. We discuss the
relation between the variables encoding the physical degrees of
freedom in the geometrical and in
 the Chern-Simons approach and show how
the embedding into the model spacetimes is obtained from the
trivialisation of the gauge field in the Chern-Simons formalism.

In Sect.~\ref{csgraftsect} we demonstrate how the construction of
evolving (2+1)-spacetimes via grafting along closed, simple
geodesics in \cite{bb,bb2} is implemented in the Chern-Simons
formalism and show that it gives rise to a transformation on phase
space, given explicitly by its action on the holonomies along a
set of generators of the fundamental group $\pi_1(S_g)$.

In Sect.~\ref{graftpoiss}, we relate this transformation to the
Poisson structure and to the Wilson loop observables. We show that
the phase space transformation obtained by grafting along a
closed, simple geodesic $\eta$ is generated via the Poisson
bracket by one of the two basic Wilson loop observables associated
to $\eta$, while the other observable acts as the Hamiltonian for
Dehn twists. We discuss the properties of the grafting
transformations and their relation to Dehn twists, which manifests
itself in a general symmetry relation for the Poisson brackets of
the associated observables.

Sect.~\ref{cosmdefsect} investigates the role of the cosmological
constant in spacetimes of Lorentzian signature. Using the results
by Benedetti and Bonsante \cite{bb,bb2}, we show that its square
root can be viewed as a deformation parameter in the geometrical
description of both static and grafted (2+1)-spacetimes. For the
Chern-Simons formulation, we establish a common framework relating
the different gauge groups by identifying their Lie algebras with
the (2+1)-dimensional Lorentz algebra over a commutative ring. The
cosmological constant then appears  in the ring's multiplication
law and can be implemented by introducing a formal parameter
$\Theta_\Lambda$ whose square is minus the cosmological constant.
We show that the grafting transformations can be viewed as Dehn
twists with this parameter $\Theta_\Lambda$. Sect.~\ref{outlook}
contains a discussion of our results and conclusions.


\section{Definitions and notation}

\label{defnotsect}

\subsection{Lie groups and Lie algebras}
\label{gaugegrdefs}

Throughout the paper we employ Einstein's summation convention.
Indices are raised and lowered either with the three-dimensional
Minkowski metric
\begin{align}
\label{minkmet} \bx\cdot\by=\eta^L(\bx,\by)=\eta_{ab}x^ay^b=-x_0
y_0+x_1 y_1+x_2 y_2
\end{align}
or with the three-dimensional Euclidean metric
\begin{align}
\label{euclmet} \bx\cdot\by=\eta^E(\bx,\by)=\delta_{ab}x^ay^b=x_0
y_0+x_1 y_1+x_2 y_2.
\end{align}
To avoid confusion, we denote the signature of the spacetime by a
variable $S$ and write $S=L$ for Lorentzian and $S=E$ for
Euclidean signature.

In the following we consider a set of six-dimensional Lie algebras
$\gothh_{\Lambda,S}$ over $\RR$ whose generators we denote by
$J_a^S$ , $P_a^S$, $a=0,1,2$. For Lorentzian signature, the Lie
algebras $\gothh_{\Lambda,S}$ depend on a parameter
$\Lambda\in\RR$, and their Lie brackets are given by
\begin{align}
\label{liealg}
[J_a^L,J_b^L]_{\Lambda}=\epsilon_{ab}^{\;\;\;c}J_c^L\qquad[J_a^L,P_b^L]_\Lambda=\epsilon_{ab}^{\;\;\;c}P_c^L\qquad[P_a^L,P_b^L]_\Lambda=\Lambda\epsilon_{ab}^{\;\;\;c}J^L_c,
\end{align}
where indices are raised and lowered with the three-dimensional
Minkowski metric \eqref{minkmet} and  $\epsilon_{abc}$ is the
three-dimensional antisymmetric tensor satisfying
$\epsilon_{012}=1$. For Euclidean signature, we consider
parameters $\Lambda<0$, and the Lie algebra $\gothh_{\Lambda,E}$
has the bracket
\begin{align}
\label{liealge}
[J^E_a,J^E_b]_\Lambda=\epsilon_{ab}^{\;\;\;c}J^E_c\qquad[J^E_a,P^E_b]_\Lambda=\epsilon_{ab}^{\;\;\;c}P^E_c\qquad[P^E_a,P^E_b]_\Lambda=\Lambda
\epsilon_{ab}^{\;\;\;c}J^E_c \qquad \Lambda<0,
\end{align}
where indices are raised with the Euclidean metric
\eqref{euclmet}\footnote{Note that the parameter $\Lambda$ in
\eqref{liealg}, denoted by $\lambda$ in \cite{Witten1}, is not
equal to the cosmological constant but to minus the cosmological
constant for Lorentzian signature, while its Euclidean analogue in
\eqref{liealge} agrees with the cosmological constant. See also
the discussion at the beginning of Sect.~\ref{modelspt}.}. The
generators $J_a^E$ in \eqref{liealge} span the real Lie algebra
$su(2)$ and can be represented by the matrices
\begin{align}
\label{sigmatrix} &J_0^E=\tfrac{1}{2}\left(\begin{array}{cc} i&
0\\0 & -i\end{array}\right) &
&J_1^E=\tfrac{1}{2}\left(\begin{array}{cc} 0 & i\\i &
0\end{array}\right)
& &J_2^E=\tfrac{1}{2}\left(\begin{array}{cc} 0 & -1\\
1 & 0\end{array}\right).
\end{align}
Similarly, the bracket of the generators $J_a^L$ in \eqref{liealg}
is the Lie bracket of the three-dimensional Lorentz algebra
$\mathfrak{so}(2,1)\cong\mathfrak{sl}(2,\RR)\cong\mathfrak{su}(1,1)$.
A set of $\mathfrak{sl}(2,\RR)$-matrices representing these
generators is given by
\begin{align}
\label{sl2rj} &J_0=\tfrac{1}{2}\left(\begin{array}{cc} 0 & -1\\1 &
0\end{array}\right) & &J_1=\tfrac{1}{2}\left(\begin{array}{cc} 1 & 0\\
0
& -1\end{array}\right) & &J_2=\tfrac{1}{2}\left(\begin{array}{cc} 0 & 1\\
1 & 0\end{array}\right).
\end{align}
However, in the following we will mostly work with the Lie algebra
$\mathfrak{su}(1,1)$, which is conjugate to $\mathfrak{sl}(2,\RR)$
in $\mathfrak{sl}(2,\CC)$ via
\begin{align}
\label{suslident}
\mathfrak{su}(1,1)=\frac{1}{\sqrt 2}\left(\begin{array}{cc} 1 & i\\
i & 1\end{array}\right) \cdot \mathfrak{sl}(2,\RR) \cdot
\frac{1}{\sqrt 2}\left(\begin{array}{cc} 1 & -i\\-i &
1\end{array}\right).
\end{align}
The $\mathfrak{su}(1,1)$ matrices associated to the generators
\eqref{sl2rj} are given by
\begin{align}
\label{jmatrix} &J_0^L=\tfrac{1}{2}\left(\begin{array}{cc} i &
0\\0 &
-i\end{array}\right) & &J_1^L=\tfrac{1}{2}\left(\begin{array}{cc} 0 & -i\\
i
& 0\end{array}\right) & &J_2^L=\tfrac{1}{2}\sigma_2=\tfrac{1}{2}\left(\begin{array}{cc} 0 & 1\\
1 & 0\end{array}\right),
\end{align}
and by exponentiating linear combinations of these matrices over
$\RR$, one obtains the Lie group
\begin{align}
\label{pslpar} SU(1,1)=\left\{\left(\begin{array}{ll} a & b\\
\bar b & \bar a\end{array}\right) \;\;|\;\;
a,b\in\CC,\;|a|^2-|b|^2=1\right\}\cong SL(2,\RR).
\end{align}
The group $SU(1,1)\cong SL(2,\RR)$ is the double cover of the
proper orthochronous Lorentz group in three dimensions
$SO(2,1)^+=PSL(2,\RR)\cong PSU(1,1)= SU(1,1)/\mathbb{Z}_2$. In the
following, we will often parametrise elements of $SU(1,1)$ and
$PSU(1,1)$
 via the exponential map which in both cases we denote by $\exp: p^a J_a^L\mapsto e^{p^a
 J_a^L}$.  Using expressions \eqref{jmatrix} for the generators of
 $\mathfrak{su}(1,1)$, we find that the parametrisation of $SU(1,1)$
 in terms of a vector $\bp\in\RR^3$ is given by
\begin{align}
 &\exp:\;
\mathfrak{su}(1,1)\rightarrow SU(1,1)\nonumber\\
 \label{lorexpmap} & p^a J_a^L\mapsto e^{p^a J_a^L}=\begin{cases}
\cosh\left(\tfrac{|\bp|}{2}\right) 1
+2\sinh\left(\tfrac{|\bp|}{2}\right)\hat p^a J^L_a &
\text{for}\;\bp^2>0\\
1+p^a J_a^L &\text{for}\;\bp^2=0\\
\cos\left(\tfrac{|\bp|}{2}\right) 1
+2\sin\left(\tfrac{|\bp|}{2}\right)\hat p^a J^L_a &
\text{for}\;\bp^2<0\\
\end{cases} \qquad \hat\bp=\tfrac{1}{\sqrt{|\bp^2|}}\bp.
\end{align}
Elements $u=e^{p^a J^L_a}\in SU(1,1)$ are called elliptic,
parabolic and hyperbolic, respectively, for $\bp^2<0$, $\bp^2=0$
and $\bp^2>0$. It follows directly from expression
\eqref{lorexpmap} that the exponential map for $SU(1,1)$ is
neither surjective nor injective. The exponential map
$\exp:\mathfrak{su}(1,1)\rightarrow PSU(1,1)\cong SO(2,1)^+$ is
surjective, but again not injective, since $e^{p^aJ^L_a}=1$ for
$\bp^2=-(2\pi n)^2$, $n\in\ZZ$.  However, in the following we will
mainly consider hyperbolic elements of $PSU(1,1)$, for which the
parametrisation in terms of a vector $\bp=(p^0,p^1,p^2)\in \RR^3$
 is unique.

For $\Lambda=0$, the  six-dimensional real Lie algebra
\eqref{liealg} is the three-dimensional Poincar\'e algebra
$\gothh_{0,L}=\mathfrak{iso}(2,1)=\mathfrak{su}(1,1)\oplus\RR^3$,
and the associated Lie group obtained by exponentiation is the
semidirect product $SU(1,1)\ltimes \RR^3\cong
SL(2,\RR)\ltimes\RR^3$, where $SU(1,1)$ acts on $\RR^3\cong
\mathfrak{su}(1,1)$ via the adjoint action
\begin{align}
\label{poincpar}
(u_1,\ba_1)\cdot(u_2,\ba_2)=(u_1u_2,\ba_1+\Ad(u_1)\ba_2)\qquad
u_1,u_2\in SU(1,1), \ba_1,\ba_2\in\RR^3.
\end{align}
For $\Lambda>0$, one can introduce an alternative set of
generators $J_a^\pm$, in terms of which the Lie bracket
\eqref{liealg} takes the form of a direct sum
\begin{align}
\label{jpmbrack}
J_a^{\pm}=\tfrac{1}{2}(J^L_a\pm\tfrac{1}{\sqrt{\Lambda}}P^L_a)\qquad\Rightarrow\qquad[J_a^{\pm},
J_b^{\pm}]_{\Lambda>0}=\epsilon_{ab}^{\;\;\;c}J_c^{\pm}\qquad
[J_a^{\pm}, J_b^{\mp}]_{\Lambda>0}=0.
\end{align}
Hence, for $\Lambda>0$, the Lie algebra
$\gothh_{\Lambda>0,L}=\mathfrak{su}(1,1)\oplus \mathfrak{su}(1,1)$
is the direct sum  of two copies of $\mathfrak{su}(1,1)$ and the
 associated Lie group is $SU(1,1)\times SU(1,1)$, whose elements we will parametrise
 using an index $+$ for the first and $-$ for the second component
\begin{align}
\label{dirprod} (u_+,u_-)\cdot(v_+,v_-)=(u_+v_+, u_-v_-)\qquad
u_\pm, v_\pm \in SU(1,1).
\end{align}
For the Lie algebras $\gothh_{\Lambda<0,S}$ a set of matrices
representing the generators $J_a^S,P_a^S$ in \eqref{liealg},
\eqref{liealge} is obtained by setting $P^S_a=i\sqrt{|\Lambda|}
J^S_a$. This implies that the Lie algebras $\gothh_{\Lambda<0,L}$,
$\gothh_{\Lambda<0,E}$ are both isomorphic to
$\mathfrak{sl}(2,\CC)$. In the first case $\mathfrak{sl}(2,\CC)$
is realised as  the complexification
$\gothh_{\Lambda<0,L}=\mathfrak{sl}(2,\CC)=\mathfrak{sl}(2,\RR)\oplus
i\,\mathfrak{sl}(2,\RR)$ of its normal real form
$\mathfrak{sl}(2,\RR)$. In the second case as it is given as the
complexification
$\gothh_{\Lambda<0,E}=\mathfrak{sl}(2,\CC)=\mathfrak{su}(2)\oplus
i \,\mathfrak{su}(2)$ of its compact real form $su(2)$.

Hence, depending on the signature $S$ and on the parameter
$\Lambda$, the Lie algebras $\gothh_{\Lambda,S}$ and the
associated Lie groups $H_{\Lambda,S}$ are given by
\begin{align}
\gothh_{\Lambda,S}=\begin{cases}
\mathfrak{sl}(2,\RR)\oplus \RR^3 & \Lambda=0, S=L\\
\mathfrak{sl}(2,\RR)\oplus \mathfrak{sl}(2,\RR) & \Lambda>0,
S=L\\
\mathfrak{sl}(2,\CC) & \Lambda<0, S=L,E
\end{cases} \qquad H_{\Lambda,S}=\begin{cases}SU(1,1)\ltimes \RR^3 & \Lambda=0, S=L\\
SU(1,1)\times SU(1,1) & \Lambda>0,
S=L\\
SL(2,\CC) & \Lambda<0, S=L,E.\end{cases}\nonumber
\end{align}
For all signatures and all values of the parameter $\Lambda$, the
three-dimensional Lorentz algebra
$\mathfrak{su}(1,1)\cong\mathfrak{sl}(2,\RR)$ is a subalgebra of
the Lie algebra $\gothh_{\Lambda,S}$. The corresponding embedding
of the group $SU(1,1)$ into the groups $H_{\Lambda,S}$ is given by
\begin{align}
\label{canembedlor}  \i_{\Lambda,S}(v)=\begin{cases} (v,0)\in
SU(1,1)\ltimes \RR^3  &
\Lambda=0,\;S=L\\
(v,v)\in SU(1,1)\times SU(1,1)& \Lambda>0,\; S=L\\
v \in SL(2,\CC) & \Lambda<0, \;S=L,E.
\end{cases}
\end{align}
The embedding $\i_{\Lambda,S}(\pm 1)$ induces an action of
$\mathbb{Z}_2$ on $H_{\Lambda,S}$. The quotients of
$H_{\Lambda,S}$ by this action are the (2+1)-dimensional
Poincar\'e group  $H_{0,L}/\mathbb{Z}_2=PSU(1,1)\ltimes\RR^3$, the
group $H_{\Lambda>0,L}/\mathbb{Z}_2=SU(1,1)\times
SU(1,1)/\mathbb{Z}_2$ and the proper orthochronous Lorentz group
in (3+1)-dimensions $H_{\Lambda<0,L}/\mathbb{Z}_2=
H_{\Lambda<0,E}/\mathbb{Z}_2=SL(2,\CC)/\mathbb{Z}_2=SO(3,1)^+$. In
addition to these groups, we will need to consider the group
$PSU(1,1)\times
PSU(1,1)=H_{\Lambda>0,L}/(\mathbb{Z}_2\times\mathbb{Z}_2)$, whose
elements we parametrise as in \eqref{dirprod}. The embedding
\eqref{canembedlor} then induces an embedding of $PSU(1,1)$ into
the quotients $H_{\Lambda,S}/\mathbb{Z}_2$ and into the group
$PSU(1,1)\times PSU(1,1)$, which we will also denote by
$\i_{\Lambda,S}$.

In the following we will sometimes parametrise elements of the
groups $H_{\Lambda,S}$, $H_{\Lambda,S}/\mathbb{Z}_2$ and of
$PSU(1,1)\times PSU(1,1)$ via the exponential map, for which in
all cases we use the symbol $\exp_{\Lambda,S}$. Depending on the
value of the parameter $\Lambda$ and the signature, these
exponential maps are given by
\begin{align}
\label{exmapform} {\exp}_{\Lambda,S}(p^a J_a^S+k^a
P_a^S)=\begin{cases}(e^{p^aJ_a^L},
- T(- \bp)\bk) & \Lambda=0,\;\text{Lorentzian}\\
(e^{(p^a+\sqrt{\Lambda}k^a)J^L_a}, e^{(p^a-\sqrt{\Lambda}k^a)J^L_a})& \Lambda>0,\;\text{Lorentzian}\\
e^{(p^a+i\sqrt{|\Lambda|}k^a)J^L_a} & \Lambda<0,\;\text{Lorentzian}\\
e^{(p^a+i\sqrt{\Lambda}k^a)J^E_a} & \Lambda<0,\;\text{Euclidean},
\end{cases}
\end{align}
where expressions of the form $e^{p^a J_a^S}$ denote the image of
the exponential map \eqref{lorexpmap} or the associated
exponential map for $PSU(1,1)$, expressions $e^{(p^a+iq^a)J^S_a}$
the image of  the exponential map for $SL(2,\CC)$ and
$SL(2,\CC)/\mathbb{Z}_2$ and $T(\bp): \RR^3\rightarrow \RR^3$,
$\bp\in\RR^3$ is a bijective linear map given via the
identification $\RR^3\cong \mathfrak{su}(1,1)$ by
\begin{align}
\label{tdef} &(T(\bp)\bk)^aJ_a^L= \sum_{n=0}^\infty \frac{\ad_{p^a
J^L_a}^n(k^a J^L_a)}{(n+1)!}=k^a J^L_a+\tfrac{1}{2}[p^b J^L_b, k^a
J^L_a]+\tfrac{1}{6}[p^c J^L_c,[p^b J^L_b, k^a J^L_a]]+\ldots.
\end{align}
Note that for all values of $\Lambda$ and all signatures under
consideration, the exponential map $\exp_{\Lambda,S}:
\gothh_{\Lambda,S}\rightarrow H_{\Lambda,S}$ is neither surjective
nor injective, which follows from the corresponding statement for
the group $SU(1,1)$. For the groups $PSU(1,1)\ltimes\RR^3$,
$SL(2,\CC)/\mathfrak{Z}_2$ and $PSU(1,1)\times
PSU(1,1)=SU(1,1)\times SU(1,1)/(\mathbb{Z}_2\times\mathbb{Z}_2)$
the exponential maps are surjective but again not injective.

\subsection{Hyperbolic geometry}

\label{hypdefs}

In this section we summarise some facts and definitions from
hyperbolic geometry used in this paper. For a general reference,
we refer the reader to the book \cite{bp} by Benedetti and
Petronio, for a specialised treatment focusing on Fuchsian groups
to the book \cite{fgroups} by Katok.  In the following, we denote
by $\mathbb{H}^d_k$  the $d$-dimensional hyperbolic space of
curvature $-|k|$, realised as the hyperboloid
\begin{align}
\label{hyphyp}
\mathbb{H}^d_k=\{\bx=(x^0,x^1,\ldots,x^d)\in\RR^{d+1}\;|\;
-x_0^2+x_1^2+\ldots+x_d^2=-\tfrac{1}{|k|},\;x^0>0\}
\end{align}
with the metric induced by the $(d+1)$-dimensional Minkowski
metric. In the two-dimensional case, we also work with the disc
model, in which $\hyp=\hyp_1$ is realised as the unit disc
\begin{align}
D=\{z\in\CC\;|\; |z|<1\}\qquad \label{discmet}
ds^2=\frac{4|dz|^2}{(1-|z|^2)^2}\;,
\end{align}
and which is related to  the two-dimensional hyperboloid model
\eqref{hyphyp} via a map $z\in D\mapsto \bx(z)\in \hyp_k$
\begin{align}
\label{minkhyp}
&x_0(z)=\frac{1}{\sqrt{k}}\frac{1+|z|^2}{1-|z|^2}\quad
x_1(z)=\frac{1}{\sqrt{k}}\frac{2\text{Re}(z)}{1-|z|^2}\quad
x^2(z)=\frac{1}{\sqrt{k}}\frac{2\text{Im}(z)}{1-|z|^2}\qquad
\forall z\in D.
\end{align}
In the hyperboloid model, the geodesics of $\mathbb{H}^d_k$ are
obtained as the intersection of $\mathbb{H}^d_k$ with
$d$-dimensional hyperplanes through the origin. In the
two-dimensional disc model, the geodesics are the diameters of the
disc and arcs of circles orthogonal to its boundary. The isometry
group $\text{Isom}(\hyp_k)=\text{Isom}(D,ds^2)$ is the proper
orthochronous Lorentz group $PSL(2,\RR)\cong
PSU(1,1)=SU(1,1)/\mathbb{Z}_2$, which acts
 on the hyperboloid $\hyp_k$ via its canonical action on Minkowski space and whose action on the
disc $D$ is given by
\begin{align}
\label{discact} \left(\begin{array}{ll} a & b\\ \bar b& \bar
a\end{array}\right)\in SU(1,1):\; z\mapsto\frac{az+b}{\bar b z
+\bar a}.
\end{align}
The uniformization theorem states that every orientable
two-surface of genus $g>1$ with a metric of constant curvature
$-k$ is isometric to a quotient $\hyp_k/\Gamma$ of $\hyp_k$ by the
action of a cocompact Fuchsian group $\Gamma$ with $2g$ hyperbolic
generators
\begin{align}
\label{Fuchsgroup} \Gamma=\langle v_{A_1},v_{B_1},\ldots,
v_{A_g},v_{B_g}\;|\; [v_{B_g},
v_{A_g}^\inv]\cdots[v_{B_1},v_{A_1}^\inv]=1\rangle\subset
PSU(1,1).
\end{align}
The group $\Gamma$ induces a tessellation of $\hyp_k$ by geodesic
arc $4g$-gons, which are mapped into each other by the elements of
$\Gamma$.  Hence, for each polygon in the tessellation, there
exist $4g$ elements of $\Gamma$ which map this polygon into its
$4g$ neighbours and identify its sides pairwise. The surface
$\hyp_k/\Gamma$ is obtained by glueing these pairs of sides of a
polygon in the tessellation. In particular, there exists a
polygon, in the following referred to as fundamental polygon and
denoted by $P_\Gamma$, which is mapped into its $4g$ neighbours by
a fixed set of generators of $\Gamma$ and their inverses. If we
label the sides of $P_\Gamma$ as in Fig.~\ref{poly1}, the
generators $v_\ai,v_\bi$ in \eqref{Fuchsgroup} identify the sides
of this fundamental polygon $P_\Gamma$ according to
\begin{align}
v_\ai: a_i\mapsto a'_i\qquad v_\bi: b_i\mapsto b'_i.
\end{align}
The geodesics on the surface $\hyp_k/\Gamma$ are obtained by
projecting the geodesics on $\hyp_k$. In particular, closed
geodesics $\eta:[0,1]\rightarrow \hyp_k/\Gamma$, $\eta(0)=\eta(1)$
on $\hyp/\Gamma$ arise as the projections of geodesics $
c_\eta:[0,1]\rightarrow\hyp_k$  for which there exists an element
of $\Gamma$ that maps these geodesics to itself
\begin{align}
\label{geodpar} c_\eta(1)= v_\eta c_\eta(0)\qquad
v_\eta=e^{n_\eta^a J_a}\in\Gamma.
\end{align}
In the following we will refer to this element as the translation
element of $\eta$ and to the associated vectors $\bn_\eta$ and
$\hat \bn_\eta=\bn_\eta/{\sqrt{|\bn_\eta^2|}}$ as the translation
vector and unit translation vector of $\eta$. Closed geodesics on
the surface $\hyp_k/\Gamma$ are therefore in one-to-one
correspondence with elements of the cocompact Fuchsian group
$\Gamma$, which is isomorphic to the surface's fundamental group
$\pi_1(\hyp_k/\Gamma)\cong \Gamma$. In the following we will often
not distinguish notationally between such geodesics, their
homotopy equivalence classes in $\pi_1(\hyp_k/\Gamma)$ and general
curves on the surface which represent these homotopy equivalence
classes.

\section{(2+1)-dimensional gravity: the geometrical formulation}

\label{geomsect}

\subsection{Model spacetimes}
\label{modelspt}

In this section, we summarise the geometrical description of
(2+1)-spacetimes as quotients of certain model spacetimes. A
general reference for (2+1)-spacetimes is the book
\cite{Carlipbook} by Carlip. A more specific treatment focusing on
the construction of (2+1)-spacetimes via grafting is given in the
papers  by Benedetti and Bonsante \cite{bb,bb2}.

(2+1)-dimensional gravity is a theory without local gravitational
degrees of freedom. As the curvature tensor of a three-dimensional
manifold is determined completely by its Ricci tensor, vacuum
solutions of the (2+1)-dimensional Einstein equations are flat or
of constant  curvature. This implies that they are locally
isometric to a three-dimensional model spacetime. In this paper,
we consider Lorentzian (2+1)-gravity  with general cosmological
constant and the Euclidean case with negative cosmological
constant. In the following, we work with a parameter
$\Lambda\in\RR$ which is identified with minus the cosmological
constant for Lorentzian spacetimes and agrees with the
cosmological constant in the Euclidean case. The choice of this
convention is motivated by the conventions in the  Chern-Simons
formulation of the theory and leads to notational simplifications
there. The model spacetimes for Lorentzian signature are then
  three-dimensional Anti de Sitter space $\text{AdS}_\Lambda$,
three-dimensional Minkowski space $\mathbb{M}^3$,
three-dimensional de Sitter space
 ${\text{dS}}_\Lambda$, respectively, for $\Lambda>0$ (negative cosmological
constant), $\Lambda=0$ (vanishing cosmological constant) and
$\Lambda<0$ (positive cosmological constant). The model spacetime
for Euclidean signature and negative cosmological constant
($\Lambda<0$) is three-dimensional hyperbolic space
$\mathbb{H}_\Lambda^3$.

In the following, we parametrise these
 spacetimes in terms of matrices, which is convenient for establishing a link with
 their description in the
Chern-Simons formalism. For the Lorentzian case with vanishing
cosmological constant,
 the relevant model spacetime is (2+1)-dimensional
Minkowski space $\mathbb{X}_{0,L}=\mathbb{M}^3$ and the group of
orientation and time orientation preserving isometries is the
(2+1)-dimensional Poincar\'e group
$\text{Isom}({\mathbb{X}_{0,L}})=PSU(1,1)\ltimes
\RR^3=H_{0,L}/\mathbb{Z}_2$. In the canonical identification of
Minkowski space with the set of $\mathfrak{su}(1,1)$-matrices
\begin{align}
\label{minkxemb}
\mathbb{M}^3\ni\bx=(x^0,x^1,x^2)\mapsto \Bx= 2x^a J^L_a= \left(\begin{array}{cc} i x^0 &-i(x^1+ix^2)\\
 i(x^1-ix^2) & -ix^0\end{array}\right)\in\mathfrak{su}(1,1),
\end{align}
the (2+1)-dimensional Minkowski metric agrees with the Killing
form of $\mathfrak{su}(1,1)$
\begin{align}
\label{minkkill} -x_0 y_0+x_1 y_1+x_2
y_2=\tfrac{1}{2}\text{Tr}\left(\Bx\cdot \By\right),
\end{align}
and the action of $\text{Isom}({\mathbb{X}_{0,L}})=PSU(1,1)\ltimes
\RR^3$ is given by
\begin{align}
 \label{poincact2} (u,\ba)\in PSU(1,1)\ltimes \RR^3:\;\Bx\mapsto u \Bx u^\inv+ 2 a^b
 J^L_b\qquad\quad\forall\Bx\in \mathfrak{su}(1,1).
\end{align}
The model spacetime for negative cosmological constant
($\Lambda>0$) and Lorentzian signature is three-dimensional Anti
de Sitter space $\text{AdS}_\Lambda$. We adopt the conventions of
\cite{bb,bb2} in which Anti de Sitter space is realised as a
quotient of the universal cover $\widetilde{\text{AdS}}_\Lambda$
by the action of $\mathbb{Z}_2$. The universal cover
$\widetilde{\text{AdS}}_\Lambda$ is the manifold
\begin{align}
 &\mathbb{X}_{\Lambda>0,L}=\widetilde{\text{AdS}}_\Lambda=\{(t_1,t_2,x_1,x_2)\in\RR^4\;|\;
t_1^2+t_2^2-x_1^2-x_2^2=\tfrac{1}{\Lambda},\;\}\nonumber \\
\label{adsmet} &d\bx^2=-(dt_1)^2-(dt_2)^2+(dx_1)^2+(dx_2)^2.
\end{align}
Via the map
\begin{align}
\label{adsxemb}
\widetilde{\text{AdS}}_\Lambda\ni\bx=(t_1,t_2,x_1,x_2)\mapsto {\Bx}=\left(\begin{array}{cc} t_1+it_2 &-i(x_1+ix_2)\\
i(x_1-ix_2) & t_1-it_2\end{array}\right)\in SU(1,1)
\end{align}
 it can be identified with the group $SU(1,1)$ such that its metric is given by minus the determinant
\begin{align}
\label{adsdetmet} \widetilde{\text{AdS}}_\Lambda=\{
\tfrac{1}{\sqrt{\Lambda}}A\;|\; A\in SU(1,1)\}\qquad
d\bx^2=-\det\,(\, d{\Bx}\,).
\end{align}
The group of orientation and time orientation preserving
isometries of $\widetilde{\text{AdS}}_\Lambda$ is the group
$SU(1,1)\times SU(1,1)/\mathbb{Z}_2=H_{\Lambda>0,L}/\mathbb{Z}_2$,
whose action  is given by the action of $SU(1,1)\times SU(1,1)$
via
\begin{align}
\label{adsact} (G_+,G_-)\in SU(1,1)\times SU(1,1):\; \Bx\mapsto
G_+ \Bx G_-^\inv\qquad\forall \Bx\in SU(1,1).
\end{align}
Anti de Sitter space $\text{AdS}_\Lambda$ is the quotient of
$\widetilde{\text{AdS}}_\Lambda$ by the action of the elements
$(\pm 1,\mp 1)$ via \eqref{adsact} and its isometry group is the
quotient $PSU(1,1)\times PSU(1,1)/\mathbb{Z}_2$
\begin{align}
\label{adsquot}
&\mathbb{X}_{\Lambda>0,L}=\text{AdS}_\Lambda=\widetilde{\text{AdS}}_\Lambda/\mathbb{Z}_2
& &\text{Isom}(\text{AdS}_\Lambda)=PSU(1,1)\times PSU(1,1).
\end{align}

For positive cosmological constant ($\Lambda<0$) and Lorentzian
signature, the model spacetime is three-dimensional de Sitter
space
\begin{align}
\mathbb{X}_{\Lambda<0,L}={\text{dS}}_\Lambda=\{\bx=(x^0,x^1,x^2,x^3)
\in \MM^4\;|\; -x_0^2+x_1^2+x_2^2+x_3^2=\tfrac{1}{|\Lambda|}\},
\end{align}
and for the Euclidean case with negative cosmological constant
($\Lambda<0$), it is three-dimensional hyperbolic space
\begin{align}
\mathbb{X}_{\Lambda<0,E}=\mathbb{H}^3_\Lambda=\{\bx=(x^0,x^1,x^2,x^3)
\in \MM^4\;|\; -x_0^2+x_1^2+x_2^2+x_3^2=-\tfrac{1}{|\Lambda|},
x^0>0\}.
\end{align}
In both cases the metric is the one induced by the
four-dimensional Minkowski metric, and the group of orientation
and, in the de Sitter case, time orientation preserving isometries
is the proper orthochronous Lorentz group in three dimensions
$\text{Isom}(\mathbb{X}_{\Lambda<0,L})=\text{Isom}(\mathbb{X}_{\Lambda<0,E})=SO(3,1)^+=SL(2,\CC)/\mathbb{Z}_2$.
The parametrisation of these spacetimes in terms of matrices is
obtained by identifying vectors in four-dimensional Minkowski
space with certain sets of matrices in $GL(2,\CC)$.
 The first identification
is the standard identification of Minkowski space with the set of
hermitian $GL(2,\CC)$ matrices
\begin{align}
\label{hypid}\mathbb{M}^4\ni \bx=(x^0,x^1,x^2,x^3)\mapsto {\Bx}=
x^0\,1+ x^i\sigma_i=\left(\begin{array}{ll} x^0+x^3 &
x^1+ix^2\\x^1-ix^2 & x^0-x^3\end{array}\right),
\end{align} in which the four-dimensional Minkowski metric takes
the form
\begin{align}
\label{hypdet}-\det\,(\,{\Bx}\,)=-x_0^2+x_1^2+x_2^2+x_3^2.
\end{align}
The action of the isometry group $SO(3,1)^+\cong
SL(2,\CC)/\mathbb{Z}_2$ on the set of hermitian $2\times
2$-matrices is given by the action of $SL(2,\CC)$ via
\begin{align}
\label{hypslact} G\in SL(2,\CC): \;\Bx\mapsto G \Bx
G^\dag\qquad\quad\forall \Bx\in GL(2,\CC), \Bx^\dag=\Bx.
\end{align}
As this action has kernel $\pm 1$ and leaves the determinant
invariant, it induces an action of $SO(3,1)^+$ which preserves the
metric \eqref{hypdet}. Hence, three-dimensional hyperbolic space
$\mathbb{H}^3_\Lambda$ can be identified with the set of hermitian
$SL(2,\CC)$ matrices with metric \eqref{hypdet} and the action of
the isometry group $\text{Isom}(\mathbb{H}^3_\Lambda)=SO(3,1)^+$
is given by \eqref{hypslact}
\begin{align}
\mathbb{H}^3_\Lambda=\{ \tfrac{1}{\sqrt{|\Lambda|}}A\;|\; A\in
SL(2,\CC),\,A^\dag=A\}.
\end{align}

The matrix representation of ${\text{dS}}_\Lambda$ is similar, but
instead of the identification \eqref{hypid}, one uses the
identification
\begin{align}
\label{dsid0}\mathbb{M}^4\ni \bx=(x^0,x^1,x^2,x^3)\mapsto
{\Bx}=\left(\begin{array}{cc} x^3+x^0 & x^1+ix^2\\-(x^1-ix^2) &
x^3-x^0\end{array}\right),
\end{align}
which assigns to each vector in Minkowski space a matrix in the
set
\begin{align}
\label{circspace0} &CL(2,\CC)=\{ A\in GL(2,\CC)\;|\; A^\circ=A\}\\
\label{dsinv} &\left(\begin{array}{ll} a & b\\ c&
d\end{array}\right)^\circ =\left(\begin{array}{rr} \bar a & -\bar c\\
-\bar b &\bar d\end{array}\right)=\left(\begin{array}{ll} i & 0\\
0 & -i\end{array}\right) \left(\begin{array}{ll} a & b\\ c&
d\end{array}\right)^\dag \left(\begin{array}{rr} -i & 0\\
0 & i\end{array}\right),
\end{align}
such that the four-dimensional Minkowski metric is given by the
determinant
\begin{align}
\label{dsdetmet}\det\,(\,d{\Bx}\,)=-x_0^2+x_1^2+x_2^2+x_3^2.
\end{align}
As the map $\circ: GL(2,\CC)\rightarrow GL(2,\CC)$ satisfies
$(A^\circ)^\circ=A$, $(AB)^\circ=B^\circ A^\circ$,
$\det(A^\circ)=\overline{\det A}$, one obtains an action of the
group $SL(2,\CC)$  on  $CL(2,\CC)$ via
\begin{align}
\label{dsslact} G\in SL(2,\CC): \; M\mapsto G MG^\circ
\end{align}
which has kernel $\pm 1$, preserves the determinant, and thus
induces an action of $SO(3,1)^+=SL(2,\CC)/\mathbb{Z}_2$ which
preserves the metric \eqref{dsdetmet}. Hence,  de Sitter space
${\text{dS}}_\Lambda$   can be realised as the set of $SL(2,\CC)$
matrices invariant under the operation $^\circ$ with metric
\eqref{dsdetmet}, and with isometry group $SL(2,\CC)/\mathbb{Z}_2$
 whose action is given by \eqref{dsslact}
\begin{align}
\label{circspace} {\text{dS}}_\Lambda=\{
\tfrac{1}{\sqrt{|\Lambda|}}A\;|\; A\in SL(2,\CC),\,A^\circ=A\}.
\end{align}
Hence, depending on the cosmological constant and the signature,
the model spacetimes $\mathbb{X}_{\Lambda,S}$ can be identified
with the sets of matrices
\begin{align}
\mathbb{X}_{\Lambda,S}=&\begin{cases} \mathbb{M}^3\cong
\mathfrak{sl}(2,\RR) & \Lambda=0,\; S=L\\
\text{AdS}_\Lambda\cong \tfrac{1}{\sqrt{\Lambda}}PSU(1,1) & \Lambda>0,\; S=L \\
\text{dS}_\Lambda\cong \tfrac{1}{\sqrt{|\Lambda|}}\{A\in
SL(2,\CC):\; A=A^\circ\} & \Lambda<0,\;S=L \\
\mathbb{H}^3\cong \tfrac{1}{\sqrt{|\Lambda|}}\{A\in SL(2,\CC):\;
A=A^\dag\} & \Lambda<0,\; S=E \end{cases}
\end{align}
with metrics given by \eqref{minkkill}, \eqref{adsdetmet},
\eqref{hypdet}, \eqref{dsdetmet} and their groups of (orientation
and time orientation) preserving isometries
\begin{align}
\text{Isom}(\mathbb{X}_{\Lambda,S})=&\begin{cases} PSU(1,1)\ltimes
\RR^3  & \Lambda=0,\; S=L\\ PSU(1,1)\times PSU(1,1) &
\Lambda>0,\; S=L\\ SL(2,\CC)/\mathbb{Z}_2 & \Lambda<0,\; S=L\\
SL(2,\CC)/\mathbb{Z}_2 & \Lambda>0,\; S=E
\end{cases}
\end{align}
act via \eqref{poincact2}, \eqref{adsact}, \eqref{hypslact},
\eqref{dsslact}.

\subsection{Static universes and the embedding of hyperbolic space $\mathbb{H}^2$}

\label{statuniv}

The defining characteristic of the model spacetime introduced in
the last subsection is that their topology is trivial.
(2+1)-spacetimes with nontrivial topology are obtained as the
quotients of domains $U_{\Lambda,S}\subset\mathbb{X}_{\Lambda,S}$
in the model spacetimes
 by the action of certain subgroups of the isometry groups
$\text{Isom}(\mathbb{X}_{\Lambda,S})$. In this paper we restrict
attention to spacetimes for which these subgroups are cocompact
Fuchsian groups $\Gamma$ with $2g>2$ generators and act via group
homomorphisms $h_{\Lambda,S}:
\Gamma\rightarrow\text{Isom}(\mathbb{X}_{\Lambda, S})$. The
resulting spacetimes have topology $\RR\times S_g$, where $S_g$ is
an oriented two-surface of genus $g>1$.

The simplest such spacetimes are the static spacetimes associated
to a cocompact Fuchsian group $\Gamma$, for a detailed discussion
see for example \cite{Carlipbook}. For Lorentzian signature, the
associated domain $U_{\Lambda,L}\subset \mathbb{X}_{\Lambda,L}$ in
the model spacetime is the interior of a forward lightcone,
i.~e.~the set of points connected to a given point
$\bx_{\Lambda,L}\in\mathbb{X}_{\Lambda,L}$ by timelike geodesics.
In the Euclidean case, it is the whole model spacetime
$\mathbb{H}^3_\Lambda$.
 In each model spacetime $\mathbb{X}_{\Lambda, S}$, this domain is foliated by two-surfaces
$U_{\Lambda, S}(T)$ of constant cosmological time $T$,
i.~e.~surfaces of constant geodesic distance $T$ from a given
point $\bx_{\Lambda,S}$, which represents a singularity of the
spacetime. For all values of the cosmological constant and all
signatures under consideration, the surfaces $U_{\Lambda, S}(T)$
are surfaces of constant curvature and can be identified with
copies of two-dimensional hyperbolic space. The action of the
(2+1)-dimensional Lorentz group $PSU(1,1)$ via its canonical
embedding $\i_{\Lambda,S}:
PSU(1,1)\rightarrow\text{Isom}(\mathbb{X}_{\Lambda,S})$ preserves
the surfaces $U_{\Lambda,S}(T)$ and agrees with the action induced
by \eqref{discact}. This induces an action of the cocompact
Fuchsian group $\Gamma$ and a tessellation of
 each surface $U_{\Lambda,S}(T)$  by geodesic arc $4g$-gons as described in Sect.~\ref{hypdefs}. The
static spacetimes $M_{\Lambda,\Gamma, S}^{st}$ associated to
$\Gamma$ are then obtained by identifying on each surface of
constant cosmological time the points related by this action of
$\Gamma$
\begin{align}
M_{\Lambda,\Gamma,S}^{st}=U_{\Lambda,S}/\Gamma.
\end{align}

To obtain explicit expressions for the static domains
$U_{\Lambda,S}\subset \mathbb{X}_{\Lambda,S}$ and their foliation
by copies of hyperbolic space, we consider timelike geodesics
$c_{\Lambda,L}$ in the Lorentzian model spacetimes
$\mathbb{X}_{\Lambda,L}$ and an associated geodesic
$c_{\Lambda,E}$ in $\mathbb{H}^3_\Lambda$
\begin{align}
\label{timegeod}
c_{\Lambda,S}(T)=\begin{cases}  2T J_0^L\,,\,T\in(0,\infty)  & \Lambda=0, \;\text{Lorentzian}\\
\tfrac{1}{\sqrt\Lambda}e^{ \sqrt{\Lambda}TJ_0^L}\,,\,T\in(0,\pi/\sqrt\Lambda) & \Lambda>0,\; \text{Lorentzian}\\
\tfrac{1}{\sqrt{|\Lambda|}} e^{-i\sqrt{|\Lambda|}T
J_0^L}\,,\,T\in(0,\infty) & \Lambda<0,\; \text{Lorentzian
 and Euclidean},
\end{cases}
\end{align}
which are parametrised by arclength and based at the identity.
Furthermore we introduce a map
\begin{align}
\label{gmap}g: \hyp\rightarrow SU(1,1),\quad
z\mapsto g(z)=\frac{1}{\sqrt{1-|z|^2}}\left(\begin{array}{ll} 1 & z\\
\bar z & 1\end{array}\right).
\end{align}
A brief calculation shows that - up to right-multiplication with a
phase - the action of $SU(1,1)$ on the disc via \eqref{discact}
corresponds to left-multiplication of the image $g(z)$
\begin{align}
\label{Mact} g(Mz)=M\cdot g(z)\cdot e^{\psi(M,z)
J^L_0}\qquad\psi(M,z)\in\RR\qquad \forall M\in PSU(1,1), z\in
\hyp.
\end{align}
As the phase commutes with $J_0^L$ and is mapped to its inverse by
the operations $^\circ$, $\dag$, one finds the map
$\Phi_T^{\Lambda,S}: \hyp\rightarrow \mathbb{X}_{\Lambda,S}$
defined by
\begin{align}
\label{tembed}
\Phi_T^{\Lambda,S}(z)= \i_{\Lambda,S}\circ g(z)\,  c_{\Lambda,S}(T)=\begin{cases}   g(z) c_{0,L}(T) g(z)^\inv & \Lambda=0, \;\text{Lorentzian}\\
g(z) c_{\Lambda>0,L}(T) g(z)^\inv & \Lambda>0,\; \text{Lorentzian}\\
g(z) c_{\Lambda<0,L}(T) g(z)^\circ & \Lambda<0,\; \text{Lorentzian}\\
g(z) c_{\Lambda<0,E}(T) g(z)^\dag & \Lambda<0,\; \text{Euclidean}
\end{cases}
\end{align}
 satisfies the covariance condition
\begin{align}
\label{phiid}
\Phi_T^{\Lambda,S}(Mz)=\i_{\Lambda,S}(M) \,\Phi_T^{\Lambda,S}(z)= \begin{cases}M\Phi_T^{\Lambda,L}(z)M^\inv & \Lambda=0,\;\text{Lorentzian}\\ M\Phi_T^{\Lambda,L}(z)M^\inv & \Lambda>0,\;\text{Lorentzian}\\
M\Phi_T^{\Lambda,L}(z)M^\circ & \Lambda<0,\; \text{Lorentzian}\\
M\Phi_T^{\Lambda,E}(z)M^\dag & \Lambda<0,\;
\text{Euclidean}\end{cases}.
\end{align}
The action of the (2+1)-dimensional Lorentz group $PSU(1,1)$ via
its canonical embedding into $\text{Isom}(\mathbb{X}_{\Lambda,S})$
therefore preserves the images of $\Phi^{\Lambda,S}_T(\hyp)$ and
agrees with its action induced by its action on the Poincar\'e
disc via \eqref{discact}. Furthermore, as the geodesics
\eqref{timegeod} are parametrised by arclength, all points in the
image $\Phi^{\Lambda,S}_T(\hyp)$ have constant geodesic distance
$T$ from the initial singularity $c_{\Lambda,S}(0)$. Hence, one
obtains a foliation of the forward lightcone or, for Euclidean
signature, of $\mathbb{H}^3_{\Lambda}$ by surfaces of constant
cosmological time
\begin{align}
\label{concfol}
&U_{\Lambda,S}=\begin{cases} \bigcup_{T\in(0,\infty)} U_{0,L}(T) & \Lambda=0, \;\text{Lorentzian}\\
\bigcup_{T\in(0,\pi/\sqrt{\Lambda})} U_{\Lambda,L}(T) & \Lambda>0,\; \text{Lorentzian}\\
\bigcup_{T\in(0,\infty)} U_{\Lambda,L}(T) &
\Lambda<0,\;\text{Lorentzian}\\
\bigcup_{T\in(0,\infty)} U_{\Lambda,E}(T) &
\;\Lambda<0,\;\text{Euclidean}\end{cases} \quad\qquad \quad
U_{\Lambda,S} (T)=\phi_T^{\Lambda,S}(\hyp).
\end{align}
To obtain concrete expressions for the matrices in \eqref{tembed},
one evaluates \eqref{timegeod} using expression \eqref{lorexpmap}
for the exponential map. For Lorentzian signature and vanishing
cosmological constant, this yields
\begin{align}
\label{minkfol} \Phi^{0,L}_T(z)=\left(\begin{array}{cc}iT\frac{1+|z|^2}{1-|z|^2} &  -T\frac{2iz}{1-|z|^2}\\
 T\frac{2i\bar z}{1-|z|^2} &
 -iT\frac{1+|z|^2}{1-|z|^2}\end{array}\right).
\end{align}
By comparing with \eqref{minkxemb}, we recover the formula
\eqref{minkhyp} which relates the disc model of hyperbolic space
to the hyperboloids $\hyp_{{1}/{T^2}}$ of curvature $1/T^2$. For
Lorentzian signature and $\Lambda>0$, we consider the associated
geodesic in the double cover $\widetilde{\text{AdS}}_\Lambda$ and
find that the parameters in \eqref{adsxemb} and the metric
\eqref{adsmet} take the form
\begin{align}
\label{adspars} &t_1=\tfrac{\cos
(\sqrt{\Lambda}T)}{\sqrt{\Lambda}}\quad
t_2=\tfrac{\sin(\sqrt{\Lambda}
T)}{\sqrt{\Lambda}}\frac{1+|z|^2}{1-|z|^2}\quad
x_1=\tfrac{\sin(\sqrt{\Lambda}
T)}{\sqrt{\Lambda}}\frac{2\text{Re}(z)}{1-|z|^2}\quad
x_2=\tfrac{\sin(\sqrt{\Lambda}
T)}{\sqrt{\Lambda}}\frac{2\text{Im}(z)}{1-|z|^2}\\
&d\bx^2=-(dt_1)^2-(dt_2)^2+(dx_1)^2+(dx_2)^2=-dT^2+\frac{4\sin^2(\sqrt{\Lambda}
T)}{\Lambda}\frac{|dz|^2}{(1-|z|^2)^2}.\nonumber
\end{align}
The surfaces $U_{\Lambda>0,L}(T)\subset
\mathbb{X}_{\Lambda>0,L}=\text{AdS}_\Lambda$ therefore have
constant curvature $-{\Lambda}/{\sin^2(\sqrt{\Lambda}T)}$. For
Lorentzian signature and $\Lambda<0$, the coordinates
parametrising ${dS}_\Lambda$ in \eqref{dsid0} and the metric are
given by
\begin{align}
\label{dspars} &x^0=\tfrac{\sinh
(\sqrt{|\Lambda|}T)}{\sqrt{|\Lambda|}}
\frac{1+|z|^2}{1-|z|^2}\quad x^1=-\tfrac{\sinh(
\sqrt{|\Lambda|}T)}{\sqrt{|\Lambda|}}\frac{2\text{Re}(z)}{1-|z|^2}\quad
x^2=-\tfrac{\sinh
(\sqrt{|\Lambda|}T)}{\sqrt{|\Lambda|}}\frac{2\text{Im}(z)}{1-|z|^2}\\
&x^3=\tfrac{\cosh (\sqrt{|\Lambda|}T)}{\sqrt|\Lambda|}\qquad
\nonumber
d\bx^2=-dT^2+\frac{4\sinh^2(\sqrt{|\Lambda|}T)}{|\Lambda|}\frac{|dz|^2}{(1-|z|^2)^2},
\end{align}
and the surfaces $U_{\Lambda<0,L}(T)$ have constant curvature
$-{|\Lambda|}/{\sinh^2(\sqrt{|\Lambda|}T)}$. For Euclidean
signature and $\Lambda<0$, the curvature of the surfaces
$U_{\Lambda<0,E}(T)$ is $-{\Lambda}/{\cosh^2(\sqrt{\Lambda}T)}$,
since the parameters in \eqref{hypid} and the metric
\eqref{hypdet} take the form
\begin{align}
\label{hyppars} &x^0=\tfrac{\cosh
(\sqrt{|\Lambda|}T)}{\sqrt{|\Lambda|}}
\frac{1+|z|^2}{1-|z|^2}\quad x^1=\tfrac{\cosh
(\sqrt{|\Lambda|}T)}{\sqrt{|\Lambda|}}\frac{2\text{Re}(z)}{1-|z|^2}\quad
x^2=\tfrac{\cosh (\sqrt{|\Lambda|}T)}{\sqrt{|\Lambda|}}\frac{2\text{Im}(z)}{1-|z|^2}\quad x^3=\tfrac{\sinh (\sqrt{|\Lambda|}T)}{\sqrt{|\Lambda|}}\nonumber\\
&d\bx^2=dT^2+4\frac{\cosh^2
(\sqrt{|\Lambda|}T)}{|\Lambda|}\frac{|dz|^2}{(1-|z|^2)^2}.
\end{align}
The cocompact Fuchsian group $\Gamma\subset PSU(1,1)$ acts on the
domains $U_{\Lambda,S}$ freely
 and properly discontinuously via the canonical inclusion $\i_{\Lambda,S}: PSU(1,1)\rightarrow
 \text{Isom}(\mathbb{X}_{\Lambda,S})$ induced by \eqref{canembedlor}.
It follows from the identity \eqref{phiid} that this action
preserves the surfaces
 $U_{\Lambda,S}(T)$ of constant cosmological time and agrees with the action induced by the identification of these surfaces with hyperbolic space.
 The static (2+1)-spacetimes $M^{st}_{\Lambda,S,\Gamma}$ associated
to $\Gamma$
 are given as the quotients of the domains $U_{\Lambda,S}$ by
 this
 action of $\Gamma$
\begin{align}
\label{statsp} M^{st}_{\Lambda,S,\Gamma}=U_{\Lambda,S}/\Gamma
=\bigcup_T U_{\Lambda,S}(T)/\Gamma.
\end{align}

\subsection{The construction of evolving (2+1)-spacetimes via grafting}

\label{graftingsect}

After discussing the static spacetimes associated to a cocompact
Fuchsian group $\Gamma$, we will now summarise the construction of
evolving (2+1)-spacetimes via grafting following the presentation
in  \cite{bb, bb2}.

\begin{figure}
\vskip .3in \protect\input epsf \protect\epsfxsize=12truecm
\protect\centerline{\epsfbox{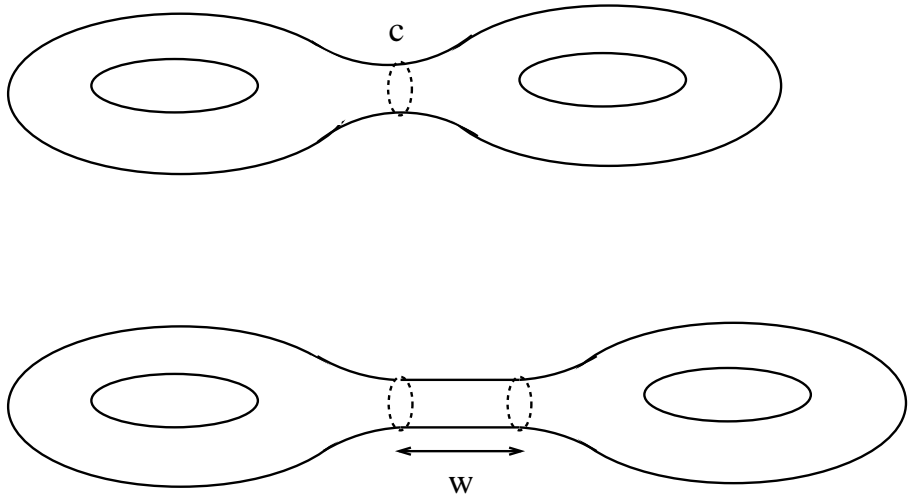}} \caption{Grafting along a
closed simple geodesic $c$ with weight $w$ on a genus 2 surface.}
\label{graft}
\end{figure}

Grafting along measured geodesic laminations is a method for
constructing two-surfaces. The simplest case are geodesic
laminations which are sets of non-intersecting closed, simple
geodesics on a two-surface. Grafting along closed, simple
geodesics was first investigated in the context of complex
projective structures and Teichm\"uller theory \cite{gol,hej,msk}.
General geodesic laminations were first considered by Thurston
\cite{th1, th2}, for historical remarks see for instance
\cite{mcmull}. The role of geodesic laminations in
(2+1)-dimensional gravity was first explored by Mess \cite{mess}
who investigated the characterisation of (2+1)-dimensional
spacetimes in terms of holonomies. More recent work on grafting in
the context of (2+1)-dimensional gravity  are the papers   by
Benedetti and Bonsante \cite{bb,bb2}, which relate the
construction of (2+1)-spacetimes via grafting for different values
of the cosmological constant.  The ingredients of the grafting
construction are a cocompact Fuchsian group $\Gamma$ and a
measured geodesic lamination on the associated two-surface $
\hyp_k/\Gamma$. In the following, we restrict attention to the
case where this geodesic lamination is a weighted multicurve on
$\hyp_k/\Gamma$, i.~e.~a set of non-intersecting closed, simple
geodesics $\eta_i$ on $\hyp_k/\Gamma$, each equipped with a weight
$w_i>0$.
\begin{align}
\label{multcurve} G=\{(\eta_i,w_i)\;|\; i\in I\}.
\end{align}
Geometrically, grafting along the multicurve \eqref{multcurve}
amounts to cutting the surface $\hyp_k/\Gamma$ along each geodesic
$c_i$, and inserting a strip of width $w_i$ as shown in
Fig.~\ref{graft}.

 In the construction of (2+1)-spacetimes via grafting, the
 grafting
 procedure is applied to each two-surface
$U_{\Lambda,S}(T)/\Gamma$ in  \eqref{statsp}. The construction is
performed on their universal covers, i.~e.~the constant
cosmological time surfaces $U_{\Lambda,S}(T)$, which are
identified with copies of hyperbolic space via \eqref{tembed} and
foliate the static domains $U_{\Lambda,S}\subset
\mathbb{X}_{\Lambda,S}$ as in \eqref{concfol}. The first step in
the grafting construction is to lift each geodesic $\eta_i$ in the
multicurve \eqref{multcurve} to a geodesic $c_{\eta_i}$ on the
universal cover $\hyp_k$. By acting on these geodesics with the
cocompact Fuchsian group $\Gamma$, one obtains a
$\Gamma$-invariant multicurve on $\hyp_k$
\begin{align}
\label{etamult} G^{\hyp_k}=\{ ( v c_{\eta_i},w_i) \;|\; i\in I,
v\in\Gamma\},
\end{align}
i.~e.~a set of non-intersecting geodesics on $\hyp_k$ with
associated weights $w_i>0$, which are mapped into each other by
the elements of $\Gamma$. Via the maps $\Phi^{\Lambda,S}_T:
\hyp\rightarrow U_{\Lambda,S}(T)\subset \mathbb{X}_{\Lambda,S}$ in
\eqref{tembed}, which identify hyperbolic space with the constant
cosmological time surfaces $U_{\Lambda,S}(T)$, one then obtains a
$\Gamma$-invariant set of non-intersecting geodesics  on each
surface $U_{\Lambda,S}(T)$.

Grafting along the multicurve \eqref{multcurve} assigns to each
surface $U_{\Lambda,S}(T)$ a deformed surface $U^G_{\Lambda,S}(T)$
constructed as follows.
 One selects a
basepoint $q_0\in \hyp$ outside of the geodesics in the multicurve
\eqref{multcurve} and considers the images
$\Phi^{\Lambda,S}_T(q_0)$ on the surfaces $U_{\Lambda,S}(T)$. One
then cuts each surface $U_{\Lambda,S}(T)$  along the images
$\Phi_T^{\Lambda,S}( v c_{\eta_i})$, $i\in I$, $v\in\Gamma$ of the
geodesics in the multicurve \eqref{etamult} on $U_{\Lambda,S}(T)$.
The resulting pieces which do not contain the images
 of the basepoint
 are then shifted away from the basepoint in
the direction determined by the geodesics' unit translation
vectors and by a distance given by the geodesic's weight. Finally,
one inserts strips, which connect the shifted pieces of each
constant cosmological time surface $U_{\Lambda,S}(T)$, and thus
obtains a connected deformed surface $U^G_{\Lambda,S}(T)$

The union of these deformed surfaces for all values of the
cosmological time $T$ then forms a simply connected regular domain
in $\mathbb{X}_{\Lambda,S}$
\begin{align}
U_{\Lambda,S}^G=\begin{cases} \bigcup_{T\in(0,\infty)} U_{0,L}^G(T) & \Lambda=0,\;\text{Lorentzian}\\
\bigcup_{T\in(0,\pi/\sqrt{\Lambda})} U_{\Lambda,L}^G(T) & \Lambda>0,\; \text{Lorentzian}\\
\bigcup_{T\in(0,\infty)} U_{\Lambda,L}^G(T) & \Lambda<0,
\;\text{Lorentzian }\\
\bigcup_{T\in(0,\infty)} U_{\Lambda,E}^G(T) &
\Lambda<0,\;\text{Euclidean}.
\end{cases}
\end{align}
Under the grafting construction, the initial singularity of the
static domains $U_{\Lambda,S}$ is mapped to a graph in
$\mathbb{X}_{\Lambda,S}$. It is shown in \cite{bb,bb2} that the
deformed surfaces $U_{\Lambda,S}^G(T)$ are surfaces of constant
geodesic distance $T$ from this graph and therefore again surfaces
of constant cosmological time $T$.

It is discussed in \cite{bb,bb2} that the cocompact Fuchsian group
$\Gamma$  acts on the grafted domain $U_{\Lambda,S}^G(T)$ via a
group homomorphism $h_{\Lambda,S}^G:\Gamma \rightarrow
\text{Isom}(\mathbb{X}_{\Lambda,S})$. This action is free and
properly discontinuous and preserves each surface
$U_{\Lambda,S}^G(T)$. Hence, by taking the quotient
$U_{\Lambda,S}^G(T)/h_{\Lambda,S}^G(\Gamma)$ of the deformed
constant cosmological time surfaces by this action of $\Gamma$ one
obtains a two-surface of genus $g$. The grafted spacetimes
$M_{\Lambda,S}^{\Gamma, G}$ associated to the cocompact Fuchsian
group $\Gamma$ and the multicurve \eqref{multcurve} on
$\hyp_k/\Gamma$ are then given as the union of these surfaces for
all values of the cosmological time or, equivalently, as the
quotient of the regular domains $U_{\Lambda,S}^G$ by this action
of $\Gamma$
\begin{align}
\label{grsptime}
M_{\Lambda,S}^{\Gamma,G}=U_{\Lambda,S}^G/h^G_{\Lambda,S}(\Gamma)=\bigcup_T
U_{\Lambda,S}^G(T)/h_{\Lambda,S}^G(\Gamma).
\end{align}

\begin{figure}
\vskip .3in \protect\input epsf \protect\epsfxsize=12truecm
\protect\centerline{\epsfbox{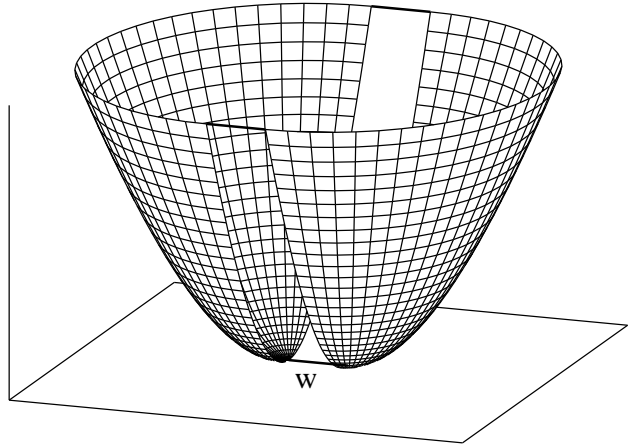}} \caption{Grafting along a
 geodesic with weight $w$ in hyperbolic space.} \label{hyper}
\end{figure}

The procedure is most easily visualised in Lorentzian
(2+1)-gravity with vanishing cosmological constant, where the
surfaces of constant cosmological time are the hyperboloids
$\hyp_{1/T^2}$ which foliate the interior of the forward
lightcone. Geodesics on the hyperboloids $\hyp_{1/T^2}$ are given
as the intersection of $\hyp_{1/T^2}$ with planes through the
origin, whose unit normal vector is the unit translation vector of
the geodesic given in \eqref{geodpar}. Cutting each surface
$U_{0,L}(T)$ along these geodesics therefore amounts to cutting
the interior of the forward lightcone along the associated planes.
The resulting pieces are then shifted away from the basepoint in
the direction of the plane's normal vector by a distance given by
the weight of the associated geodesic  as shown in
Fig.\ref{hyper}.
The strips connecting the different pieces of a surface
$U_{0,L}(T)$ are obtained by connecting the points the different
pieces of $U_{0,L}(T)$ which correspond to a single point on a
geodesic by straight lines.

For the other model spacetimes the construction is similar but its
description is more involved. As we will not need the details of
the construction, we refer the reader to the papers \cite{bb,bb2},
which give an explicit parametrisation of the resulting surfaces
and relate these surfaces for different values of the cosmological
constant. In the following, we will only make use of a formula for
the translation of the images $\Phi^{\Lambda,S}_T(x)\in
U_{\Lambda,S}(T)$ of points outside of the geodesics in the
multicurve \eqref{etamult}. The relative shift of such points
under the grafting construction is determined by their position
relative to the geodesics in \eqref{etamult} and given by a map
$B_{G,\Lambda,S}: \hyp\times
\hyp\rightarrow\text{Isom}(\mathbb{X}_{\Lambda,S})$.
 To determine the value of
$B_{G,\Lambda,S}(p,q)$ for two points $p,q\in \hyp$ outside the
geodesics in \eqref{etamult}, one connects them with a geodesic
$a_{pq}$ on $\hyp$ oriented towards $q$. One then determines the
geodesics in the multicurve \eqref{etamult} which intersect this
geodesic as well as the associated oriented intersection numbers.
It is shown in \cite{bb,bb2}, see in particular Sect.~4.2.1,
4.4.1, 4.6.1 and 4.7.2, that if these geodesics are labelled by
$c_i$, $i=1,\ldots, m$, such that the intersection point of
$a_{pq}$ with $c_i$ occurs before the one with $c_j$ for $i<j$ and
if $\epsilon_i$ are the associated oriented intersection numbers
with the convention $\epsilon_i=1$ if $c_i$ crosses $a_{pq}$ from
the left to the right, then the relative shift
$B_{G,\Lambda,S}(p,q)$ is given by\footnote{The factors
$\sqrt{\Lambda},\sqrt{|\Lambda|}$ are not present in
\cite{bb,bb2}, where only spacetimes with cosmological constant
$\Lambda\in\{0,\pm 1\}$ are considered. However, this
normalisation is suggested by the fact that the associated
spacelike geodesics should be parametrised by arc length.}

\begin{align}
\label{multmin} &B_{G,0,L}(p,q)=\sum_{i=1}^nm \epsilon_i
w_i\hat\bn_i \;\in\RR^3\subset PSU(1,1)\ltimes\RR^3\\
 &B_{G,\Lambda>0,L}(p,q)=(B^+_{G,\Lambda>0,L}(p,q),B^-_{G,\Lambda>0,L}(p,q))\in PSU(1,1)\times PSU(1,1)\nonumber\\
&\qquad\qquad\qquad\qquad B^\pm_{G,\Lambda>0,L}(p,q)= e^{ \pm
\sqrt{\Lambda}{\epsilon_1w_1} \hat n_1^a J_a^L}e^{{\pm
\sqrt{\Lambda}\epsilon_2w_2} \hat n_2^a J_a^L}\cdots e^{\pm
{\sqrt{\Lambda}\epsilon_nw_n}\hat n_m^a J_a^L}\nonumber\\
&B_{G,\Lambda<0,S}(p,q)=B_{G,\Lambda<0,E}(p,q)=e^{{i\sqrt{|\Lambda|}\epsilon_1w_1}\hat
n_1^a J_a^L}e^{i\sqrt{|\Lambda|}\epsilon_2w_2\hat n_2^a
J_a^L}\cdots e^{{i\sqrt{|\Lambda|}\epsilon_nw_n}\hat n_m^a
J_a^L}\in SL(2,\CC)/\mathbb{Z}_2,\nonumber
\end{align}
where $w_i$ is the weight of the geodesic $c_i$ and $\hat\bn_i$
its unit translation vector as defined  in \eqref{geodpar}. It is
shown in \cite{bb,bb2} that map $B_{G,\Lambda,S}: \hyp\times
\hyp\rightarrow\text{Isom}(\mathbb{X}_{\Lambda,S})$ satisfies the
identities
\begin{align}
\label{cocycprops} &B_{G,\Lambda,S}(p,q)\cdot
B_{G,\Lambda,S}(q,r)=B_{G,\Lambda,S}(p,r) & &\forall
p,q,r\in \hyp\\
 &B_{G,\Lambda,S}(v p,v q)=\i_{\Lambda,S}(v)\cdot
 B_{G,\Lambda,S}(p,q) \cdot \i_{\Lambda,S}(v)^\inv & &\forall p,q\in \hyp,\,v\in
 PSU(1,1)\nonumber,
\end{align}
which reflect the geometrical properties of the grafting
procedure. This allows one to define a group homomorphism
$h^G_{\Lambda,S}:\Gamma\rightarrow
\text{Isom}(\mathbb{X}_{\Lambda,S})$ from the cocompact Fuchsian
group $\Gamma$ into the isometry group of the model spacetime by
setting
\begin{align}
\label{glogroupact} h_{\Lambda,S}^G(v)=B_{G,\Lambda,S}(q_0, v
q_0)\cdot \i_{\Lambda,S}(v)\qquad\forall v\in\Gamma,
\end{align}
where $\i_{\Lambda,S}: PSU(1,1)\rightarrow
\text{Isom}(\mathbb{X}_{\Lambda,S})$ is the canonical embedding of
$PSU(1,1)$ into the isometry group of the model spacetime given by
\eqref{canembedlor} and $q_0\in\hyp$ the basepoint. It is
discussed in \cite{bb,bb2} that this group homomorphism defines a
free and properly discontinuous action of the group $\Gamma$ on
the grafted domains $U^G_{\Lambda,S}$ which  maps each surface
$U^G_{\Lambda,S}(T)$ to itself. Furthermore, for any two points
$\Phi^\Lambda_T(x)$, $\Phi^\Lambda_T(x') \in U_{\Lambda,S}(T)$
outside the geodesics which are related by the canonical action
\eqref{phiid} of an element $v\in\Gamma$, the corresponding points
on the grafted surface $U_{\Lambda,S}^G(T)$ are related by the
action of $v$ via \eqref{glogroupact}
\begin{align}
\label{pointident} &\Phi_{\Lambda,S}^T(x')=\i_{\Lambda,S}(v)\cdot
\Phi_{\Lambda,S}^T(x)\\
\Rightarrow \quad &B_{G,\Lambda,S}(q_0,x')
\Phi_{\Lambda,S}^T(x')=h_{\Lambda,S}^G(v) \cdot
B_{G,\Lambda,S}(q_0,x) \Phi_{\Lambda,S}^T(x).\nonumber
\end{align}
The quotient \eqref{grsptime} of the domains
$U^G_{\Lambda,S}\subset\mathbb{X}_{\Lambda,S}$ by this action of
$\Gamma$ is therefore well-defined and gives rise to a spacetime
of topology $\RR\times S_g$.


\section{(2+1)-dimensional gravity: the Chern-Simons formulation}
\label{CSsect}

\subsection{(2+1)-dimensional gravity as a Chern-Simons gauge theory}
\label{CSintro}

 The absence of local gravitational degrees of freedom in
(2+1)-dimensional gravity allows one to formulate the theory as a
Chern-Simons gauge theory \cite{AT, Witten1}. The Chern-Simons
formulation of (2+1)-dimensional gravity is derived from Cartan's
description, in which a spacetime manifold $M$ is characterised in
terms of a dreibein of one forms $e_a$, $a=0,1,2$, and spin
connection one-forms $\omega_a$, $a=0,1,2$, on $M$. The metric on
$M$ is given by the dreibein
\begin{align}
\label{metdreib} g=\eta_S^{ab} e_a\otimes e_b,
\end{align}
where $\eta^{ab}_S$ denotes the Minkowski metric \eqref{minkmet}
or the Euclidean metric \eqref{euclmet}., while the one-forms
$\omega_a$ are the coefficients of the spin connection $\omega =
\omega^a J^S_a$. Einstein's equations of motion then take the form
of
 the requirements of vanishing torsion and constant curvature
\begin{align}
\label{notor} &T^a = de^a+ \epsilon^a_{\;\;bc} \omega^b
e^c=0\qquad F_\omega^a = d\omega^a  + \frac{1}{2}
\epsilon^a_{\;\;bc} \omega^b\wedge \omega^c =-\tfrac{\Lambda}{2}
\epsilon^a_{\;\;bc}e^b\wedge e^c.
\end{align}
To obtain the Chern-Simons formulation of (2+1)-dimensional
gravity, one combines dreibein and spin connection into the
 Cartan connection \cite{Sharpe} or
Chern-Simons gauge field
\begin{align} \label{Cartan} A = \omega^a J^{S}_a + e^a P^{S}_a,
\end{align}
where $J_a^{S}$, $P_a^{S}$, $a=0,1,2$, denote the generators of
the six-dimensional Lie algebras $\gothh_{\Lambda,S}$ with bracket
\eqref{liealg},\eqref{liealge}. Hence, depending on the signature
and the cosmological constant, the Chern-Simons gauge field is
 a one-form on $M$ with values in the Lie algebra
$\gothh_{\Lambda,S}$. The choice of the Lie algebra determines the
gauge group of the associated Chern-Simons theory up to coverings,
and in the following, we will take  the isometry groups
$\text{Isom}(\mathbb{X}_{\Lambda,S})$ of the associated model
spacetimes as the gauge groups.

When expressed in terms of the one-form \eqref{Cartan},  the
Einstein-Hilbert action in Cartan's formulation of the theory
takes the form of a Chern-Simons action
\begin{align}
\label{CSact0} S_{CS}[A]=\int_M\langle A\wedge dA+\tfrac{2}{3}
A\wedge A\wedge A\rangle,
\end{align}
where $\langle\,,\,\rangle$ is an $\Ad$-invariant, non-degenerate
bilinear form on the Lie algebra  $\gothh_{\Lambda,S}$ given by
\begin{align}
\label{inprod} &\langle
J_a^S,P^S_b\rangle=\eta^S_{ab}\qquad\langle
J^S_a,J^S_b\rangle=\langle P^S_a,P^S_b\rangle=0.
\end{align}
The equations of motion derived from \eqref{CSact0} are a flatness
condition on the gauge field
\begin{align}
\label{flatcond} F=dA+A\wedge A=0,
\end{align}
which combines the requirements \eqref{notor} of vanishing torsion
and constant curvature
 \begin{align}
\label{decomp} F=  T^a  P^S_a + (F_\omega^a+\tfrac{\Lambda}{2}
\epsilon^a_{\;\;bc} e^b\wedge e^c)\,J^S_a.
\end{align}
The Chern-Simons action \eqref{CSact0} is invariant under
Chern-Simons gauge transformations
\begin{align}
\label{CSgt1} A\mapsto \gamma A\gamma^\inv +\gamma d\gamma^\inv
\qquad\gamma: M\rightarrow \text{Isom}(\mathbb{X}_{\Lambda,S}).
\end{align}
It has been shown by Witten \cite{Witten1} that  infinitesimal
Chern-Simons gauge transformations are on-shell equivalent to
infinitesimal diffeomorphisms. The space of metrics solving
Einsteins's equation modulo infinitesimally generated
diffeomorphisms  is therefore isomorphic to the space of flat
Chern-Simons gauge fields modulo infinitesimally generated
Chern-Simons gauge transformations.

Note, however, that some caution should be applied when
identifying the phase space of (2+1)-dimensional gravity in its
geometrical formulation with the phase space of the associated
Chern-Simons theory. First, the equivalence between
diffeomorphisms and Chern-Simons gauge transformations does not
hold for large diffeomorphisms, which are not infinitesimally
generated, and for the large gauge transformations arising in
Chern-Simons theory with non-simply connected gauge groups.
Second, in order to define a metric of Lorentzian or Euclidean
signature via \eqref{metdreib}, the dreibein $e_a$ has to be
non-degenerate, which is not required in the Chern-Simons
formalism. It is discussed in \cite{Matschull2} for the Lorentzian
case with vanishing cosmological constant  and spacetimes
containing particles that this  leads to differences in the global
structure of the phase spaces. A similar result for a spacetime
with three particles is derived in \cite{bb2}, Sect.~4.9, where it
is argued that such problems arise generically when the spacetime
is of topology $\RR\times S$, where $S$ is a non-compact
two-surface. However, as this paper restricts attention to
spacetimes with compact spatial surfaces and is mainly concerned
with the local properties of the phase space, we will not address
these issues in the following.

On the spacetime manifolds of topology $M\cong\RR\times S_g$
considered in this paper, it is possible to give a Hamiltonian
formulation of the theory. For this, one introduces coordinates
$x^0,x^1,x^2$ on $M\approx\RR\times S_g$ such that $x^0$
parametrises $\RR$ and $x^1,x^2$ are coordinates on $S_g$ and
splits the gauge field \eqref{Cartan} as
\begin{align}
\label{gfsplit} A=A_0 dx^0 +A_S,
\end{align}
where $A_0:\RR\times S_g\rightarrow \gothh_{\Lambda,S}$ is a
function with values in the Lie algebras $\gothh_{\Lambda,S}$ and
$A_S$ a gauge field on $S_g$. The Chern-Simons action
 \eqref{CSact0}on $M$ then takes the form
\begin{align} \label{CSact}
 S[A_S,A_0]=
\int_\RR dx^0\int_{S_g} \tfrac{1}{2}\langle\partial_0 A_S\wedge
A_S\rangle +\langle A_0\,,\, F_S \rangle,\end{align} where $F_S$
is the curvature of the spatial gauge field $A_S$
\begin{align} \label{curvv} F_S=d_S A_S + A_S\wedge A_S \end{align} with $d_S$
denoting differentiation on the surface $S_g$. The function $A_0$
plays the role of a Lagrange multiplier. Varying it leads to the
flatness constraint \begin{align} \label{spaceflat} F_S=0,
\end{align} while variation of $A_S$ results in the evolution equation
\begin{align}
 \partial_0 A_S=d_SA_0+[A_S,A_0].\label{asvary}\end{align}
The phase space of the theory is therefore the moduli space
$\mathcal{M}_g^{\text{Isom}(\mathbb{X}_{\Lambda,S})}$ of flat
$\text{Isom}(\mathbb{X}_{\Lambda,S})$-connections $A_S$ modulo
gauge transformations on the spatial surface $S_g$.

\subsection{Trivialisation and holonomies}
\label{trivembed}

As discussed in Sect.~\ref{geomsect}, the absence of local
gravitational degrees of freedom in (2+1)-dimensional gravity
implies that each (2+1)-spacetime is locally isometric to one of
the model spacetimes $\mathbb{X}_{\Lambda,S}$. In the Chern-Simons
formalism, this absence of local degrees of freedom manifests
itself in the fact that gauge fields solving the equations of
motions are flat and can be trivialised, i.~e.~written as pure
gauge on any simply connected region $R\subset \RR\times S_g$
\begin{align}
\label{triv} A=\gamma d\gamma^\inv\qquad \gamma: R\rightarrow
\text{Isom}(\mathbb{X}_{\Lambda,S}).
\end{align}
Given a function $\gamma: R\rightarrow
\text{Isom}(\mathbb{X}_{\Lambda,S})$ which trivialises a flat
gauge field $A$ on $R$, the associated functions
$\gamma(x^0,\cdot)$ trivialise the corresponding flat spatial
gauge fields $A_s$ for all values of  $x^0$
\begin{align}
A_S(x^0,\cdot)=\gamma(x^0,\cdot)d_S\gamma^\inv(x^0,\cdot).
\end{align}
To simplify notation, we will often neglect the dependence on the
parameter $x^0$ in the following and denote this function also by
$\gamma$.

\begin{figure}
\vskip .3in \protect\input epsf \protect\epsfxsize=12truecm
\protect\centerline{\epsfbox{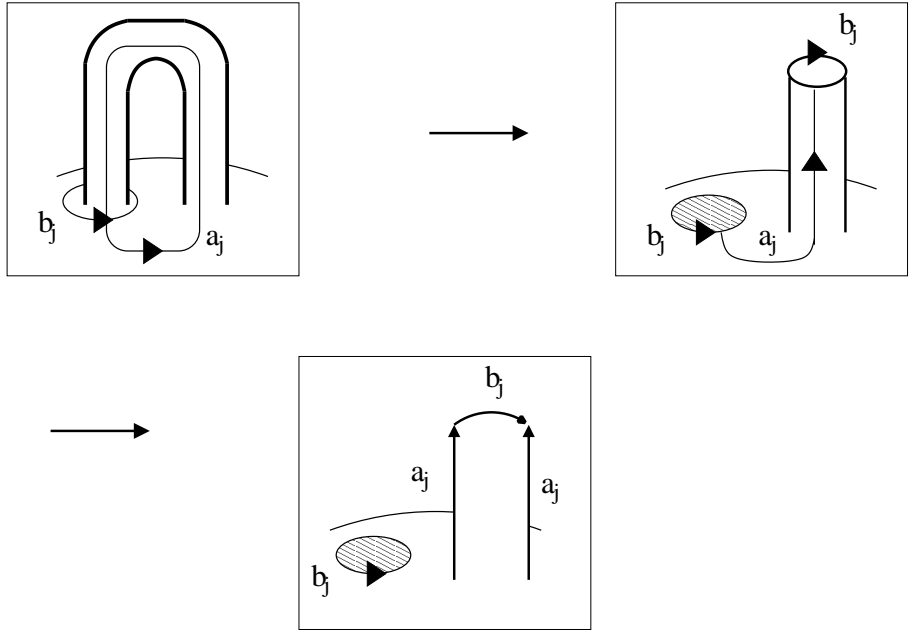}} \caption{Cutting the
surface $S_g$ along the generators of $\pi_1(S_g)$}
\label{cutting}
\end{figure}

\begin{figure}
\vskip .3in \protect\input epsf \protect\epsfxsize=12truecm
\protect\centerline{\epsfbox{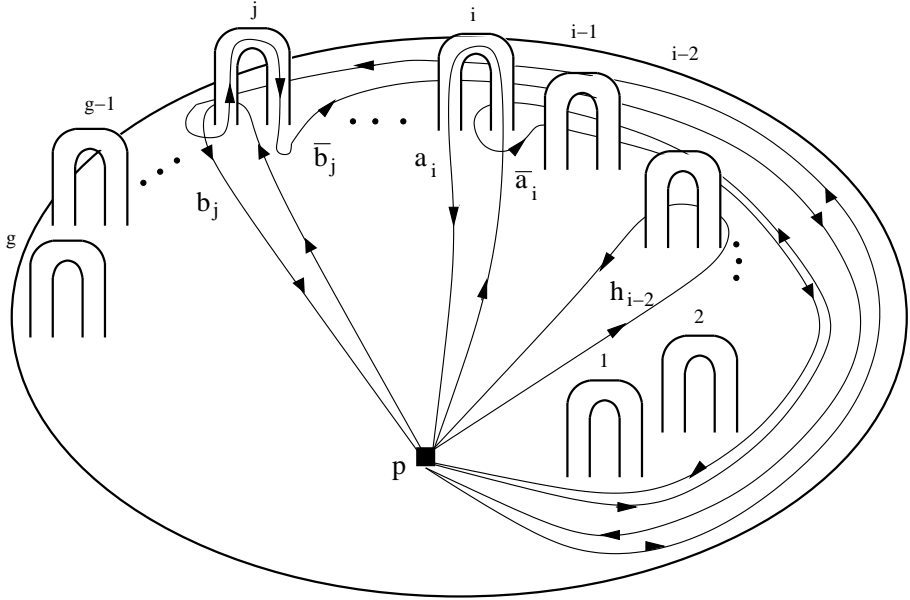}} \caption{Generators and
dual generators of the fundamental group $\pi_1(S_g)$}
\label{pi1gr}
\end{figure}

\begin{figure}
\vskip .3in \protect\input epsf \protect\epsfxsize=12truecm
\protect\centerline{\epsfbox{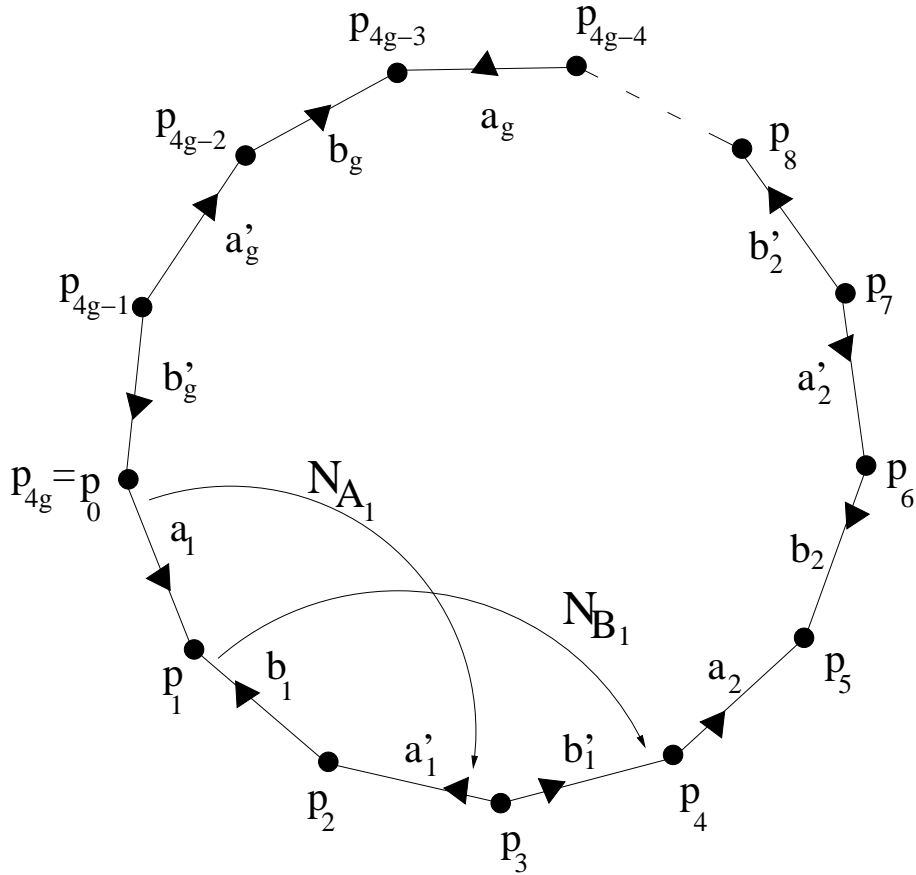}} \caption{The polygon
$P_g$} \label{poly1}
\end{figure}

 A maximal simply connected region in $\RR\times S_g$ is obtained by cutting the
spatial surface $S_g$ along a set of generators of the fundamental
group $\pi_1(S_g)$ as  in Fig.~\ref{cutting}.
As discussed in Sect.~\ref{hypdefs}, the fundamental group of a
genus $g$ surface $S_g$ is isomorphic to a cocompact Fuchsian
group with $2g$ generators, which are subject to a single defining
relation
\begin{align}
\label{fundgroup} \pi_1(S_g)=\langle a_1,b_1,\ldots, a_g,b_g\,;\,
[b_g,a_g^\inv]\cdots[b_1,a_1^\inv]=1\rangle\qquad
[b_i,a_i^\inv]=b_i\circ a_i^\inv\circ b_i^\inv \circ a_i.
\end{align}
Throughout the paper, we work with a fixed system of generators
 $a_i, b_i$, $i=1,\ldots,g$, which are the homotopy equivalence classes of two loops around each handle and based at a point $p\in S_g$
 as shown in Fig.~\ref{pi1gr}.
Cutting  the surface along each of the curves representing these
generators results in a $4g$-gon $P_g$ pictured in
Fig.~\ref{poly1}.

As discussed by Alekseev and Malkin \cite{AMII}, a function
$\gamma: P_g\rightarrow \text{Isom}(\mathbb{X}_{\Lambda,S})$ on
$P_g$ defines a flat gauge field on $S_g$ if and only if it
satisfies an overlap condition relating its value on the two sides
 which correspond to a given generator of the fundamental group. For
any $y\in\{a_1,b_1,\ldots, a_g,b_g\}$, $y'\in\{a'_1,b'_1,\ldots,
a'_g,b'_g\}$ one must have
\begin{align}
\label{csident} & A_S|_{y'}=\gamma d_S\gamma^\inv|_{y'}=\gamma
d_S\gamma^\inv|_{y}=A_S|_y,\end{align} which is the case if and
only if there exist constant elements $N_{Y}\in
\text{Isom}(\mathbb{X}_{\Lambda,S})$ such that
\begin{align}
\label{csident2} &\gamma^\inv|_{y'}=N_{Y} \gamma^\inv|_{y}.
\end{align}
The elements $N_Y$, $Y\in\{A_1,B_1,\ldots,A_g,B_g\}$ are the
Chern-Simons analogue of the group isomorphisms
\eqref{glogroupact} in the geometrical formulation. They
 contain all
 information about the physical state and are
closely related to the holonomies $A_i,B_i$ along the generators
$a_i,b_i$ of the fundamental group. These holonomies  are given by
the value of the trivialising function on the corners of the
polygon $P_g$ \cite{AMII}
\begin{align}
\label{polyhols}
&\ai\!=\!\gamma(p_{4i-3})\gamma(p_{4i-4})^\inv\!\!=\!\gamma(p_{4i-2})\gamma(p_{4i-1})^\inv\;\;\bi\!=\!\gamma(p_{4i-3})\gamma(p_{4i-2})^\inv\!\!\!=\gamma(p_{4i})\gamma(p_{4i-1})^\inv\!\!\!\!,
\end{align}
and satisfy a single relation arising from the defining relation
of the fundamental group $\pi_1(S_g)$
\begin{align}
\label{holconst} [B_g,A_g^\inv]\cdots[B_1,A_1^\inv]\approx 1\qquad
[B_i,A_i^\inv]=B_i\cdot \ai^\inv\cdot \bi^\inv \cdot\ai.
\end{align}
Via the overlap condition \eqref{csident}, one can relate the
value of the trivialising function $\gamma$ at the corners of the
polygon $P_g$ to its value at a given corner $p_0$
\begin{align}
\label{cornergf} &\gamma^\inv(p_{4i})= N_{H_i}N_{H_{i-1}}\cdots
N_{H_1}\gamma^\inv(p_0)=\gamma^\inv(p_0)H_1^\inv\cdots
H_{i-1}^\inv H_i^\inv\\
&\gamma^\inv(p_{4i+1})= N_{A_{i+1}}^\inv N_{B_{i+1}}^\inv
N_{A_{i+1}} N_{H_i}\cdots
N_{H_1}\gamma^\inv(p_{0})=\gamma^\inv(p_0)H_1^\inv\cdots
H_{i-1}^\inv H_i^\inv A_{i+1}^\inv
\nonumber\\
&\gamma^\inv(p_{4i+2})=  N_{B_{i+1}}^\inv N_{A_{i+1}}
N_{H_i}\cdots
N_{H_1}\gamma^\inv(p_{0})=\gamma^\inv(p_0)H_1^\inv\cdots
H_{i-1}^\inv H_i^\inv A_{i+1}^\inv B_{i+1}\nonumber\\
&\gamma^\inv(p_{4i+3})=  N_{A_{i+1}} N_{H_i}\cdots
N_{H_1}\gamma^\inv(p_{0})=\gamma^\inv(p_0)H_1^\inv\cdots
H_{i-1}^\inv H_i^\inv A_{i+1}^\inv B_{i+1} A_{i+1}\nonumber\\
&H_i=[B_i,A_i^\inv] \qquad N_{H_i}=[N_\bi, N_\ai^\inv],\nonumber
\end{align}
which allows one to express the holonomies $A_i,B_i$ along the
generators $a_i,b_i\in\pi_1(S_g)$  in terms of the Poincar\'e
elements $N_\ai,N_\bi$ in the overlap condition \eqref{csident}
and vice versa
\begin{align}
\label{holexp3} &A_i= \gamma(p_0)N_{H_1}^\inv\cdots
N_{H_{i-1}}^\inv N_{H_i}^\inv \cdot
N_{\bi}\cdot N_{H_{i-1}}\cdots N_{H_1}\gamma^\inv(p_0)\\
&\bi=\gamma(p_0)N_{H_1}^\inv\cdots N_{H_{i-1}}^\inv
N_{H_i}^\inv\cdot  N_{\ai}\cdot N_{H_{i-1}}\cdots
N_{H_1}\gamma^\inv(p_0)\nonumber\\
 \label{holexp4}
&N_{A_i}\!\!=\!\!\gamma^\inv(p_0) H_1^\inv\!\!\!\!\!\cdots\!
H_{i-1}^\inv H_i^\inv \bi H_{i-1}\!\cdots\! H_1\gamma(p_0)\\
&N_{B_i}\!\!=\!\!\gamma^\inv(p_0) H_1^\inv\!\!\!\!\!\cdots\!
H_{i-1}^\inv H_i^\inv \ai H_{i-1}\!\cdots\!
H_1\gamma(p_0).\nonumber
\end{align}
Up to conjugation with the value $\gamma(p_0)$ of the trivialising
function at the basepoint, the expressions \eqref{holexp3} and
\eqref{holexp4} relating the holonomies $\ai,\bi$ and the group
elements $N_\ai,N_\bi$ are of the same form. This reflects the
fact that, up to conjugation with $\gamma(p_0)$, the elements
$N_\ai, N_\bi$ are the holonomies along another system of
generators $\dca_i, \dcb_i\in\pi_1(S_g)$ pictured in
Fig.~\ref{pi1gr} and given by
\begin{align}
\label{dualgens} &\dca_i\!=\! h_1^\inv\circ...\circ h_i^\inv\circ
b_i\circ h_{i-1}\circ...\circ h_1\qquad \dcb_i\!=\!
h_1^\inv\circ...\circ h_i^\inv\circ a_i\circ h_{i-1}\circ... \circ
h_1.
\end{align}
These generators are  investigated in detail in \cite{ich2}, where
it is shown that their representatives can be viewed as a dual
graph for the curves representing  $a_i,b_i\in\pi_1(S_g)$ and that
they can be used to determine the intersection points of a general
embedded curve on $S_g$ with the generators $a_i,b_i$. In the
following we will therefore refer to the generators
$\dca_i,\dcb_i$ as the dual generators.

As the elements $N_\ai,N_\bi\in
\text{Isom}(\mathbb{X}_{\Lambda,S})$ in the overlap condition or,
equivalently, the holonomies $\ai,\bi\in
\text{Isom}(\mathbb{X}_{\Lambda,S})$ contain all information about
the physical state, they can be used to parametrise the phase
space of the theory. Taking into account that these variables are
subject to a constraint \eqref{holconst} and that gauge
transformations on the surface $S_g$
 act on the holonomies $A_i,B_i$ by simultaneous conjugation, one
 finds that  the moduli space of flat $\text{Isom}(\mathbb{X}_{\Lambda,S})$-connections on $S_g$ is given as the quotient
 \begin{align}
 \label{modpar}
 \mathcal{M}_g^{\text{Isom}(\mathbb{X}_{\Lambda,S})}\!\!\!\!\!\!\!\!=\{(A_1,B_1,\ldots, A_g,B_g)\in \text{Isom}(\mathbb{X}_{\Lambda,S})^{2g}\;|\;
 [B_g,A_g^\inv]\cdots[B_1,A_1^\inv]=1\}/\text{Isom}(\mathbb{X}_{\Lambda,S}).
 \end{align}

\subsection{Phase space and Poisson structure}

The advantage of the Chern-Simons formulation of (2+1)-dimensional
gravity is that it allows one to give a rather simple description
of the Poisson structure on the phase space, which is based on its
parametrisation \eqref{modpar} in terms of the holonomies along a
set of generators $a_i,b_i\in\pi_1(S_g)$  \cite{AMII, FR}. In the
following we will use the formalism by Fock and Rosly \cite{FR},
which is defined for Chern-Simons theory with a general gauge
group $H$ and parametrises the Poisson structure on the moduli
space $\mathcal{M}_g^H$  in terms of an auxiliary Poisson
structure on the manifold $H^{2g}$. The description is summarised
in the following theorem.

\begin{theorem}(Fock,Rosly \cite{FR})
\label{frtheorem}

Consider Chern-Simons theory with gauge group $H$ on a manifold
$\RR\times S_g$. Denote by $T_a$, $a=1,\ldots,\text{dim}\;H$, a
basis of the Lie algebra $\gothh=\text{Lie}\;H$ and by $t_{ab}$
the matrix representing the $\Ad$-invariant symmetric bilinear
form in the Chern-Simons action \eqref{CSact0} with respect to
this basis
\begin{align}
\label{frpair} t_{ab}=\langle T_a,T_b\rangle\qquad
t^{ab}t_{bc}=\delta^a_c.
\end{align}
Let $r=r^{ab}T_a\otimes T_b$ be  a classical $r$-matrix for the
gauge group $H$, i.~e.~an element $r\in \gothh\otimes\gothh$ which
satisfies the classical Yang Baxter equation (CYBE)
\begin{align}
\label{cybe} &[[r,r]]=[r_{12},r_{13}]+[r_{12},r_{23}]+[r_{13},
r_{23}]=0\\
& r_{12}=r^{ab} T_a\otimes T_b\otimes 1,\; r_{13}=r^{ab}
T_a\otimes 1\otimes  T_b, r_{23}=r^{ab}1\otimes T_a\otimes
T_b,\nonumber
\end{align}
and whose symmetric part is dual to the bilinear form
\eqref{frpair}
\begin{align}
&r=r_{(s)}+r_{(a)}\qquad
r_{(a)}=\tfrac{1}{2}(r^{ab}-r^{ba})T_a\otimes T_b\;,\quad
r_{(s)}=\tfrac{1}{2}(r^{ab}+r^{ba})T_a\otimes
T_b=\tfrac{1}{2}t^{ab}T_a\otimes T_b.
\end{align}
Consider the manifold $H^{2g}$, where  the different copies of $H$
are identified with the holonomies $\ai,\bi\in H$ along a set of
generators of the fundamental group $\pi_1(S_g)$ and denote by
$R_a^{X}$, $L_a^{X}$, $X\in\{A_1,B_1,\ldots, A_g,B_g\}$ the
left-and right-invariant vector fields associated to a basis of
$\gothh$ and the different components of $H^{2g}$
\begin{align}
&L_a^Xf(A_1,...,B_g)=\frac{d}{dt}|_{t=0}f(..., e^{-tT_a}\cdot
X,...)\quad &R_a^Xf(A_1,...,B_g)=\frac{d}{dt}|_{t=0}f(..., X\cdot
e^{tT_a},...).
\end{align}
Then, the bivector $B\in \text{Vec}(H)\otimes \text{Vec}(H)$
\begin{align}
\label{frbivect2} B=&r^{ab}_{(a)}\!\!\left(\!\sum_{j=1}^g
R^{\aj}_a\!+\!L^{\aj}_a\!+\!R^{\bjj}_a\!+\!L^{\bjj}_a\right)\!\!\otimes\!\!\left(\sum_{j=1}^gR^{\aj}_b\!+\!L^{\aj}_b\!+\!R^{\bjj}_b\!+\!L^{\bjj}_b\right)\nonumber\\
+&\tfrac{1}{2}t^{ab}\!\!\!\!\sum_{i,j=1,\;i<j}^g
(R^{\ai}_a\!+\!L^{\ai}_a\!+\!R^{\bi}_a\!+\!L^{\bi}_a)\wedge(R^{\aj}_b\!+\!L^{\aj}_b\!+\!R^{\bjj}_b\!+\!L^{\bjj}_b)\nonumber\\
+& \tfrac{1}{2}t^{ab}\sum_{i=1}^g R^\ai_a\wedge(R^\bi_b+L^\ai_b
+L^\bi_b)+R^\bi_a\wedge (L^\ai_b+L^\bi_b)+L^\ai_a\wedge L^\bi_b,
\end{align}
 defines a Poisson structure on $H^{2g}$.
After imposing the constraint \eqref{holconst} and dividing by the
associated gauge transformations, which act by simultaneous
conjugation of all components with $H$, this Poisson structure
agrees with the canonical Poisson structure on the moduli space
\begin{align}
\mathcal{M}^H_g=\{(A_1,B_1,\ldots,A_g,B_g)\in H^{2g}\;|\; [B_g,
A_g^\inv]\cdots[B_1,A_1^\inv]=1\}/ H.
\end{align}
\end{theorem}
In Fock and Rosly's formalism, physical observables are given by
functions on the manifold $H^{2g}$ which are invariant under
simultaneous conjugation of all components with the gauge group
$H$. Note that the Poisson bracket of such observables with a
general function $g\in\cif(H^{2g})$ does not depend on the
particular choice of the classical $r$-matrix but only on the
matrix $t^{ab}$ representing the $\Ad$-invariant, symmetric
bilinear form in the Chern-Simons action. As the component of the
bivector \eqref{frbivect2} which depends on the antisymmetric
component $r_{(a)}$ is proportional to terms of the form
$\sum_{i=1}^g R^\ai_a+L^\ai_a+R^\bi_a+L_\bi^a$, this contribution
vanishes if one of the function is invariant under simultaneous
conjugation of its arguments with $H$.

A particular set of physical observables, in the following
referred to as Wilson loop observables, are conjugation invariant
functions of the holonomies along closed  curves on $S_g$. As the
equations of motion are a flatness condition on the gauge field,
these observables do not depend on the curve itself but only on
its homotopy equivalence class in $\pi_1(S_g)$ and are invariant
under a change of the basepoint. In Fock and Rosly's formalism,
these observables are described by expressing the holonomy along
an element $\eta\in\pi_1(S_g)$ as a
 product in the holonomies along the generators
$a_i,b_i\in\pi_1(S_g)$
\begin{align}
\label{holdecomp} &\eta=x_r^{\alpha_r}\cdots x_1^{\alpha_1}\,,\;
x_k\in\{a_1,\ldots,b_g\},\;\alpha_k\in\{\pm
1\}\quad\Rightarrow\quad
 &H_\eta=X_r^{\alpha_r}\cdots X_1^{\alpha_1}.
\end{align}
The Wilson loop observable $f_\eta\in\cif(H^{2g})$ associated to
$\eta\in\pi_1(S_g)$ and a general conjugation invariant function
$f\in\cif(H)$ on the gauge group is then given by
\begin{align}
\label{wloopdef} f_\eta:(A_1,\ldots,B_g)\mapsto f(H_\eta)
\end{align}
with $H_\eta$ given as a product in the elements $\ai,\bi$ and
their inverses as in \eqref{holdecomp}. It follows directly that
the Wilson loop observables are invariant under simultaneous
conjugation of all holonomies $A_i,B_i$ with elements of $H$ and
satisfy
\begin{align}
f_{\tau\circ\eta\circ\tau^\inv}=f_\eta\qquad \forall
\eta,\tau\in\pi_1(S_g).
\end{align}
In order to apply Fock and Rosly's description \cite{FR} of phase
space and Poisson structure  to the Chern-Simons formulation of
(2+1)-dimensional gravity, one needs classical $r$-matrices for
 the Lie algebras $\gothh_{\Lambda,S}$
such that the  symmetric components of these $r$-matrices agree
with the $\Ad$-invariant, symmetric bilinear forms \eqref{inprod}.
For Lorentzian signature, such a classical $r$-matrix is given by
\begin{align}
\label{rmatmink} r=P^L_a\otimes J_L^a+
n_a\epsilon^{abc}J^L_b\otimes J^L_c\
\end{align} with a constant vector $\bn=(n^0,n^1,n^2)\in\RR^3$
satisfying $\eta^L(\bn,\bn)=\Lambda$. The corresponding $r$-matrix
for the Euclidean case with $\Lambda<0$ has the form
\begin{align}
\label{rmateucl} r^E=P^E_a\otimes J_E^a+
n_a\epsilon^{abc}J^E_b\otimes J^E_c
\end{align}
with a constant vector  $\bn=(n_0,n_1,n_2)\in\RR^3$ satisfying
$\eta^E(\bn,\bn)=|\Lambda|$.
 Note that the choice of the classical $r$-matrix and hence
the Poisson structure \eqref{frbivect2} is not necessarily unique
- for a list of classical $r$-matrices for the (2+1)-dimensional
Poincar\'e algebra see \cite{Stachura}. However, in the following
we will only consider Poisson brackets where at least one of the
functions is a Wilson loop observable so that our results do not
depend on this choice.

\section{Trivialisation and embedding}
 \label{trivsect}

As discussed in Sect.~\ref{geomsect}
 and Sect.~\ref{trivembed},
 the absence of local gravitational degrees of freedom  manifests itself in the geometrical and the Chern-Simons description
of (2+1)-spacetimes in, respectively, in the embedding of simply
connected regions into the model spacetimes
$\mathbb{X}_{\Lambda,S}$ and in the trivialisation of the
Chern-Simons gauge field. In this section, we discuss the relation
between these concepts and show how the embedding of spacetime
regions can be constructed from the function trivialising the
Chern-Simons gauge field.

We consider a simply connected region $R\subset \RR\times S_g$ in
the spacetime manifold with a metric $g$ and a flat Chern-Simons
gauge field $A$, related to the metric via \eqref{metdreib}. We
denote by   $\Bx_{\Lambda,S}: R\rightarrow \mathbb{X}_{\Lambda,S}$
the embedding into the model spacetime $\mathbb{X}_{\Lambda,S}$
and by $\gamma: R\rightarrow \text{Isom}(\mathbb{X}_{\Lambda,S})$
a function which trivialises the gauge field as in \eqref{triv}.
The decomposition \eqref{Cartan} of the gauge field in terms of
the generators $J_a^S,P_a^S$ then implies that the dreibein $e_a$
and the spin connection  $\omega_a$ are given by
\begin{align}
\label{trivdrsp} e_a=\langle\gamma d \gamma^\inv,
J^{S}_a\rangle\qquad \omega_a=\langle\gamma d\gamma^\inv,
P_a^{S}\rangle,
\end{align}
where $J_a^{S}$, $P_a^{S}$ denote the generators  of the Lie
algebras \eqref{liealg}, \eqref{liealge}  and
$\langle\,,\,\rangle$ the $\Ad$-invariant bilinear form
\eqref{inprod} in the Chern-Simons action. The expression
\eqref{metdreib} for the metric in terms of the dreibein relating
 then implies that the metric $g$ on $R$ takes the form
\begin{align} \label{metcond} &g=\eta^S_{ab}e^a\otimes
e^b=\eta_S^{ab}\langle\gamma d\gamma^\inv\!\!\!, J_a^S\rangle
\langle\gamma d\gamma^\inv\!\!\!, J^S_b\rangle=-4\det(
\eta_S)\det(\langle \gamma
 d\gamma^\inv, J_S^a\rangle J_a^S).
\end{align}

On the other hand, the metric $g$ must agree with the pull-back of
the metric in the model spacetime $\mathbb{X}_{\Lambda,S}$ via the
embedding $\Bx_{\Lambda,S}$. To relate the trivialising function
$\gamma:R\rightarrow\text{Isom}(\mathbb{X}_{\Lambda,S})$ to the
embedding $\Bx_{\Lambda,S}: R\rightarrow\mathbb{X}_{\Lambda,S}$,
one therefore has to construct a function $\Pi_{\Lambda,S}:
\text{Isom}(\mathbb{X}_{\Lambda,S})\rightarrow
\mathbb{X}_{\Lambda,S}$ from the isometry group into the model
spacetime such that the pull-back of the metric in the model
spacetime via $\Pi_{\Lambda,S}\circ\gamma^\inv$ agrees with the
metric \eqref{metcond}
\begin{align}
\label{geommet} d(\Bx_{\Lambda,S}\circ\gamma^\inv)^2=\eta_S^{ab}
\langle\gamma d\gamma^\inv\!\!\!, J_a^S\rangle \langle\gamma
d\gamma^\inv\!\!\!, J^S_b\rangle.
\end{align}
Furthermore, as the embedding of the region $R$ into the model
spacetime $\mathbb{X}_{\Lambda,S}$ is only defined up to a global
action of the isometry group
$\text{Isom}(\mathbb{X}_{\Lambda,S})$, two embeddings related by
such an action of the isometry group should correspond to the same
gauge field on $R$. This the case if and only if the action of
$\text{Isom}(\mathbb{X}_{\Lambda,S})$ on $\mathbb{X}_{\Lambda,S}$
corresponds to left-multiplication of the trivialising function
$\gamma^\inv\mapsto N \gamma^\inv$,
$N\in\text{Isom}(\mathbb{X}_{\Lambda,S})$, i.~e.~if the function
$\Pi_{\Lambda,S}: \text{Isom}(\mathbb{X}_{\Lambda,S})\rightarrow
\mathbb{X}_{\Lambda,S}$ satisfies the condition
\begin{align}
\label{gaminvtrafo}
\Pi_{\Lambda,S}(N\gamma^\inv)=N\Pi_{\Lambda,S}(\gamma^\inv)\qquad
\forall N\in\text{Isom}(\mathbb{X}_{\Lambda,S}).
\end{align}

This suggests that the functions $\Pi_{\Lambda,S}:
\text{Isom}(\mathbb{X}_{\Lambda,S})\rightarrow
\mathbb{X}_{\Lambda,S}$ should be defined as
\begin{align}
\label{embeddef} &\Pi_{0,L}(v,\bx)=\bx & &\forall(v,\bx)\in
PSU(1,1)\ltimes\RR^3\\
&\Pi_{\Lambda>0,L}(v_+,v_-)=\tfrac{1}{\sqrt{\Lambda}}\,v_+v_-^\inv
& &\forall (v_+,v_-)\in
PSU(1,1)\times PSU(1,1)\nonumber\\
&\Pi_{\Lambda<0,L}(v)= \tfrac{1}{\sqrt{|\Lambda|}}\,v v^\circ  &
&\forall v\in
SL(2,\CC)/\mathbb{Z}_2\nonumber\\
&\Pi_{\Lambda<0,E}(v)= \tfrac{1}{\sqrt{|\Lambda|}}\,v v^\dag  &
&\forall v\in SL(2,\CC)/\mathbb{Z}_2\nonumber,
\end{align}
since this ensures that identity \eqref{gaminvtrafo} is satisfied.
It remains to show  that these functions yield the right metric on
$R$, i.~e.~that identity \eqref{geommet} holds for each value of
the cosmological constant and each signature of the spacetime.

The simplest case is the one with Lorentzian signature and
vanishing cosmological constant, which is investigated in
\cite{we1,ich}. Parametrising the trivialising function as
$\gamma^\inv=(v,\bx)$ with $v: R\rightarrow PSU(1,1)$, $\bx:
R\rightarrow \RR^3$ and using the group multiplication law
\eqref{poincpar}, one finds
\begin{align}
\gamma d\gamma^\inv=\omega^a J^L_a+e^a P^L_a=v^\inv dv+ v^\inv
d\bx\,\quad e^a P^L_a=v^\inv d\bx,\quad\omega^aJ^L_a=v^\inv dv.
\end{align}
The Lorentz invariance of the (2+1)-dimensional Minkowski metric
then implies that the metric given by \eqref{metcond} agrees
 with the pull-back of the (2+1)-dimensional Minkowski metric via
 $\Pi_{0,L}\circ\gamma^\inv$
\begin{align}
g=\eta^{ab}_L (v^\inv d\bx)_a(v^\inv d\bx)_b=
-dx_0^2+dx_1^2+dx_2^2.
\end{align}

For Lorentzian signature and $\Lambda>0$, the trivialising
function can be parametrised as $\gamma^\inv=(\gamma_+,\gamma_-):
R\rightarrow PSU(1,1)\times PSU(1,1)$.  Using the relation between
the generators $J_a,P_a$ of the Lie algebra \eqref{liealg} and the
alternative generators $J_a^\pm$ defined by \eqref{jpmbrack}, we
find that the gauge field is given by
\begin{align}
\label{omeads} &\gamma d\gamma^\inv=(\omega^a+e^a) J_a^+
+(\omega^a-e^a) J^-_a=\gamma_+^\inv d\gamma_+ +\gamma_-^\inv
d\gamma_-\;,
\end{align}
and the expression for the dreibein in terms of $\gamma$ takes the
form
\begin{align}
 \label{adse} &e^a=\tfrac{1}{2\sqrt{\Lambda}}\langle
\gamma_+^\inv d\gamma_+-\gamma_-^\inv d\gamma_-\,,\,
J^L_a\rangle=\langle\gamma_-^\inv \cdot (
\gamma_+\gamma_-^\inv)^\inv d(\gamma_+\gamma_-^\inv)\cdot
\gamma_-\,,\, J^L_a\rangle.
\end{align}
To prove that the metric defined by this dreibein via
\eqref{metcond} agrees with the one induced by the metric
 on the model spacetime $\mathbb{X}_{\Lambda,S}$,
we use the following lemma, which can be proved by direct
calculation.
\begin{lemma}
For general $M\in SU(1,1)$ parametrised as in \eqref{pslpar}, we
have
\begin{align}
\label{pslid} M^\inv dM=2e^a J^L_a\quad\Rightarrow\quad \det
dM=\det(M^\inv dM)=e^2_0-e_1^2-e_2^2=|da|^2-|db|^2.
\end{align}
\end{lemma}
Applying this lemma to $\gamma_+\gamma^\inv_-$ together with
expression \eqref{adse} for the dreibein and using the
$\Ad$-invariance of the determinant, we find that the metric
defined by \eqref{metcond} agrees with the pull-back of the
AdS-metric via the embedding $\Pi_{\Lambda<0,L}\circ\gamma^\inv$
\begin{align}
g=-\tfrac{1}{\Lambda}\det(d(\gamma_+\gamma_-^\inv))=\eta_{ab}\langle\gamma
d\gamma^\inv, J^L_a\rangle\langle\gamma d\gamma^\inv,
J^L_b\rangle.
\end{align}
For Lorentzian and Euclidean signature and $\Lambda<0$, the gauge
field takes the form
\begin{align}
\label{dsgf} \gamma d\gamma^\inv=\omega^a
J^S_a+e^aP^S_a=(\omega^a+i\sqrt{|\Lambda|} e^a) J^S_a.
\end{align}
To prove that the metric defined via \eqref{metcond} agrees with
the pull-back of the metric on the model spacetime via
$\Pi_{\Lambda,S}\circ\gamma^\inv$ we apply the following lemma to
$M=\gamma^\inv(\gamma^\inv)^\circ$ and
$M=\gamma^\inv(\gamma^\inv)^\dag$.
\begin{lemma}
\label{dsemblem} For general $M\in SL(2,\CC)$ we have
\begin{align}
\label{dsid} &M^\inv dM=(\omega^a+i\sqrt{|\Lambda|}e^a) J^L_a &
&\Rightarrow &
-&e_0^2+e_1^2+e_2^2=\tfrac{1}{|\Lambda|}\det ( d(MM^\circ))\\
\label{hyperid} &M^\inv dM=(\omega^a+i\sqrt{|\Lambda|}e^a) J_a^E &
&\Rightarrow &  &e_0^2+e_1^2+e_2^2=-\tfrac{1}{{|\Lambda|}}\det
(d(MM^\dag))
\end{align}
\end{lemma}
{\bf Proof:} The proof is a straightforward calculation.
Parametrising the matrix $M\in SL(2,\CC)$ as
\begin{align}
\label{mparam} M=\left(\begin{array}{ll} u & v\\ w &
z\end{array}\right)\qquad uz-vw=1,
\end{align}
we find that the $SL(2,\CC)$ matrices $MM^\circ$ and  $MM^\dag$
are given by
\begin{align}
MM^\circ=\left(\begin{array}{ll} |u|^2-|v|^2 & v \bar z-u\bar w\\
-(\bar v z-\bar u w) & |z|^2-|w|^2\end{array}\right)\qquad MM^\dag=\left(\begin{array}{ll} |u|^2+|v|^2 & v \bar z+u\bar w\\
\bar v z+\bar u w & |z|^2+|w|^2\end{array}\right),
\end{align}
and, after some computation
\begin{align}
\det(d(MM^\circ))&=|wdu|^2+|udw|^2+|zdv|^2+|vdz|^2-2|udz-vdw|^2\\
+&2\text{Re}(dudz-dvdw-v\bar zd\bar vd\bar z-u\bar w d\bar u dw)\nonumber\\
\det(d(MM^\dag))&=-(|wdu|^2+|udw|^2+|zdv|^2+|vdz|^2)-2|udz-vdw|^2\\
+&2\text{Re}(dudz-dvdw+v\bar zd\bar vd\bar z+u\bar w d\bar u
dw)\nonumber.
\end{align}
On the other hand, expanding $M^\inv dM=(\omega^a+ie^a)J_a^L$
yields
\begin{align}
\label{dstriad}
&\sqrt{|\Lambda|}e^0\!\!=\!2\text{Re}(udz\!-\!wdv)\\
&\sqrt{|\Lambda|}e^1\!\!=\!-\text{Re}(udw\!-\!wdu)\!+\!\text{Re}(zdv\!-\!vdz)\nonumber\\
&\sqrt{|\Lambda|}e^2\!\!=\!\text{Im}(udw\!-\!wdu)\!+\!\text{Im}(zdv\!-\!vdz)\nonumber,
\end{align}
while the corresponding expressions for the Euclidean case  are
given by
\begin{align}
\label{hyptriad} &\sqrt{|\Lambda|}e^0\!\!=\!-2\text{Re}(udz\!-\!wdv)\\
&\sqrt{|\Lambda|}e^1\!\!=\!\text{Re}(udw\!-\!wdu)\!+\!\text{Re}(zdv\!-\!vdz)\nonumber\\
&\sqrt{|\Lambda|}e^2\!\!=\!-\text{Im}(udw\!-\!wdu)\!+\!\text{Im}(zdv\!-\!vdz)\nonumber.
\end{align}
After some further computation using $udz-wdv=vdw-zdu$ we obtain
\eqref{dsid}, \eqref{hyperid}. \hfill $\Box$

Hence, we have shown for all values of the cosmological constant
and all signatures under consideration that the maps
$\Pi_{\Lambda,S}: \text{Isom}(\mathbb{X}_{\Lambda,S})\rightarrow
\mathbb{X}_{\Lambda,S}$ defined in \eqref{embeddef} satisfy the
identities \eqref{geommet} and \eqref{gaminvtrafo}. The embedding
 into the model spacetimes $\mathbb{X}_{\Lambda,S}$ characterised by these conditions is thus given by
composing these maps with the trivialising function $\gamma^\inv:
R\rightarrow\text{Isom}(\mathbb{X}_{\Lambda,S})$
 \begin{align}
 \Bx_{\Lambda,S}=\Pi_{\Lambda,S}\circ\gamma^\inv: R\rightarrow
 \mathbb{X}_{\Lambda,S}.
 \end{align}


\section{Grafting in the Chern-Simons formalism}
\label{csgraftsect}

\subsection{Embedding into the regular domain and action of the group $\Gamma$}
\label{embedgamma}

After deriving explicit expressions which relate the embedding of
a spacetime region into the model spacetimes
$\mathbb{X}_{\Lambda,S}$ to the function trivialising the
Chern-Simons gauge field, we will now apply these results to
investigate the construction of (2+1)-spacetimes via grafting from
the Chern-Simons viewpoint. The reasoning is similar to the one in
\cite{ich} but does not make use of the simplifications specific
to Lorentzian spacetimes with vanishing cosmological constant. To
see how grafting manifests itself on the phase space of the
associated Chern-Simons theory, one needs to determine how the
variables parametrising the phase space, the holonomies $\ai,\bi$
along a set of generators of the fundamental group $\pi_1(S_g)$,
transform under the grafting construction. This requires relating
these holonomies to the variables which encode the physical
degrees of freedom in the geometrical description.

For this, we recall that in the geometrical formulation of
(2+1)-dimensional gravity, spacetimes are given as quotients of
regular domains $U_{\Lambda,S}\subset\mathbb{X}_{\Lambda,S}$ by
the action of a cocompact Fuchsian group $\Gamma$ via a
homomorphism $h_{\Lambda,S}^G: \Gamma\rightarrow
\text{Isom}(\mathbb{X}_{\Lambda,S})$. This action leaves the
surfaces $U_{\Lambda,S}(T)$ of constant cosmological time $T$
invariant, and the spacetime is given by identifying on each
surface $U_{\Lambda,S}(T)$ the points related by this action of
$\Gamma$. The physical degrees of freedom are therefore encoded in
the cocompact Fuchsian group $\Gamma$ and the group homomorphism
$h_{\Lambda,S}^G: \Gamma\rightarrow
\text{Isom}(\mathbb{X}_{\Lambda,S})$.

In the Chern-Simons formalism,  the physical degrees of freedom
are given  by the holonomies $\ai,\bi\in
\text{Isom}(\mathbb{X}_{\Lambda,S})$ or, equivalently, the
elements $N_\ai,N_\bi\in \text{Isom}(\mathbb{X}_{\Lambda,S})$
which arise in the overlap condition \eqref{csident2}. By cutting
the manifold along the representatives of the generators of the
fundamental group and trivialising the gauge field on the
resulting region, one obtains a set of functions
$\gamma(x^0,\cdot): P_g\rightarrow
\text{Isom}(\mathbb{X}_{\Lambda,S})$ on a $4g$-gon $P_g$. The
values of $\gamma^\inv$ at the two sides corresponding to a given
generator are related by left-multiplication with the elements
$N_\ai,N_\bi$, and it is shown in Sect.~\ref{trivsect} that the
left multiplication of the trivialising function $\gamma^\inv$
with elements of the isometry group
$\text{Isom}(\mathbb{X}_{\Lambda,S})$ corresponds to the action of
this group on the model space time $\mathbb{X}_{\Lambda,S}$.

This suggests identifying the parameter $x^0$ in the splitting
\eqref{gfsplit} of the Chern-Simons gauge field with the
cosmological time $T$, identifying the generators
$a_i,b_i\in\pi_1(S_g)$ with the projection of the geodesics on
$U_{\Lambda,S}(T)$ which are identified by the action of the
generators $v_\ai,v_\bi \in\Gamma$ and to take the corner $p_0$ of
the resulting polygon $P_g$ as the basepoint for the grafting.
With these identifications, the embedding constructed from the
trivialising function $\gamma(x^0,\cdot): P_g\rightarrow
\text{Isom}(\mathbb{X}_{\Lambda,S})$ for constant $x^0=T$ then
maps the polygon $P_g$ into the surface $U_{\Lambda,S}(T)$ of
constant cosmological time $T$. As the sides of the embedded
polygon in $U_{\Lambda,S}(T)$ are identified pairwise by the
action $h_{\Lambda,S}(v_Y)$, $Y\in\{A_1,B_1,\ldots,A_g,B_g\}$, of
the generators of $\Gamma$, this implies that the group elements
in the overlap condition \eqref{csident} must agree with the image
of these generators under the group homomorphism
$h_{\Lambda,S}:\Gamma\rightarrow\text{Isom}(\mathbb{X}_{\Lambda,S})$
\begin{align}
\label{identcond} N_Y=h_{\Lambda,S}(v_Y)\qquad\forall
Y\in\{A_1,B_1,\ldots,A_g,B_g\}.
\end{align}
Using the formula \eqref{cornergf} which expresses the value of
the trivialising function at the corners $p_i$ of the polygon
$P_g$ in terms of the elements $N_\ai,N_\bi$ and the value of
$\gamma$ at a point $p_0\in P_g$ we can then determine the
holonomies $A_i,B_i$ and find that they are given by
\begin{align}
\label{aibiexp} &A_i=\gamma(p_0)\cdot
h_{\Lambda,S}(v_{H_1}^\inv\cdots v_{H_{i}}^\inv
v_\bi v_{H_{i-1}}\cdots v_{H_1})\cdot \gamma(p_0)^\inv\\
&B_i=\gamma(p_0)\cdot h_{\Lambda,S}(v_{H_1}^\inv\cdots
v_{H_{i}}^\inv v_\ai v_{H_{i-1}}\cdots
v_{H_1})\cdot\gamma(p_0)^\inv.\nonumber
\end{align}

\subsection{The transformation of the holonomies under grafting}
\label{holtrafo}

We will now use the relation between the phase space variables
$N_\ai,N_\bi$ in the Chern-Simons formalism and the group
homomorphism $h_{\Lambda,S}:\Gamma\rightarrow
\text{Isom}(\mathbb{X}_{\Lambda,S})$ in the geometrical
description to derive the transformation of the holonomies
$\ai,\bi$ along a set of generators of the fundamental group under
grafting.

We start by considering the static universes associated to the
cocompact Fuchsian group $\Gamma$. As discussed in
Sect.~\ref{statuniv}, the surfaces of constant cosmological time
are then copies of hyperbolic space $\hyp$ embedded into the model
spacetime via the maps $\Phi_T^{\Lambda,S}:\hyp\rightarrow
\mathbb{X}_{\Lambda,S}$ defined in \eqref{tembed}. The polygon
$P_g$ obtained by cutting the spatial surface $S_g$ is embedded
into the image fundamental polygon $\Phi^T_{\Lambda,S}(P_\Gamma)$
in the tessellation of $U_{\Lambda,S}(T)$ induced by $\Gamma$
\begin{align}
\Bx^{st}_{\Lambda,S}(T, \cdot): \;P_g\rightarrow
\Phi^T_{\Lambda,S}(P_\Gamma)\subset U_{\Lambda,S}(T).
\end{align}
The group homomorphism $h_{\Lambda,S}^{st}:
\Gamma\rightarrow\text{Isom}(\mathbb{X}_{\Lambda,S})$ is given by
the canonical embedding of $PSU(1,1)$ into the isometry group of
the model spacetime $h^{st}_{\Lambda,S}(v)=\i_{\Lambda,S}(v)$.
Hence, using the identities \eqref{identcond} and \eqref{aibiexp},
we find that, up to conjugation with the value of $\gamma$ at the
basepoint $p_0\in S_g$, all holonomies $A_i,B_i$ are purely
Lorentzian
\begin{align}
\label{aibiexp2} &A_i=\gamma(p_0)\cdot
\i_{\Lambda,S}(v_{H_1}^\inv\cdots v_{H_{i}}^\inv
v_\bi v_{H_{i-1}}\cdots v_{H_1})\cdot \gamma(p_0)^\inv\\
&B_i=\gamma(p_0)\cdot \i_{\Lambda,S}(v_{H_1}^\inv\cdots
v_{H_{i}}^\inv v_\ai v_{H_{i-1}}\cdots
v_{H_1})\cdot\gamma(p_0)^\inv.\nonumber
\end{align}

We now consider the spacetimes obtained from these static
spacetimes by grafting along a closed, simple geodesic $\eta$ on
$\hyp_k/\Gamma$ with weight $w$. Again, the identification of the
parameter $x^0$ in \eqref{gfsplit} with the cosmological time
implies that for each value of the parameter $x^0=T$ the polygon
$P_g$ is embedded into a surface $U_{\Lambda,S}^\eta(T)$ of
constant cosmological time. However, these surfaces are no longer
copies of hyperbolic space but the deformed surfaces obtained by
inserting a strip along each geodesic in the $\Gamma$-invariant
multicurve on $U_{\Lambda,S}(T)$ associated to $\eta$. The
cocompact Fuchsian group $\Gamma$ acts on these surfaces of
constant cosmological time via the group homomorphism
$h_{\Lambda,S}^\eta:\Gamma\rightarrow
\text{Isom}(\mathbb{X}_{\Lambda,S})$ defined by
\eqref{multmin},\eqref{glogroupact}, and the group elements in the
overlap condition are given by $N_Y=h_{\Lambda,S}^\eta(v_Y)$.

To derive a formula for the transformation of the holonomies
$\ai,\bi$ along the generators of the fundamental group, we
consider a generic side $y$ of the polygon $P_g$ with starting
point and endpoint $p_i^Y,p_f^Y\in\{p_0,\ldots p_{4g}\}$. Denoting
by $v_i^Y,v_f^Y\in\Gamma$ the elements of the cocompact Fuchsian
group that relate the value of the trivialising function
$\gamma^\inv$ at the points $p_i^Y,p_f^Y$ to its value at $p_0$ in
the static case
\begin{align}
\gamma_{st}^\inv(p_i^Y)=\i_{\Lambda,S}(v_i^Y)\gamma_{st}^\inv(p_0)\qquad
\gamma_{st}^\inv(p_f^Y)=\i_{\Lambda,S}(v_f^Y)\gamma_{st}^\inv(p_0),
\end{align}
we can express the holonomy along $y$ as
\begin{align}
Y=\gamma(p_f^Y)\gamma^\inv(p_i^Y)=&\gamma(p_0)
h_{\Lambda,S}^\eta(v_f^Y)^\inv h_{\Lambda,S}^\eta(v_i^Y)
\gamma(p_0)^\inv.
\end{align}
Using identity \eqref{glogroupact} for the group homomorphism
$h^\eta_{\Lambda,S}$ and the identities \eqref{cocycprops} we then
obtain
\begin{align}
\label{ytrafo} Y= &\gamma(p_0)\i_{\Lambda,S}(v_f^Y)^\inv
B_{\eta,\Lambda,S}(q_0,v_f^Y q_0)^\inv
B_{\eta,\Lambda,S}(q_0,v_i^Yq_0)
\i_{\Lambda,S}(v_i^Y)\gamma(p_0)^\inv\\
=&\gamma_{st}(p_f^Y) B_{\eta,\Lambda,S}(v_i^Y q_0,v_f^Yq_0)^\inv
\gamma_{st}(p_i^Y)^\inv =Y_{st}\cdot \gamma_{st}(p_i^Y)
B_{\eta,\Lambda,S}(v_i^Yq_0,v_f^Y
q_0)^\inv\gamma_{st}(p_i^Y)^\inv,\nonumber
\end{align}
where $\gamma_{st}$, $Y_{st}$ denote, respectively, the
trivialising function and the holonomy along $y$ in the static
universe associated to $\Gamma$, $q_0\in\hyp$ the basepoint for
the grafting, which we took to coincide with the embedding of the
corner $p_0$ and $B_{\eta,\Lambda,S}$ is given by \eqref{multmin}.
Setting $p_i^\ai=p_{4(i-1)}$, $p_f^\ai=p_{4i-3}$ and
$p_i^\bi=p_{4i-2}$, $p_f^\bi=p_{4i-3}$ in \eqref{ytrafo} and using
expression \eqref{cornergf} for the value of the trivialising
function $\gamma$ at the corners of the polygon $P_g$, we find
that the group elements $v_i^\ai,v_i^\bi,v_f^\ai,v_f^\bi$ are
given by
\begin{align}
\label{aibigroupels} &v_i^\ai=v_{H_{i-1}}\cdots v_{H_1} &
&v_f^\ai=v_\ai^\inv
v_\bi^\inv v_\ai v_{H_{i-1}}\cdots v_{H_1}\\
&v_i^\bi=v_\bi^\inv v_\ai v_{H_{i-1}}\cdots v_{H_1} &
&v_f^\bi=v_\ai^\inv v_\bi^\inv v_\ai v_{H_{i-1}}\cdots
v_{H_1}\nonumber,
\end{align}
and that the transformation of the holonomies $A_i,B_i$ under
grafting along $\eta$ takes the form
\begin{align}
\label{graibitr} &A_i^{st}\mapsto A_i^{st} \cdot\;
(H^{st}_{i-1}\cdots H_1^{st}) \;\gamma(p_0)
B_{\eta,\Lambda,S}(v_i^\ai q_0, v_f^\ai q_0)^\inv
\gamma^\inv(p_0)\;
(H_{i-1}^{st}\cdots H_1^{st})^\inv\\
&B_i^{st}\mapsto B_i^{st} \cdot (B^{st\;-1}_i
A^{st}_iH^{st}_{i-1}\cdots H^{st}_1)\;\gamma(p_0)
B_{\eta,\Lambda,S}(v_i^\bi q_0, v_f^\bi q_0)^\inv
\gamma^\inv(p_0)\; (B^{st\;-1} A^{st}_iH^{st}_{i-1}\cdots
H^{st}_1)^\inv\nonumber.
\end{align}
We will now evaluate this formula for the case of a simple, closed
geodesic $\eta$ with weight $w$ on $\hyp_k/\Gamma$ using the
concrete expression \eqref{multmin} for the map
$B_{\eta,\Lambda,S}: \hyp\times \hyp\rightarrow
\text{Isom}(\mathbb{X}_{\Lambda,S})$. As discussed in
Sect.~\ref{geomsect}, the geodesic $\eta$ lifts to a
$\Gamma$-invariant multicurve on $\hyp_k$
\begin{align}
\label{cmultcurve} G^{\hyp_k}_{\eta,\Gamma}=\{ \Ad(v)c_\eta\;|\;
v\in\Gamma\},
\end{align}
where $c_\eta: \RR\rightarrow\hyp_k$, $c_\eta(0)\in P_\Gamma$ is
the lift of $\eta$ with basepoint in the fundamental polygon
$P_\Gamma\subset \hyp_k$ and with translation element
\begin{align}
\label{ctransl} c_\eta(1)=v_\eta c_\eta(0)=e^{n_\eta^a J_a}\qquad
c_\eta(t)\neq v c_\eta(0)\quad\forall v\in\Gamma, t\in(0,1).
\end{align}
In order to evaluate the formula \eqref{multmin} for the
multicurve \eqref{cmultcurve}, we need to determine which
geodesics in \eqref{cmultcurve} intersect a given side of the
fundamental polygon $P_\Gamma$ and to derive their translation
vectors. For this, we note that a geodesic $c'= v c_\eta$,
$v\in\Gamma$, intersects a side of $P_\Gamma$ if and only if
$c_\eta$ intersects the corresponding side of the polygon $P'=
v^\inv P_\Gamma$. Hence, the intersection points of geodesics in
the multicurve \eqref{cmultcurve} with the sides of $P_\Gamma$ are
in one-to-one correspondence with intersection points of
$c_\eta|_{[0,1)}: [0,1)\rightarrow \hyp_k$ with polygons in the
tessellation of $\hyp_k$ induced by $\Gamma$. Furthermore, these
intersection points are labelled by the factors in the expression
of the translation element \eqref{ctransl} as a product in the
generators $v_\ai,v_\bi$ of $\Gamma$, which can be seen as
follows.

Because $c_\eta$ is the lift of closed, simple geodesic $\eta$ on
$\hyp_k/\Gamma$, it
 traverses a sequence of polygons in the tessellation of $\hyp_k$
 induced by $\Gamma$
\begin{align}
\label{polyseq} P_1\!=\!P_\Gamma,\;\;\; P_2\!=\!v_{r}
P_\Gamma,\;\;\; P_3\!=\!v_{r-1} v_r P_\Gamma,\;\;\;\ldots,\;\;\;
P_{r+1}\!=\!v_1\cdots v_r P_\Gamma\!=\!v_\eta P_\Gamma,
\end{align}
which are mapped into each other by group elements $v_i\in\Gamma$,
until it
 reaches the point $c_\eta(1)=v_1\cdots v_r c_\eta(0)=v_\eta c_\eta(0)\in
 P_{r+1}$ identified with $c_\eta(0)$.
As the elements of the Fuchsian group $\Gamma$  which map the
polygon $P_\Gamma$ into its neighbours are the generators
$v_\ai,v_\bi\in\Gamma$ and their inverses, we find that the group
element $v_r$ is of the form $v_r=v_{X_r}^{\alpha_r}$, with
$v_{X_r}\in\{v_{A_1},\ldots, v_{B_g}\}$, $\alpha_r\in\{\pm 1\}$.
Similarly, for a general polygon $P'=v P_\Gamma$ the elements of
$\Gamma$ which map this polygon into its neighbours are given by
$v v_\ai^{\pm 1} v^\inv$, $v v_\bi^{\pm 1} v^\inv$, which implies
that the group elements $v_k$ in \eqref{polyseq} are of the form
\begin{align}\label{polpar}&v_k= v_{X_r}^{\alpha_r}\cdots
v_{X_{k+1}}^{\alpha_{k+1}} v_{X_{k}}^{\alpha_k}
v_{X_{k+1}}^{-\alpha_{k+1}}\cdots v_{X_r}^{-\alpha_r} \end{align}
with $v_{X_i}\in\{v_{A_1},\ldots v_{B_g}\}$, $\alpha_i=\in\{\pm
1\}$. In particular,  the translation element $v_\eta$ in
\eqref{ctransl} and the associated translation vector $\bn_\eta$
are given by
\begin{align}  \label{polyidentpar}&v_\eta=v_1\cdots
v_r=v_{X_r}^{\alpha_r}\cdots v_{X_1}^{\alpha_1}=e^{n_\eta^a J_a}.
\end{align}

Hence, intersection points of the geodesics in the multicurve
$G^{\hyp_k}_{\eta,\Gamma}$ with a given side
$y\in\{a_1,b_1,\ldots,a_g,b_g\}$ of $P_\Gamma$ are in one-to-one
correspondence with factors $v_{X_k}^{\alpha_k}$, $X_k=Y$, in the
expression \eqref{polyidentpar} of the translation element
$v_\eta$ in terms of the generators of $\Gamma$. The geodesics in
 \eqref{cmultcurve} which intersect
 the fundamental polygon $P_\Gamma$ are therefore given by
\begin{align}
\label{cdef} c_1=c_\eta,\quad c_2=v_{X_1}^{\alpha_1}\;c_\eta,\quad
c_3=v_{X_2}^{\alpha_2}v_{X_1}^{\alpha_1}\;c_\eta,\quad\ldots,\quad
c_{r}=v_{X_{r-1}}^{\alpha_{r-1}}\cdots v_{X_1}^{\alpha_1}\;c_\eta,
\end{align}
and the associated translation vectors take the form
\begin{align}
\bn_{c_k}=\Ad(v_{X_{k-1}}^{\alpha_{k-1}}\cdots
v_{X_1}^{\alpha_1})\bn_\eta.
\end{align}
Furthermore, we note that the geodesic in \eqref{cdef} which
intersects the side $y=x_k$ is $c_k$ if $\alpha_k=1$ and $c_{k+1}$
if $\alpha_k=-1$. Taking into account the orientation of the sides
$a_i,b_i$ in the polygon $P_g$, see Fig.~\ref{poly1}, we find that
intersections with sides $a_i$ have positive intersection number
for $\alpha_k=1$ and negative intersection number for
$\alpha_k=-1$, while the intersection numbers for sides $b_i$ are
positive and negative, respectively, for $\alpha_k=-1$ and
$\alpha_k=1$. With the definition $\epsilon(Y)=1$ for $Y=\ai$,
$\epsilon(Y)=-1$ for $Y=\bi$ and
 \begin{align} \label{bnkdef}\bn_{k}=\begin{cases}
\Ad(v_{X_{k-1}}^{\alpha_{k-1}}\cdots v_{X_1}^{\alpha_1})\bn_{\eta}
& \text{for}\;\alpha_k=1\\ \Ad(v_{X_{k}}^{\alpha_k}\cdots
v_{X_1}^{\alpha_1})\bn_{\eta} & \text{for}\;\alpha_k=-1,
\end{cases}
\end{align}
we can express the group elements $B_{\eta,\Lambda,S}(v_i^Y q_0,
v_f^Y q_0)$ in \eqref{graibitr} as
\begin{align}
\label{rhoconc} &B_{\eta,\Lambda=0,L}(v_i^Y q_0, v_f^Y q_0
)=-w\epsilon(Y)\sum_{X_k=Y} \alpha_k\hat\bn_k\\
&B_{\eta,\Lambda>0,L}(v_i^Y q_0, v_f^Y q_0) =(\prod_{X_k=Y} e^{-
w\sqrt{\Lambda}\epsilon(Y) \alpha_k\hat
n^a_{k}J^L_a},\prod_{X_k=Y}
e^{w\sqrt{\Lambda}\epsilon(Y) \alpha_k\hat n^a_{k}J^L_a})\nonumber\\
 &B_{\eta,\Lambda<0,L}(v_i^Y q_0, v_f^Y q_0)=B_{\eta,\Lambda<0,E}(v_i^Y q_0, v_f^Y q_0)=\prod_{X_k=Y}
e^{-i \sqrt{|\Lambda|}w\epsilon(Y)\alpha_k\hat
n^a_{k}J_a^{M}},\nonumber
\end{align}
where the factors  are ordered from the left to the right in the
order in which the intersection points occur on the generator
$y\in\{a_1,\ldots,b_g\}$. Inserting formula \eqref{rhoconc}  into
\eqref{graibitr},  then yields an expression for the
transformation
  of the
holonomies under grafting along $\eta$ in terms of the translation
vector of $c_\eta$. We obtain the following theorem.
\begin{theorem}
\label{geomgrftth} Consider a
 closed, simple geodesic
$\eta:[0,1]\rightarrow \hyp_k/\Gamma$ with weight $w>0$ and its
lift $c_\eta:[0,1]\rightarrow \hyp_k$ with basepoint $c_\eta(0)
\in P_\Gamma$. Let $v_\eta\in\Gamma$ be the translation element of
$c_\eta$ defined as in \eqref{ctransl} and given in terms of the
generators $v_\ai,v_\bi\in\Gamma$ by \eqref{polyidentpar} and
denote by $v_\eta^k$ the associated cyclic permutations of
\eqref{polyidentpar}
\begin{align}
\label{holpers} v_\eta^k=e^{ n_k^a J_a}=\begin{cases}
v_{X_{k-1}}^{\alpha_{k-1}}\cdots v_{X_1}^{\alpha_1}
v_{X_r}^{\alpha_r}\cdots v_{X_{k}}^{\alpha_{k}}
& \alpha_k=1\\
 v_{X_{k}}^{\alpha_{k}}\cdots
v_{X_1}^{\alpha_1} v_{X_r}^{\alpha_r}\cdots
v_{X_{k+1}}^{\alpha_{k+1}} & \alpha_k=-1.
\end{cases}
\end{align}
Then the transformation of the holonomies $A_i,B_i$ under grafting
along $\eta$ with weight $w$ is given by
\begin{align}
\label{graibi2} Gr_{\eta,\Lambda,S}^w:\;&A^{st}_i\mapsto A^{st}_i
\cdot(H^{st}_{i-1}\cdots H^{st}_1)\gamma(p_0)\left(\prod_{X_k=\ai}
F_{\Lambda,S}^{-w\alpha_k}(v_\eta^k)\right)\gamma^\inv(p_0)(H^{st}_{i-1}\cdots
H^{st}_1)^\inv\\
&B^{st}_i\mapsto B^{st}_i \cdot(B_i^{st\;\inv}
A^{st}_iH^{st}_{i-1}\cdots
H^{st}_1)\gamma(p_0)\left(\prod_{X_k=\bi}
F_{\Lambda,S}^{w\alpha_k}(v_\eta^k)\right)\gamma^\inv(p_0)(B_i^{st\;\inv}
A^{st}_iH^{st}_{i-1}\cdots H^{st}_1)^\inv\nonumber,
\end{align}
where the factors are ordered from the right to the left in the
order in which the associated intersection point occur on the
generators $a_i,b_i$ and $F_\Lambda^w: PSU(1,1)\rightarrow
\text{Isom}(\mathbb{X}_{\Lambda,S})$ is given by
\begin{align}
\label{ftrafodef} &F_{\Lambda,S}^w(e^{n^a J^L_a})=
\begin{cases} (1, w\hat\bn) & \Lambda=0,\;\text{Lorentzian}\\
(e^{{w}\sqrt{\Lambda} \hat n^aJ^L_a}, e^{-w \sqrt{\Lambda}\hat n^a J^L_a} ) & \Lambda>0,\;\text{Lorentzian}\\
e^{i\sqrt{|\Lambda|}w\hat n^a J^L_a} &
\Lambda<0,\;\text{Lorentzian and Euclidean}.
\end{cases}\\
&\hat\bn=\tfrac{1}{\sqrt{|\bn^2|}}\bn\in\RR^3.\nonumber
\end{align}
\end{theorem}


\section{Grafting and Poisson structure}
\label{graftpoiss}

\subsection{The transformations generated by the physical observables}
\label{obsgen}

 Theorem \ref{geomgrftth} gives a formula for the transformation
of the holonomies $A_i,B_i$ for a static spacetime under grafting
along a closed simple geodesic $\eta$ on $\hyp_k/\Gamma$ with
weight $w$. In this section, we will demonstrate that the
associated  one-parameter group of diffeomorphisms on the phase
space is generated via the Poisson bracket by a gauge invariant
Hamiltonian. We find that for all values of the cosmological
constant and all signatures under consideration this Hamiltonian
is a Wilson loop observable associated to $\eta$ and constructed
from an $\Ad$-invariant, symmetric bilinear form on the Lie
algebra $\gothh_{\Lambda,S}$.

In the Chern-Simons formulation of (2+1)-dimensional gravity, the
Poisson brackets of Wilson loop observables were first
investigated in the work of Regge and Nelson \cite{RN, RN1,
RN2,RN3, RN4} and by Ashtekar, Husain, Rovelli and Smolin
\cite{ahrss}, for the case of a punctured disc see also Martin
\cite{martin}. In a mathematical context, the Poisson brackets of
Wilson loop observables and the associated flows on phase space
were first derived in the classical paper \cite{goldman} by
Goldman, who considers the moduli space of flat $H$-connections on
surfaces $S_g$ and for general groups $H$. However, the
formulation in \cite{goldman} is rather abstract and does not
characterise these flows in terms of the holonomies of a set of
generators of the fundamental group $\pi_1(S_g)$. It is shown in
\cite{ich2} that Fock and Rosly's
 description of the phase space \cite{FR} allows one to obtain concrete
expressions for the transformations of these holonomies under the
flow generated by the Wilson loop observables associated to a
general simple curve on $S_g$ and a general conjugation invariant
function of the gauge group. The results are valid for
Chern-Simons theory with a general gauge group $H$ and can be
summarised as follows.

\begin{theorem} (\cite{ich2}, see also  \cite{goldman})

Consider Fock and Rosly's description of the moduli space
$\mathcal{M}_g^H$ of flat $H$-connections on $S_g$ with the
notation introduced in Theorem \ref{frtheorem}. Let $\eta,\lambda$
be closed, simple curves on $S_g$, whose homotopy equivalence
classes are given, respectively, as a product of the dual
generators $\dca_i,\dcb_i$  defined in \eqref{dualgens} and as a
product in the generators $a_i,b_i$ and their inverses
\begin{align}
\label{etaexp}
&\eta=\dcx_r^{\alpha_r}\circ\dcx_{r-1}^{\alpha_{r-1}}\circ\ldots\circ
\dcx_1^{\alpha_1} &
&\dcx_k\in\{\dca_1,\dcb_1,\ldots,\dca_g,\dcb_g\},\;\alpha_k\in\{\pm
1\}\\
&\lambda=y_s^{\beta_s}\circ y_{s-1}^{\beta_{s-1}}\circ \ldots\circ
y_1^{\beta_1} & &y_k\in\{a_1,\ldots,b_g\}, \beta_k\in\{\pm
1\}.\nonumber
\end{align}
Denote
 by $\eta_k,\lambda_k$ the cyclic permutations of the products in \eqref{etaexp}
\begin{align}
\label{etakdef} &\eta_k=\begin{cases}
\dcx_{k-1}^{\alpha_{k-1}}\circ\ldots\circ\dcx_1^{\alpha_1}\circ\dcx_r^{\alpha_r}\circ\ldots\circ\dcx_{k}^{\alpha_{k}}
&
\alpha_k=1\\\dcx_{k}^{\alpha_{k}}\circ\ldots\circ\dcx_1^{\alpha_1}\circ\dcx_r^{\alpha_r}\circ\ldots\circ\dcx_{k+1}^{\alpha_{k+1}}
& \alpha_k=-1\end{cases}\\ \label{lambdakdef}
&\lambda_k=\begin{cases} y_{k-1}^{\beta_{k-1}}\circ\ldots\circ
y_1^{\beta_1}\circ y_s^{\beta_s}\circ\ldots\circ y_{k}^{\beta_{k}}
& \beta_k=1\\y_{k}^{\beta_{k}}\circ\ldots\circ y_1^{\beta_1}\circ
y_s^{\beta_s}\circ\ldots\circ y_{k+1}^{\beta_{k+1}} &
\beta_k=-1\end{cases}
\end{align}
and by $H_{\eta_k}$, $H_{\lambda_k}$ the associated holonomies.
Then, the intersection points of $\eta$ and $\lambda$,
respectively, with curves representing the generators
$a_i,b_i\in\pi_1(S_g)$ and $\dca_i,\dcb_i$ are in one-to-one
correspondence with factors $\dcx_k=\dca_i,\dcx_k=\dcb_i$ in
\eqref{etakdef} and factors $y_k=a_i,y_k=b_i$  in
\eqref{lambdakdef} and the exponents $\alpha_k$, $\beta_k$
determine the associated oriented intersection numbers.
Furthermore, let $f_\eta$, $h_\lambda$ be the Wilson loop
observables associated to $\eta,\lambda$ and to conjugation
invariant functions $f,h\in\cif(H)$ of the gauge group. Then, the
following statements hold.
\begin{enumerate}
\item The Poisson bracket of the Wilson loop observables $f_\eta,
h_\lambda$ is given by
\begin{align}
\label{obspb} &\{h_\lambda,
f_\eta\}(A_1,\ldots,B_g)=\sum_{i=1}^g\sum_{\dcx_k=\dca_i, y_l=a_i}
\alpha_k\beta_l \;\langle\, g_f(H_{\eta_k})\,,\, g_h(
H_{i-1}\cdots H_1 H_{\lambda_l} H_1^\inv\cdots H_{i-1}^\inv
)\,\rangle\nonumber\\
&-\sum_{i=1}^g\sum_{\dcx_k=\dcb_i, y_l=b_i} \alpha_k\beta_l
\;\langle\, g_f(H_{\eta_k})\,,\, g_h(B_i^\inv A_i H_{i-1}\cdots
H_1 H_{\lambda_l} H_1^\inv\cdots H_{i-1}^\inv A_i^\inv B_i)
)\,\rangle,
\end{align}
where $g_f, g_h: H\rightarrow \gothh$ are defined by the action of
the left invariant vector fields $R_a\in\text{Vec}(H)$ on
$f,h\in\cif(H)$
\begin{align}
\label{gfunct} \langle g_f(u), T_a\rangle=R_af(u) \qquad \langle
g_h(u), T_a\rangle=R_a h(u)\qquad\forall u\in H.
\end{align}

 \item The one-parameter group of diffeomorphisms $T^t_{f,\eta}:
H^{2g}\rightarrow H^{2g}$ generated by the Wilson loop observable
$f_\eta$ via the Poisson bracket acts on the holonomies $A_i,B_i$
according to
\begin{align}
\nonumber \{f_\eta, h\}&=\frac{d}{dt}|_{t=0} h\circ
T^t_{f,\eta}\qquad\forall h\in\cif(H^{2g}),\\
\label{gentrafosaibi} T^t_{f,\eta}:\;&A_i\mapsto A_i
\;\cdot\;(H_{i-1}\cdots H_1)\cdot \prod_{\dcx_k=\dca_i}
G^{-t\alpha_k}_f(H_{\eta_k}) \cdot (H_{i-1}\cdots
H_1)^\inv\nonumber\\
&B_i\mapsto B_i \;\cdot\;(B_i^\inv A_iH_{i-1}\cdots H_1)\cdot
\prod_{\dcx_k=\dcb_i} G^{-t\alpha_k}_f(H_{\eta_k}) \cdot(B_i^\inv
A_iH_{i-1}\cdots H_1)^\inv,
\end{align}
where the factors are ordered from the right to the left in the
order in which the associated intersection points occur on the
curves representing the generators $a_i,b_i$ and the function
$G^t_f: H\rightarrow H$ is obtained by exponentiating the function
$g_f: H\rightarrow \gothh$
\begin{align}
\label{gfunct2} G^t_f(u)=e^{t g_f(u)}\qquad\forall u\in H .
\end{align}
\end{enumerate}
\end{theorem}
A particular set of Wilson loop observables which will be relevant
in the following are the observables associated to
$\Ad$-invariant, symmetric bilinear forms
$\kappa\in\gothh^*\otimes\gothh^*$ on the Lie algebra of the gauge
group. These observables are constructed using the parametrisation
via the exponential map $\exp:\gothh\rightarrow H$ by setting
\begin{align}
\label{formfunct} \tilde \kappa(e^{k^a
T_a})=\tfrac{1}{2}\kappa(k^aT_a, k^a
T_a)=\tfrac{1}{2}\kappa_{ab}k^ak^b.
\end{align}
As it is supposed in \cite{ich2} that the exponential map is
surjective, but not necessarily bijective, one has to restrict the
range of admissible vectors $\bk\in\RR^{\text{dim}\,H}$ to obtain
a well-defined expression. The resulting function $\tilde\kappa:
H\rightarrow \RR$ is then not necessarily continuous everywhere.
However, in the following we are interested in the local
properties of the phase space.  As the exponential maps is locally
bijective, any element $u\in H$  has a neighbourhood in which the
Wilson loop observable $\tilde\kappa: H\rightarrow \RR$ is a
$\cif$-function. It has been shown by Goldman \cite{goldman} that
the associated functions $g_{\tilde\kappa}: H\rightarrow \gothh$,
$G_{\tilde\kappa}^t: H\rightarrow H$ defined in \eqref{gfunct},
\eqref{gfunct2} then take the form
\begin{align}
\label{kappags} g_{\tilde
\kappa}(e^{k^aT_a})=\bl_\kappa(k^aT_a)\qquad
G_{\tilde\kappa}^t(e^{k^a T_a})=e^{t\bl_\kappa(k^a T_a)}
\end{align}
with a linear map $\bl_\kappa: \gothh\rightarrow\gothh$ given by
\begin{align}
\label{smapdef} \langle \bl_\kappa(\bx),\by\rangle=\langle
\bx,\bl_\kappa(\by)\rangle=\kappa(\bx,\by)\qquad\forall
\bx,\by\in\gothh.
\end{align}
In particular, one obtains a generic set of observables
constructed from the $\Ad$-invariant, non-degenerate, symmetric
bilinear form $\langle\,,\,\rangle$ in the Chern-Simons action
with associated conjugation invariant function $\tilde
t\in\cif(H)$
\begin{align}
\label{pairobs} \tilde t(e^{k^a T_a})=\tfrac{1}{2}\langle k^a T_a,
k^a T_a\rangle=\tfrac{1}{2} t_{ab} k^ak^b.
\end{align}
In this case, the linear map
$\bl_{\langle\,,\,\rangle}:\gothh\rightarrow \gothh$ is the
identity $\bl_{\langle\,,\,\rangle}=\text{id}_\gothh$ and the
associated functions $g_{\tilde t}$, $G_{\tilde t}^t$ in
\eqref{gfunct}, \eqref{gfunct2} are given by
\begin{align}
\label{pairfuncts} g_{\tilde t}(e^\bx)=\bx\qquad G^t_{\tilde
t}(e^{\bx})=e^{t\bx}\qquad\forall\bx\in\gothh.
\end{align}
The transformations \eqref{gentrafosaibi} the associated
observables $\tilde t_\lambda$  generate via the Poisson bracket
are investigated in \cite{ich2}, for the case of semidirect
product gauge groups $G\ltimes\gothg^*$ see also \cite{we3}, where
it is shown that they have the interpretation of infinitesimal
Dehn twists along $\lambda$. We will discuss these observables and
the associated flows on phase space in more detail in
Sect.~\ref{grdtsect} and Sect.~\ref{ccgrdt}, where we investigate
the relation between grafting and infinitesimal Dehn twists.

\subsection{Hamiltonians for Grafting}

\label{grafthamilt}  Using the results from \cite{ich2, goldman},
summarised in the last subsection, we can now investigate the
transformations on phase space generated by observables associated
to certain $\Ad$-invariant, symmetric bilinear forms in the
Chern-Simons formulation of (2+1)-dimensional gravity and show
that they agree with the grafting transformation \eqref{graibi2}.

The first step is to identify the $\Ad$-invariant bilinear forms
on the Lie algebras $\gothh_{\Lambda,S}$. It is shown in
\cite{Witten1} that for all values of the cosmological constant
and all signatures under consideration, the space of
$\Ad$-invariant, symmetric bilinear forms on $\gothh_{\Lambda,S}$
is two-dimensional. Besides the pairing $\langle\,,\,\rangle$ in
the Chern-Simons action given by \eqref{inprod}, the Lie algebras
$\gothh_{\Lambda,S}$ admit another $\Ad$-invariant symmetric
bilinear form, which is given in terms of the generators
$J_a^S,P_a^S$ by \begin{align} \label{bilin} &\kappa(
J^S_a,J^S_b)=\eta^S_{ab} & &\kappa(J^S_a,P^S_b)=0 &
&\kappa(P^S_a,P^S_b)=\Lambda \eta^S_{ab}.
\end{align}
It follows that the associated linear maps $\bl_\kappa:
\gothh_{\Lambda,S}\rightarrow\gothh_{\Lambda, S}$ defined in
\eqref{smapdef} take the form
\begin{align}
\label{blmaps}
&\bl_\kappa(J^S_a)=P^S_a\quad\bl_\kappa(P^S_a)=\Lambda J^S_a.
\end{align} The
associated map $G^t_{\tilde\kappa,\Lambda,S}:
\text{Isom}(\mathbb{X}_{\Lambda,S})\rightarrow
\text{Isom}(\mathbb{X}_{\Lambda,S})$  defined in \eqref{gfunct2}
is obtained by exponentiating this map $\bl_\kappa:
\gothh_{\Lambda,S}\rightarrow\gothh_{\Lambda,S}$ as in
\eqref{kappags}. As the exponential maps $\exp_{\Lambda,S}:
\gothh_{\Lambda,S}\rightarrow \text{Isom}(\mathbb{X}_{\Lambda,S})$
are surjective and therefore locally bijective for all isometry
groups $\text{Isom}(\mathbb{X}_{\Lambda,S})$, see the remark after
\eqref{tdef},  we obtain a one-parameter group of transformations
$G^t_{\tilde\kappa,\Lambda,S}:
\text{Isom}(\mathbb{X}_{\Lambda,S})\rightarrow
\text{Isom}(\mathbb{X}_{\Lambda,S})$ given by
\begin{align}\label{biling} G^t_{\tilde\kappa,\Lambda,S}(\exp_{\Lambda,S}(p^a
J_a^S+k^a P_a^S))=&\exp_{\Lambda,S}(t(p^a P_a^s+\Lambda k^a
J_a^S))\\
 =&
\begin{cases}(1,t\bp) &\Lambda=0,\;\text{Lorentzian}\\
(e^{t\sqrt{\Lambda}(p^a+\sqrt{\Lambda}k^a)J^L_a},
e^{-t\sqrt{\Lambda}(p^a-\sqrt{\Lambda}k^a)J^L_a}) &
\Lambda>0,\;\text{Lorentzian}\\
e^{it\sqrt{|\Lambda|}(p^a +i\sqrt{|\Lambda|}k^a)J^L_a} &
\Lambda<0,\;\text{Lorentzian}\\
e^{it\sqrt{\Lambda}(p^a +i\sqrt{\Lambda}k^a)J_a^E} &
\Lambda<0,\;\text{Euclidean},\nonumber
\end{cases}
\end{align}
and can formally express the map $G^t_{\tilde\kappa,\Lambda,S}:
\text{Isom}(\mathbb{X}_{\Lambda,S})\rightarrow
\text{Isom}(\mathbb{X}_{\Lambda,S})$ as
\begin{align}
\label{biling2}
&G^t_{\tilde\kappa,\Lambda=0,L}(e^{p^a J_a^L},\ba)= (1, t\bp) & &\forall \bp,\ba\in\RR^3\\
&G^t_{\tilde\kappa,\Lambda>0,L}(u_+,u_-)=(u_+^{t\sqrt{\Lambda}}, u_-^{-t\sqrt{\Lambda}}) & &\forall u_\pm\in PSU(1,1)\nonumber\\
&G^t_{\tilde\kappa,\Lambda<0,L}(u)=G^t_{\Lambda<0,E}(u)
=u^{it\sqrt{|\Lambda|}} & &\forall u\in
SL(2,\CC)/\mathbb{Z}_2.\nonumber
\end{align}
The fact that the exponential map
$\exp_{\Lambda,S}:\gothh_{\Lambda,S}\rightarrow
\text{Isom}(\mathbb{X}_{\Lambda,S})$ is locally, but not globally
bijective, implies that the maps \eqref{biling}, \eqref{biling2}
are only defined locally. In order to obtain a unique
parametrisation of elements of
$\text{Isom}(\mathbb{X}_{\Lambda,S})$ in terms of elements of
$\gothh_{\Lambda,S}$, one has to restrict the range of the vectors
$\bp,\bk\in\RR^3$ appropriately, which implies that the Wilson
loop observable $\tilde\kappa$ and the map
$G^t_{\tilde\kappa,\Lambda,S}:
\text{Isom}(\mathbb{X}_{\Lambda,S})\rightarrow
\text{Isom}(\mathbb{X}_{\Lambda,S})$ are not necessarily
continuous everywhere. However, as the exponential map is locally
bijective, every element of $\text{Isom}(\mathbb{X}_{\Lambda,S})$
has a neighbourhood in which the Wilson loop observable
$\tilde\kappa$ is $\cif$ and the map
$G^t_{\tilde\kappa,\Lambda,S}$ a one-parameter group of
diffeomorphisms.  In particular, the parametrisation via the
exponential map is unique for elements of the group $PSU(1,1)$
embedded into $\text{Isom}(\mathbb{X}_{\Lambda,S})$ via
\eqref{canembedlor}, see the remark after \eqref{lorexpmap}, which
is the form the holonomies take for the static universes. Hence,
the parametrisation via the exponential map is well-defined and
$\cif$ in this case, and we can insert expressions \eqref{biling},
\eqref{biling2} into the general formula \eqref{gentrafosaibi}. We
obtain the following theorem, which generalises Theorem 5.2. in
\cite{ich}.
\begin{theorem}
\label{grafttrafo} Let $\eta$ be a closed simple curve on $S_g$
whose homotopy equivalence class is given as a product in the dual
generators \eqref{dualgens} and their inverses as in
\eqref{etaexp} and $\kappa$ the $\Ad$-invariant, symmetric
bilinear form on $\gothh_{\Lambda,S}$ defined in \eqref{bilin}.
 Then, the
one-parameter group of diffeomorphisms
$T^t_{\tilde\kappa,\eta,\Lambda,S}:
\text{Isom}(\mathbb{X}_{\Lambda,S})^{2g}\rightarrow
\text{Isom}(\mathbb{X}_{\Lambda,S})^{2g}$ generated by the Wilson
loop observable $\tilde \kappa_\eta$ is given by
\begin{align}
\{\tilde\kappa_\eta, g\}&=\frac{d}{dt}|_{t=0} \;g\circ
T^t_{\tilde\kappa,\eta,\Lambda,S}\qquad\forall
g\in\cif(H_\Lambda^{2g}) \label{conctrafosaibi}
\\T^t_{\tilde \kappa,\eta,\Lambda,S}:\;&A_i\mapsto A_i \;\cdot\;(H_{i-1}\cdots
H_1)\cdot \prod_{\dcx_k=\dca_i}
G^{-t\alpha_k}_{\tilde\kappa,\Lambda,S}(H_{\eta_k}) \cdot
(H_{i-1}\cdots
H_1)^\inv\nonumber\\
&B_i\mapsto B_i \;\cdot\;(B_i^\inv A_iH_{i-1}\cdots H_1)\cdot
\prod_{\dcx_k=\dcb_i}
G^{-t\alpha_k}_{\tilde\kappa,\Lambda,S}(H_{\eta_k}) \cdot
(B_i^\inv A_iH_{i-1}\cdots H_1)^\inv,\nonumber
\end{align}
where $H_{\eta_k}:
\text{Isom}(\mathbb{X}_{\Lambda,S})^{2g}\rightarrow
\text{Isom}(\mathbb{X}_{\Lambda,S})$ represents the holonomy along
$\eta_k$ defined by \eqref{etakdef},
$G^t_{\tilde\kappa,\Lambda,S}:
\text{Isom}(\mathbb{X}_{\Lambda,S})\rightarrow
\text{Isom}(\mathbb{X}_{\Lambda,S})$ is given by \eqref{biling},
\eqref{biling2} and the factors are ordered from the right to the
left in the order in which the corresponding intersection points
occur on the generators $a_i,b_i$.
\end{theorem}
We will now demonstrate that the one-parameter group of
transformations $T^t_{\tilde\kappa,\eta,\Lambda,S}:
\text{Isom}(\mathbb{X}_{\Lambda,S})^{2g}\rightarrow
\text{Isom}(\mathbb{X}_{\Lambda,S})^{2g}$ is related to the
transformation \eqref{graibi2} of the holonomies $A_i,B_i$ under
grafting along a closed simple geodesic $\eta$ on the surface
$\hyp_k/\Gamma$ with homotopy equivalence class \eqref{etaexp}.
For this, we evaluate \eqref{conctrafosaibi} for the static case,
where the group elements $N_\ai,N_\bi$ in the overlap condition
\eqref{csident2} are the images of the generators
$v_\ai,v_\bi\in\Gamma$ under  the canonical embedding
\eqref{canembedlor} of $PSU(1,1)$ into the isometry group of the
model spacetime $N_\ai=\i_{\Lambda,S}(v_\ai)$,
$N_\bi=\i_{\Lambda,S}(v_\bi)$. From expressions \eqref{aibiexp}
for the holonomies $\ai,\bi$ in terms of $N_\ai,N_\bi$ it then
follows that the holonomies along the elements $\eta_k$ defined in
\eqref{etakdef} take the form
\begin{align}
H_{\eta_k}=\gamma(p_0)v_\eta^k\gamma^\inv(p_0)=\gamma(p^0)e^{n_k^a
J_a}\gamma^\inv(p_0)
\end{align}
with $v_\eta^k$, $\bn_k$  given by \eqref{holpers}. As it is shown
in \cite{goldman, ich2} that the functions
$G^t_{\tilde\kappa,\Lambda,S}:
\text{Isom}(\mathbb{X}_{\Lambda,S})\rightarrow
\text{Isom}(\mathbb{X}_{\Lambda,S})$ satisfy the covariance
condition
\begin{align}
G^t_{\tilde\kappa,\Lambda,S}(gug^\inv)=g\cdot
G^t_{\Lambda,S}(u)\cdot g^\inv\qquad\forall u,g\in
\text{Isom}(\mathbb{X}_{\Lambda,S})
\end{align}
we obtain an expression in terms of the translation vectors
$\bn_k$ defined in \eqref{bnkdef}
\begin{align}
\label{biling1}
&G^{t\alpha_k}_{\tilde\kappa,\Lambda,S}(H_{\eta_k})=
\begin{cases}\gamma(p_0)(1,t \bn_k)\gamma^\inv(p_0) &\Lambda=0,\;\text{Lorentzian}\\
\gamma(p_0)(e^{{t}\sqrt{\Lambda}n_k^aJ_a}, e^{- \sqrt{\Lambda}{t}
n_k^a J_a})\gamma^\inv(p_0) &
\Lambda>0,\;\text{Lorentzian}\\
\gamma(p_0)e^{it\sqrt{|\Lambda|} n_k^a J_a}\gamma^\inv(p_0) &
\Lambda<0,\;\text{Lorentzian or Euclidean}\;.
\end{cases}\nonumber
\end{align}
By inserting this expression into
\eqref{conctrafosaibi} and comparing with the transformation of
the holonomies under grafting given by \eqref{graibi2},
\eqref{ftrafodef}, we find that these transformations agree up to
normalisation.
\begin{theorem}

Consider a static spacetime, where the holonomies $\ai,\bi$ along
the generators of the fundamental group are given by
\begin{align}
&\ai=\gamma(p_0)\cdot\,\i_{\Lambda,S}(v_{H_1}^\inv\cdots
v_{H_{i}}^\inv v_\bi v_{H_{i-1}}\cdots v_{H_1})\cdot
\gamma(p_0)^\inv\\
&\bi=\gamma(p_0)\cdot\,\i_{\Lambda,S}(v_{H_1}^\inv\cdots
v_{H_{i}}^\inv v_\ai v_{H_{i-1}}\cdots v_{H_1})\cdot
\gamma(p_0)^\inv\nonumber.
\end{align}
Then, the transformation \eqref{graibi2} of the holonomies under
grafting along a closed simple geodesic $\eta$ on $\hyp_k/\Gamma$
agrees with the transformation \eqref{conctrafosaibi} generated by
the observable $\tilde\kappa_\eta$ if the parameter $t$ in
\eqref{conctrafosaibi} is related to the weight $w$ by
$t=w|\bn_\eta|$
\begin{align}
Gr_{\eta,\Lambda,S}^{w}(A_1,\ldots,B_g)=T^t_{\tilde\kappa,\eta,\Lambda,S}(A_1,\ldots,B_g)
\qquad \text{for}\;\;t=w
\sqrt{|\kappa(\bn_\eta,\bn_\eta)}|=w|\bn_\eta|.
\end{align}
\end{theorem}

The fact that the transformation of the holonomies $\ai,\bi$ under
grafting is generated by a gauge invariant observable allows us to
directly deduce some of its properties, which are summarised in
the following corollary, for the Lorentzian case with vanishing
cosmological constant see also \cite{ich}.
\begin{corollary}$\quad$

\begin{enumerate}
\item The grafting transformations
$T^t_{\tilde\kappa,\eta,\Lambda,S}$ act by Poisson isomorphisms
\begin{align}
\{f\circ T^t_{\tilde\kappa,\eta,\Lambda,S},g\circ
T^t_{\tilde\kappa,\eta,\Lambda,S}\}=\{f,g\}\circ
T^t_{\tilde\kappa,\eta,\Lambda,S}\qquad\forall
f,g\in\cif(\text{Isom}(\mathbb{X}_{\Lambda,S})^{2g}).
\end{align}

 \item The grafting transformation $T^t_{\tilde\kappa,\eta,\Lambda,S}$ leaves the
constraint \eqref{holconst} invariant and commutes with the
associated gauge transformations by simultaneous conjugation with
$\text{Isom}(\mathbb{X}_{\Lambda,S})$
\begin{align}
&T^t_{\tilde\kappa,\eta,\Lambda,S}(g A_1 g^\inv,\ldots, gB_g
g^\inv)=g T^t_{\tilde\kappa,\eta,\Lambda,S} (A_1,\ldots,B_g)
g^\inv\quad\forall g\in \text{Isom}(\mathbb{X}_{\Lambda,S}).
\end{align}

\item In the Lorentzian case with $\Lambda=0$, all grafting
transformations commute.
\end{enumerate}
\end{corollary}

{\bf Proof:} The first two statements are a direct consequence of
the fact that the transformations $T^t_{\tilde\kappa,
\eta,\Lambda,S}$ are generated by the gauge invariant observable
$\tilde\kappa_\eta$. The last statement follows immediately from
the formula \eqref{biling}.\hfill$\Box$


\subsection{Grafting and Dehn twists}
\label{grdtsect}

 After demonstrating
that the transformation of the holonomies under grafting along a
closed, simple curve $\eta$ on $S_g$ is generated by the Wilson
loop observable $\tilde\kappa_\eta$, we will now investigate the
relation between this grafting transformation and the action of
infinitesimal Dehn twists along $\eta$.

\begin{figure}
\vskip .3in \protect\input epsf \protect\epsfxsize=12truecm
\protect\centerline{\epsfbox{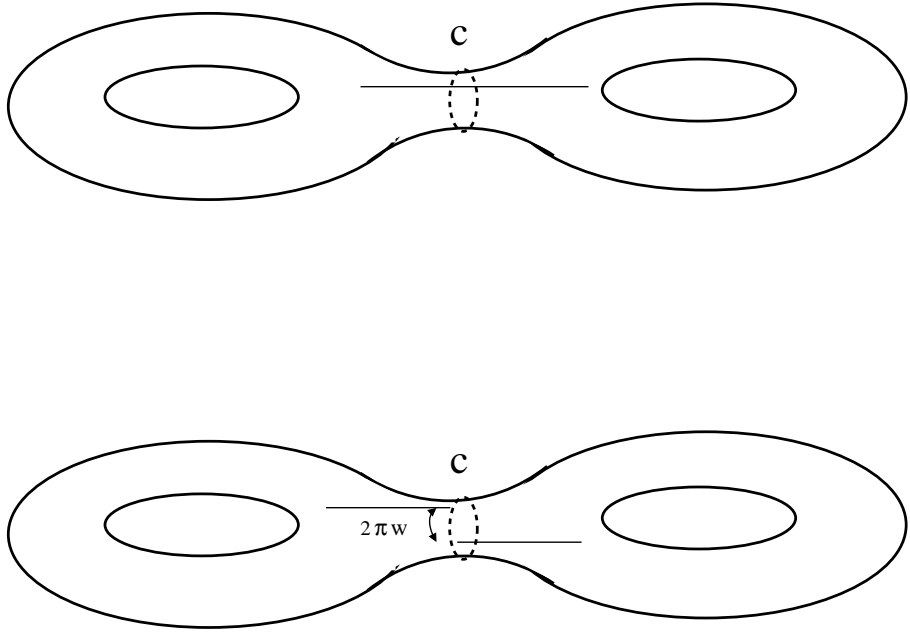}} \caption{ A Dehn twist
with parameter $w$ around a closed, simple geodesic $c$ on a genus
2 surface.} \label{dtpic}
\end{figure}

Geometrically, an infinitesimal Dehn twist along a closed geodesic
$\eta$ on a two-surface $\hyp_k/\Gamma$ with parameter $w$ amounts
to cutting the surface along $\eta$ and rotating the edges of the
cut by an angle $2\pi w$ as shown in Fig.~\ref{dtpic}.
In Chern-Simons theory, infinitesimal Dehn  twists along closed,
simple curves on the spatial surface are present generically for
any gauge group $H$ and  give rise to a transformation on the
moduli space of flat $H$-connections. The action of
(infinitesimal) Dehn twists in Fock and Rosly's description
\cite{FR} of the moduli space is investigated in \cite{ich2}, for
the case of semidirect product groups $G\ltimes\gothg^*$ see also
\cite{we3}, where it is shown that they are generated by the gauge
invariant observable constructed from the $\Ad$-invariant bilinear
form $\langle\,,\,\rangle$ in the Chern-Simons action. The results
can be summarised as follows.

\begin{theorem} \cite{ich2, we3}
\label{dttheor}

Consider Fock and Rosly's description of the moduli space
$\mathcal{M}_g^H$ in terms of the auxiliary Poisson structure
\eqref{frbivect2} on $H^{2g}$. Let $\eta$ be a  closed, simple
curve on $S_g$, whose homotopy equivalence class is given as a
product in the dual generators \eqref{dualgens} and their inverses
as in \eqref{etaexp}. Denote by  $\eta_k$ the cyclic permutations
of this expression as in \eqref{etakdef} and by $H_{\eta_k}$ the
associated holonomies. Then the transformation $T^t_{\tilde
t,\eta}: H^{2g}\rightarrow H^{2g}$ generated  by the Wilson loop
observable $\tilde t_\eta$ associated to $\eta$ and the
$\Ad$-invariant symmetric bilinear from in the Chern-Simons action
as in \eqref{pairobs} is given by
\begin{align}
\label{dtform} T^t_{\tilde t,\eta}:\;&A_i\mapsto A_i
\;\cdot\;(H_{i-1}\cdots H_1)\cdot \prod_{\dcx_k=\dca_i}
H_{\eta_k}^{-t\alpha_k}\;\cdot (H_{i-1}\cdots
H_1)^\inv\nonumber\\
&B_i\mapsto B_i \;\cdot\;(B_i^\inv A_iH_{i-1}\cdots
H_1)\cdot\prod_{\dcx_k=\dcb_i} H_{\eta_k}^{-t\alpha_k} \cdot
(B_i^\inv A_iH_{i-1}\cdots H_1)^\inv,
\end{align}
where the elements $\eta_k\in\pi_1(S_g)$ with holonomies
$H_{\eta_k}$ are defined as in \eqref{etakdef} and the products
are ordered from the right to the left in the order in which the
associated intersection points occur on the generators
$a_i,b_i\in\pi_1(S_g)$. For flow parameter $t=-1$, this
transformation agrees with the transformation of the holonomies
$\ai,\bi$ under a Dehn twist around $\eta$
\begin{align}
T^{-1}_{\tilde t,\eta}=D_\eta
\end{align}
where $D_\eta: H^{2g}\rightarrow H^{2g}$ is the diffeomorphism
induced by the action of the Dehn twist along $\eta$ on the
generators of the fundamental group $\pi_1(S_g)$.
\end{theorem}

Using the results from \cite{ich2,we3} summarised in Theorem
\ref{dttheor}, we can now compare the transformation of the
holonomies $\ai,\bi$ under the grafting transformations
$T^t_{\tilde\kappa,\eta,\Lambda,S}$ in \eqref{conctrafosaibi} with
their transformations \eqref{dtform} under infinitesimal Dehn
twists. For this, we note that the formulas
\eqref{conctrafosaibi}, \eqref{dtform}  differ only in the fact
that instead of the factor $H_{\eta_k}^{t\alpha_k}$ in
\eqref{dtform}, the grafting transformation \eqref{conctrafosaibi}
contains a factor
\begin{align}
G^{t\alpha_k}_{\tilde\kappa,\Lambda,S}(H_{\eta_k})=G_{\tilde\kappa,\Lambda,S}^1(H_{\eta_k}^{t\alpha_k}),
\end{align}
with the map $G^t_{\tilde\kappa,\Lambda,S}:
\text{Isom}(\mathbb{X}_{\Lambda,S})\rightarrow\text{Isom}(\mathbb{X}_{\Lambda,S})
$ given by \eqref{biling2}. For the  Lorentzian case with
$\Lambda=0$, the map $G^t_{\tilde\kappa, 0,L}$ assigns to each
element of the gauge group $PSU(1,1)\ltimes \RR^3$ the translation
vector associated to its Lorentz component
\begin{align}
G_{\tilde\kappa,0,L}^t(\exp_{\Lambda=0,L}((p^aJ^L_a+k^a P^L_a)))=
G^t_{\tilde\kappa,0,L}((e^{p^a J_a}, T(\bp)\bk))=(1,
t\bp)=\exp_{0,L}(t p^a P^L_a).
\end{align}
For static universes the holonomies $H_{\eta_k}$ in
\eqref{conctrafosaibi} and \eqref{dtform} are conjugated to
elements of the (2+1)-dimensional Lorentz group $PSU(1,1)$ and
grafting acts on the holonomies $\ai,\bi$ by right-multiplication
with the associated elements of the Lie algebra
$\mathfrak{su}(1,1)$. Hence, for the Lorentzian case with
vanishing cosmological constant, grafting along a closed, simple
curve $\eta$ can be viewed as the derivative or first-order
approximation of a Dehn twist along $\eta$. For Lorentzian
signature with $\Lambda>0$ and gauge group $PSU(1,1)\times
PSU(1,1)$, identity \eqref{biling2} allows one to express the
factors $G^{t\alpha_k}_{\tilde\kappa,\Lambda>0,L}(H_{\eta_k})$ in
\eqref{conctrafosaibi}  as
\begin{align}
&G_{\tilde\kappa,\Lambda>0,L}^{t\alpha_k}(H_{\eta_k})=((H_{\eta_k}^{\alpha_k})_+^{t\sqrt{\Lambda}},
(H_{\eta_k}^{\alpha_k})_-^{-t\sqrt{\Lambda}})\qquad\text{where}\;
H_{\eta_k}^{\alpha_k}=((H_{\eta_k}^{\alpha_k})_+,(H_{\eta_k}^{\alpha_k})_-)
\end{align}
Hence, we find  that the grafting transformation
\eqref{conctrafosaibi} with parameter $t\sqrt{\Lambda}$ along
$\eta$ acts as an infinitesimal Dehn twist along $\eta$ with
parameter $t\sqrt{\Lambda}$ on the first component and as an
infinitesimal Dehn twist with parameter $-t\sqrt{\Lambda}$ on the
second component of $PSU(1,1)\times PSU(1,1)$. In Lorentzian and
Euclidean (2+1)-gravity with $\Lambda<0$, the factor
$G^t_{\tilde\kappa,\Lambda,S}(H_{\eta_k}^{\alpha_k})$ in the
expression \eqref{conctrafosaibi} for the transformation of the
holonomies under grafting is
\begin{align}
\label{gridt}
G_{\tilde\kappa,\Lambda,S}^t(H_{\eta_k}^{\alpha_k})=H_{\eta_k}^{it\sqrt{|\Lambda|}\alpha_k}.
\end{align}
and the grafting transformation
$T^t_{\tilde\kappa,\eta,\Lambda,S}$ with parameter $t$ given by
\eqref{conctrafosaibi} can therefore be viewed as a Dehn twist
\eqref{dtform} with parameter $it\sqrt{|\Lambda|}$. We will come
back to this relation between grafting and Dehn twist in
Sect.~\ref{ccgrdt} where we discuss the role of the cosmological
constant as a deformation parameter.

These relations between the transformation of the holonomies under
grafting and Dehn twists for the different values of the
cosmological constant and different signatures are mirrored in a
relation for the Poisson brackets of the  Wilson loop observables
$\tilde t_\eta$, $\tilde\kappa_\eta$ associated to closed, simple
curves $\eta$ on $S_g$. By inserting the  maps $g_{\tilde t},\,
g_{\tilde\kappa}: \text{Isom}(\mathbb{X}_{\Lambda,S})\rightarrow
\gothh_{\Lambda,S}$ into the formula \eqref{obspb} for the Poisson
brackets of the Wilson loop observables and using the identity
$\bl_\kappa^2=\Lambda \;\text{id}_{\gothh_{\Lambda,S}}$ we obtain
the following Theorem which generalises Theorem 5.4. in \cite{ich}
for the case of Lorentzian signature with $\Lambda=0$.
\begin{theorem} (Symmetry relation for the observables)
\label{symth}

For any two closed, simple curves  $\lambda,\eta$ on $S_g$, the
associated Wilson loop observables $\tilde t_\lambda,
\tilde\kappa_\lambda$ and $\tilde t_\eta, \tilde\kappa_\eta$
satisfy the symmetry relations
\begin{align}
\label{minkrel} &\{\tilde t_\eta,\tilde\kappa_\lambda\}=\{\tilde
\kappa_\eta, \tilde t_\lambda\}\qquad
\{\tilde\kappa_\eta,\tilde\kappa_\lambda\}=\Lambda\{\tilde
t_\eta,\tilde t_\lambda\}. \end{align}
\end{theorem}
Theorem \ref{symth} establishes a relation between the
transformation of the gauge invariant observables $\tilde t_\eta$,
$\tilde\kappa_\eta$ associated to a closed, simple curve $\eta$ on
$S_g$ under infinitesimal Dehn twist and grafting along another
closed simple curve $\lambda$. The first identities in
\eqref{minkrel} imply that, infinitesimally, the transformation of
the observable $\tilde t_\eta$ under grafting along $\lambda$ is
the same as the transformation of $\tilde \kappa_\eta$ under a
Dehn twist along $\lambda$. The second identity states that the
transformation of the observable $\tilde\kappa_\eta$ under
infinitesimal grafting along $\lambda$ corresponds to the
transformation of $\tilde t_\eta$ under an infinitesimal Dehn
twist along $\lambda$, rescaled by a factor $\Lambda$.

\section{The cosmological constant as a deformation parameter}
\label{cosmdefsect}

\subsection{The geometrical description}

In this section we restrict attention to (2+1)-spacetimes with
Lorentzian signature  and investigate the role of the cosmological
constant as a deformation parameter in both the geometrical and
the Chern-Simons formulation of (2+1)-dimensional gravity.

In the geometrical formulation, the cosmological constant appears
as a parameter in the model spacetime $\mathbb{X}_{\Lambda,L}$,
the domains $U^G_{\Lambda,L}\subset\mathbb{X}_{\Lambda,L}$ and in
the group isomorphism $h^G_{\Lambda,L}:\Gamma\rightarrow
\text{Isom}(\mathbb{X}_{\Lambda,L})$ which determines the action
of the cocompact Fuchsian group $\Gamma$ on the domain
$U^G_{\Lambda,L}$. The dependence of the domains $U^G_{\Lambda,S}$
on the sign of the cosmological constant is investigated in detail
in the papers by Benedetti and Bonsante \cite{bb,bb2} who show
that Lorentzian spacetimes with $\Lambda\in\{0,\pm 1\}$ and the
Euclidean case with $\Lambda=-1$ can be related by rescalings and
a Wick rotation compatible with the associated actions of the
cocompact Fuchsian group $\Gamma$. In contrast, our focus is on
the role of the cosmological constant as a continuous parameter
deforming these domains, the associated model spacetimes and group
isomorphisms.

We start by considering the static spacetimes associated to
$\Gamma$. As discussed in Sect.~\ref{statuniv}, these spacetimes
are given as quotients of the interior of the forward lightcone
$U_{\Lambda,L}\subset\mathbb{X}_{\Lambda,L}$ by the canonical
action of $\Gamma$ via the embedding $\i_{\Lambda,S}:
PSU(1,1)\rightarrow\text{Isom}(\mathbb{X}_{\Lambda,S})$ into the
isometry group of the model spacetime. The parametrisation of
these lightcones in terms of matrices and their foliation by
copies of hyperbolic space are given by expressions
\eqref{minkxemb}, \eqref{adspars} and \eqref{dspars},
respectively, for $\Lambda=0$, $\Lambda>0$ and $\Lambda<0$. For
vanishing cosmological constant, the entries in the
$\mathfrak{su}(1,1)$ matrix \eqref{minkxemb} parametrising the
forward lightcone are given by the identification \eqref{minkhyp}
 of the disc model of hyperbolic space with the hyperboloids. For
$\Lambda>0$, the entries in
 the $PSU(1,1)$-matrices \eqref{adsxemb} which parametrise the model spacetime
are given by equation \eqref{adspars} and involve the functions
$\sin(\sqrt{\Lambda} T)$, $\cos(\sqrt{\Lambda}T)$. Using the
identities
\begin{align}
\lim_{\sqrt{\Lambda} \rightarrow 0}\cos({\sqrt{\Lambda}
T})=1\qquad \lim_{\sqrt{\Lambda} \rightarrow
0}\tfrac{1}{\sqrt{\Lambda}}\sin({\sqrt{\Lambda} T})=T,
\end{align}
one finds that in the limit $\sqrt{\Lambda}\rightarrow 0$, these
parameters behave according to
\begin{align}&\lim_{\sqrt{\Lambda}\rightarrow 0}t_2(T,z)= T
\tfrac{1+|z|^2}{1-|z|^2}=x_0^L(T,z)\qquad
\lim_{\sqrt{\Lambda}\rightarrow
0}x_1(T,z)=T\tfrac{\text{Re}(z)}{1-|z|^2}=x_1^L(T,z)\qquad
\\
&\lim_{\sqrt{\Lambda}\rightarrow 0}x_2(T,z)=
T\tfrac{\text{Re}(z)}{1-|z|^2}=x_2^L(T,z)\qquad\lim_{\sqrt{\Lambda}\rightarrow
0}t_1(T,z)= \infty,\nonumber
\end{align}
where $x_0^L(T,z),x_1^L(T,z),x_2^L(T,z)$ denote the coordinates in
the identification  of  the unit disc with the hyperboloid
$\hyp_{1/T^2}$ given by \eqref{minkhyp}. For $\Lambda<0$, the
entries in the parametrisation \eqref{dsid0} of the forward
lightcone is given by \eqref{dspars}. Using the identities
\begin{align}
\lim_{\sqrt{|\Lambda|}\rightarrow
0}\cosh(\sqrt{|\Lambda|}T)=1\qquad
\lim_{\sqrt{|\Lambda|}\rightarrow
0}\tfrac{1}{\sqrt{|\Lambda|}}\sinh(\sqrt{|\Lambda|}T)=T,
\end{align}
in \eqref{dspars}, we obtain
\begin{align}
& \lim_{\sqrt{|\Lambda|}\rightarrow 0}x_0(T,z)= T
\tfrac{1+|z|^2}{1-|z|^2}=x_0^L(T,z)\qquad\lim_{\sqrt{|\Lambda|}\rightarrow
0} x_1(T,z)= T\tfrac{\text{Re}(z)}{1-|z|^2}=x_1^L(T,z)\qquad
\\
&\lim_{\sqrt{|\Lambda|}\rightarrow
0}x_2(T,z)=T\tfrac{\text{Re}(z)}{1-|z|^2}=x_2^L(T,z)\qquad
\lim_{\sqrt{|\Lambda|}\rightarrow 0}x_3(T,z)\rightarrow
\infty\nonumber.
\end{align}
Hence, for both positive and negative cosmological constant,  in
the limit $\sqrt{|\Lambda|}\rightarrow 0$ one coordinate in the
parametrisation of the forward lightcone in the model spacetime
$\mathbb{X}_{\Lambda,L}$ tends to infinity while the other
coordinates converge to the corresponding coordinates
parametrising the forward lightcone in Minkowski space.

To determine the role of the cosmological constant as a
deformation parameter of the domains
$U^G_{\Lambda,L}\subset\mathbb{X}_{\Lambda,L}$ associated to
grafted spacetimes, we make use of a result by Benedetti and
Bonsante \cite{bb2} concerning the dependence of the grafted
domains $U^G_{\Lambda,L}$ on the weight of the multicurve $G$. In
\cite{bb2}, Proposition 4.7.1, they consider the multicurve $tG$
on $\hyp_k$ obtained by multiplying all weights in a multicurve
$G$ with a factor $t$ and the associated domains
$U^G_{\Lambda,L}$, $U^{tG}_{\Lambda,L}\subset
\mathbb{X}_{\Lambda,L}$ for cosmological constant
$\Lambda\in\{0,\pm 1\}$. They show that if the grafted domains
$U^{tG}_{\Lambda,L}$ are rescaled by a factor $1/t$, they converge
to the corresponding domain $U^G_{0,L}$ in Minkowski space in the
limit $t\rightarrow 0$
\begin{align}
\label{rescaleid} \lim_{t\rightarrow 0} \tfrac{1}{t}U^{
tG}_{\Lambda,L}=U^G_{0,L}\qquad \Lambda\in\{0,\pm 1\},
\end{align}
where the limit is understood in terms of the coordinates in a
certain parametrisation of these domains given in \cite{bb2}.
Although they do not consider cosmological constants $\Lambda\neq
0,\pm 1$, this result can be applied to determine the dependence
of the grafted domains $U^G_{\Lambda,L}$ on the cosmological
constant $\Lambda$. For this, it is sufficient to recall from
Sect.~\ref{modelspt} that the parametrisation of the model
spacetimes $\mathbb{X}_{\Lambda,L}$ for $|\Lambda|\neq 0,1$  is
obtained by rescaling the associated matrices parametrising
$\mathbb{X}_{\pm 1, L}$ with a factor $1/\sqrt{|\Lambda|}$.
Furthermore, we found in Sect.~\ref{graftingsect}, see in
particular equation \eqref{multmin}, that the weight of the
multicurves for $|\Lambda|=1$  had to be rescaled with a factor
$\sqrt{|\Lambda|}$ to ensure that the associated geodesics are
parametrised by arclenght. The domains $U^G_{\Lambda,L}\subset
\mathbb{X}_{\Lambda,L}$ for non-vanishing cosmological constant
$\Lambda$ are therefore related to the associated domains
$U^G_{\pm 1,L}$ via the identity
\begin{align}
U^{G}_{\Lambda,L}=\tfrac{1}{\sqrt{|\Lambda|}} U^{\sqrt{|\Lambda|}
G}_{\text{sgn}({\Lambda}),L}\qquad \Lambda\neq 0,
\end{align}
and equation \eqref{rescaleid} implies
\begin{align}
\lim_{\sqrt{|\Lambda|}\rightarrow 0} U^G_{\Lambda,L}=U^G_{0,L}
\end{align}
Hence, with an appropriate identification of the coordinates
parametrising the domains
$U^G_{\Lambda,L}\subset\mathbb{X}_{\Lambda,L}$ in the model
spacetimes, the limit $\sqrt{|\Lambda|}\rightarrow 0$ is defined
and yields the coordinates parametrising the domain $U^G_{0,L}$ in
Minkowski space for both positive and negative cosmological
constant.

Finally, we consider the role of the cosmological constant as a
deformation parameter of the group homomorphism
$h^G_{\Lambda,L}:\Gamma\rightarrow
\text{Isom}(\mathbb{X}_{\Lambda,L})$ by which the cocompact
Fuchsian group $\Gamma$ acts on the domains
$U^G_{\Lambda,L}\subset\mathbb{X}_{\Lambda,L}$. For this we note
that the map
$B_{G,\Lambda,L}:\hyp\times\hyp\rightarrow\text{Isom}(\mathbb{X}_{\Lambda,L})$
in formula \eqref{multmin} satisfies
\begin{align}
B_{G,\Lambda,L}(p,q)=B_{\sqrt{|\Lambda|} G\,,\;
\text{sgn}(\Lambda)\,,\;M}(p,q)\qquad \forall p,q\in\hyp,
\Lambda\neq 0,
\end{align}
which implies
\begin{align}
\label{derivslr}
&\tfrac{d}{d\sqrt{\Lambda}}|_{\sqrt{\Lambda}=0}\;\;\quad
h^G_{\Lambda>0,L}(v)=(h^G_{0,L}(v), -h^G_{0,L}(v) )\in
\mathfrak{su}(1,1)\oplus \mathfrak{su}(1,1) & &\forall
v\in\Gamma\\
\label{derivslc}
&\tfrac{d}{d\sqrt{|\Lambda|}}|_{\sqrt{|\Lambda|}=0}\;\;
h^G_{\Lambda<0,L}(v)=i h^G_{0,L}(v)\in \mathfrak{sl}(2,\CC) &
&\forall v\in\Gamma.
\end{align}
Hence, one finds that in the geometrical description of grafted
(2+1)-spacetimes given by Benedetti and Bonsante \cite{bb2}, the
square root $\sqrt{|\Lambda|}$ of the cosmological constant plays
the role of a deformation parameter for both, the domains
$U^G_{\Lambda,L}\subset \mathbb{X}_{\Lambda,L}$ and the
isomorphisms $h_{\Lambda,L}^G:\Gamma\rightarrow
\text{Isom}(\mathbb{X}_{\Lambda,L})$ which determine how the group
$\Gamma$ acts on these domains. In the former, it appears as a
parameter in the coordinates parametrising the domains
$U^G_{\Lambda,L}$ in the model spacetimes
$\mathbb{X}_{\Lambda,L}$, and the limit
$\sqrt{|\Lambda|}\rightarrow 0$ relates these coordinates to the
parametrisation of the associated domain $U^G_{0,L}$ in Minkowski
space. In the latter, it appears as  a
 parameter in the group homomorphisms $h^G_{\Lambda,S}:\Gamma\rightarrow\text{Isom}(\mathbb{X}_{\Lambda,S})$, and one finds that the group homomorphism $h^G_{0,L}$ is
 given by
the derivatives of $h^G_{\Lambda>0,L}$ and $h^G_{\Lambda<0,L}$
with respect to $\sqrt{|\Lambda|}$.

\subsection{The Chern-Simons description}

In the Chern-Simons description of (2+1)-spacetimes, the
cosmological constant appears as a parameter in the Lie algebras
$\gothh_{\Lambda,L}$ and the associated gauge groups
$\text{Isom}(\mathbb{X}_{\Lambda,L})$. To clarify in what sense it
can be viewed as a deformation parameter, one has to introduce a
common framework which encompasses the Lie algebras \eqref{liealg}
for all signs of the cosmological constant. Such a framework can
be realised by interpreting the Lie algebras \eqref{liealg} as the
(2+1)-dimensional Lorentz algebra over a commutative ring
$R_\Lambda$ with a $\Lambda$-dependent multiplication law.

\begin{lemma} Consider the vector space $\RR^2$ with the usual
vector addition and a multiplication operation $\cdot:
\RR^2\times\RR^2\rightarrow \RR^2$ which depends on a real
parameter $\Lambda$ and is given by
\begin{align}
\label{ringmult} (a,b)\cdot (c,d)=(ac+\Lambda bd, ad+ bc)\qquad
\forall a,b,c,d\in\RR.
\end{align}
 Then, \eqref{ringmult} equips $\RR^2$ with the structure of a
 commutative ring
 $R_\Lambda=(\RR^2,+,\cdot)$ with the unit elements for the addition
 $+$ and the multiplication $\cdot$ given by, respectively, $(0,0)$, $(1,0)$.
\end{lemma}
In the following we will express elements of the ring $R_\Lambda$
in terms of a formal parameter $\Theta_\Lambda$ and write
$a+\Theta_\Lambda b$ for the element $(a,b)$. Formally, the
product of two elements $(a,b), (c,d)\in R_\Lambda$ is then given
by
\begin{align}
(a+ b\Theta_\Lambda )\cdot(c+ d\Theta_\Lambda ) =
ac+(ad+bc)\Theta_\Lambda+ bd\Theta^2_\Lambda,
\end{align}
and to obtain agreement with the multiplication law
\eqref{ringmult}, one must  set $\Theta_\Lambda^2=\Lambda$, and
the parameter $\Theta_\Lambda$ can therefore be viewed as a formal
square root of  $\Lambda$.

For  $\Lambda<0$, the commutative ring $R_\Lambda$ is isomorphic
to the field $\CC$ and the formal parameter $\Theta_\Lambda$ is
the complex number
$\Theta_\Lambda=\sqrt{\Lambda}=i\sqrt{|\Lambda|}$. For
$\Lambda=0$, the formal parameter $\Theta_\Lambda$ satisfies
$\Theta_0^2=0$ like the one occurring in supersymmetry and
corresponds to the formal parameter $\theta$ used to parametrise
the (2+1)-dimensional Poincar\'e group in \cite{martin}, for a
more detailed discussion see also \cite{ich}. This implies that
the commutative ring $R_0$ is not a field, since
 elements of the form $\Theta_\Lambda b$, $b\in\RR\setminus\{0\}$ are
zero divisors
\begin{align}
\Theta_\Lambda b\cdot \Theta_\Lambda c=0\qquad\forall b,c\in\RR.
\end{align}
For $\Lambda>0$, the formal parameter $\Theta_\Lambda$ satisfies
$\Theta_\Lambda^2=\Lambda>0$. Again, $R_\Lambda$ is not a field
and has zero divisors
 $\sqrt{\Lambda}a\pm \Theta_\Lambda a$, $a\in\RR\setminus\{0\}$, which
satisfy
\begin{align}
\label{nulltads} (\sqrt{\Lambda} a+\Theta_\Lambda
a)(\sqrt{\Lambda}a-\Theta_\Lambda a)=0\qquad (\sqrt{\Lambda}a \pm
\Theta_\Lambda a)^2=
2a\sqrt{\Lambda}(\sqrt{\Lambda}a\pm\Theta_\Lambda a).
\end{align}
The  ring $R_\Lambda$ allows one to identify all Lie algebras
$\gothh_{\Lambda,L}$ with brackets \eqref{liealg} with  the
(2+1)-dimensional Lorentz algebra, only that now this Lie algebra
is no longer considered as a Lie algebra over $\RR$ but as a Lie
algebra over the commutative ring\footnote{Definitions and
properties concerning Lie algebras over commutative rings can be
found for instance in \cite{bourbaki}, Chapter 1, but in the
following we will make only use of some basic concepts.}
$R_\Lambda$.
 This identification of the Lie algebras
$\gothh_{\Lambda,L}$ with the (2+1)-dimensional Lorentz algebra
over  $R_\Lambda$ generalises the concept of  complexification of
real Lie algebras and in the case of negative cosmological
constant yields the complexification
$\mathfrak{sl}(2,\CC)=\mathfrak{sl}(2,\RR)\oplus
i\,\mathfrak{sl}(2,\RR)$. We consider the (2+1)-dimensional
Lorentz algebra with generators $J_a$, $a=0,1,2$, and bracket
\begin{align}
\label{lorbr} [J_a,J_b]=\epsilon_{ab}^{\;\;\;c}J_c.
\end{align}
By identifying the generators $J_a$ with the
$\mathfrak{sl}(2,\RR)$ matrices in  \eqref{sl2rj}, we obtain a
$R_\Lambda$ module isomorphism from $R_\Lambda^3$ into the set
$\mathfrak{sl}(2, R_\Lambda)$ of traceless two-by-two matrices
with entries in $R_\Lambda$ or, equivalently, the set of
endomorphisms of the $R_\Lambda$-module $R_\Lambda^2$ with
vanishing trace form.
\begin{align}
\label{ringpar} \bx+\Theta_\Lambda\by\in\RR^2_\Lambda\mapsto
&(x^a+\Theta_\Lambda
y^a)J_a\\
=&\left(\begin{array}{cc} -\tfrac{1}{2}(x^1+\Theta_\Lambda y^1) &
\tfrac{1}{2}(x^0+y^0)+\tfrac{1}{2}\Theta_\Lambda(x^2+y^2)\\
-\tfrac{1}{2}(x^0+y^0)+\tfrac{1}{2}\Theta_\Lambda(x^2+y^2) &
\tfrac{1}{2}(x^1+\Theta_\Lambda y^1 )
\end{array}\right).\nonumber
\end{align}
The commutator of two $\mathfrak{sl}(2,R_\Lambda)$ matrices then
agrees with the bracket obtained by extending \eqref{lorbr}
bilinearly in $R_\Lambda$, and with the identification
\begin{align}
\label{pdef} P_a=\Theta_\Lambda J_a \end{align} one recovers the
Lie bracket \eqref{liealg} of the real Lie algebras
$\gothh_{\Lambda,L}$.

 Moreover, the
identification of the Lie algebras $\gothg_{\Lambda,L}$  with
$\mathfrak{sl}(2,R_\Lambda)$ allows one to relate the
$\Ad$-invariant, symmetric bilinear forms $\langle\,,\,\rangle$
and $\kappa$ on $\gothh_{\Lambda,L}$ defined by \eqref{inprod},
\eqref{bilin} to the Killing form of the (2+1)-dimensional Lorentz
algebra. For this one extends the Killing form $g_K$ on
$\mathfrak{sl}(2,\RR)$
\begin{align}
\label{killingf}
g_K(J_a,J_b)=\tfrac{1}{2}\eta^L_{ab}=\text{Tr}(J_a\cdot J_b)
\end{align}
bilinearly to an $\Ad$-invariant symmetric $R_\Lambda$-bilinear
form $g_K: \mathfrak{sl}(2,R_\Lambda)\times
\mathfrak{sl}(2,R_\Lambda)\rightarrow R_\Lambda$. Using the
parametrisation \eqref{ringpar} and comparing the resulting
expressions with \eqref{inprod}, \eqref{bilin} one obtains
\begin{align}
\label{formid} g_K((\bp+\Theta_\Lambda\bk), ( \bq +
\Theta_\Lambda\bm ))=&\tfrac{1}{2}(p^a q^b+\Lambda k^a m^b)
\eta^L_{ab}+ \tfrac{1}{2}\Theta_\Lambda(p^a m^b+ k^a
q^b)\eta^L_{ab}\\
=&\tfrac{1}{2} \kappa(p^a J_a+k^a P_a, q^b J_b+m^b P_b) +
\Theta_\Lambda\tfrac{1}{2}\langle p^a J_a+k^a P_a, q^b J_b+m^b
P_b\rangle.\nonumber
\end{align}
Hence, the $\Ad$-invariant symmetric forms $\kappa$
 and $\langle\,,\,\rangle$ on $\gothh_{\Lambda,L}$ are realised as
the projections of the Killing form on
$\mathfrak{sl}(2,R_\Lambda)$ on, respectively, the first and
second component of the ring $R_\Lambda=(\RR^2,+,\cdot)$. This
generalises the situation for $\Lambda<0$, where these forms can
be identified with the real and imaginary part of the Killing form
on $\mathfrak{sl}(2,\CC)$. Moreover, it sheds some light on the
distinguished role played by the $\Ad$-invariant symmetric
bilinear forms $\langle\,,\,\rangle$ and $\kappa$ on the Lie
algebra $\gothh_{\Lambda,L}$. While any linear combination of
these two forms is again an $\Ad$-invariant, symmetric bilinear
from on $\gothh_{\Lambda,L}$, the forms $\langle\,,\,\rangle$ and
$\kappa$ are the only ones that arise canonically from the Killing
form on $\mathfrak{sl}(2, R_\Lambda)$.


We will now demonstrate how the identification of the Lie algebra
$\gothh_{\Lambda,L}$ with the (2+1)-dimensional Lorentz algebra
over the commutative ring $R_\Lambda$ gives rise to the associated
matrix groups $H_{\Lambda,L}$. Although in general the exponential
of Lie algebras over commutative rings cannot be defined in a
straightforward manner, the particularly simple structure of the
ring $R_\Lambda$ allows us to obtain the groups $H_{\Lambda>0,L}=
SU(1,1)\times SU(1,1)\cong SL(2,\RR)\times SL(2,\RR)$, $H_{0,L}=
SU(1,1)\ltimes \RR^3\cong SL(2,\RR)\ltimes\RR^3$ and
$H_{\Lambda<0,L}=SL(2,\CC)$ by exponentiating matrices in
$\mathfrak{sl}(2,R_\Lambda)$. For this, we consider the formal
expression
\begin{align}
\label{formalexp}
\exp_\Lambda(\bx+\Theta_\Lambda\by)=\sum_{n=0}^\infty
\frac{((x^a+\Theta y^a) J_a)^n}{n!}\qquad\bx+\Theta_\Lambda\by\in
\mathfrak{sl}(2,R_\Lambda),
\end{align}
where $((x^a+\Theta y^a)J_a)^n$ stands for the $n^{th}$ power of
the matrices \eqref{ringpar}. In the case of negative cosmological
constant, we have $\Theta_\Lambda=i\sqrt{|\Lambda|}$ and
expression \eqref{formalexp} is the exponential map
$\exp_{\Lambda<0,L}:\mathfrak{sl}(2,\CC)\rightarrow SL(2,\CC)$.
The case of vanishing cosmological constant is investigated in
\cite{ich}. Using identity $\Theta_{0}^2=0$, one can express
\eqref{formalexp} as
\begin{align}
\label{martinexp} \exp_{\Lambda=0}((x^a +\Theta_{0}
y^a)J_a)=&\sum_{n=0}^\infty \frac{(x^a
J_a)^n}{n!}+\Theta_{0}\sum_{n=0}^\infty\sum_{m=0}^{n-1} \frac{(x^a
J_a)^L (y^b J_b)(x^c
J_c)^{n-m-1}\!\!\!\!\!\!\!\!\!\!\!\!\!\!\!}{n!}.
\end{align}
To simplify this expression further, we move the factors $y^b J_b$
in \eqref{martinexp} to the left and evaluate the resulting
commutators using the formulas
\begin{align}
\label{tident} [(x^aJ_a)^L, y^b J_b]=\sum_{k=1}^L
\left(\begin{array}{l} m\\ k\end{array}\right) \ad_{x^aJ_a}^k (y^b
J_b)\cdot (x^c J_c)^{m-k}\qquad \sum_{m=k}^{n-1} \left(\begin{array}{l} m\\
k\end{array}\right)=\left(\begin{array}{c} n\\
k+1\end{array}\right)
\end{align}
which can be proved by induction. After some further computation,
this yields
\begin{align}
\label{minkring} \exp_{\Lambda=0}((x^a+\Theta_{0} y^a)J_a)=(1+
\Theta_{0} (T(\bx)\by)^b J_b)\cdot e^{x^a J_a},
\end{align}
where $e^{x^a J_a}$ stands for the exponential of the
$\mathfrak{sl}(2,\RR)$-matrix $x^a J_a$ given by \eqref{lorexpmap}
and the linear map $T(\bx): \RR^3\rightarrow\RR^3$ is the one
defined in \eqref{tdef}. By comparing with the parametrisation of
the (2+1)-dimensional Poincar\'e group $SU(1,1)\ltimes \RR^3$
introduced in Sect.~\ref{defnotsect}, we find that
\eqref{minkring} agrees with \eqref{exmapform} if we identify
\begin{align}
(u,\by)\in SU(1,1)\ltimes \RR^3 \cong (1+\Theta_{0} y^a J_a)\cdot
u\qquad\forall u\in SL(2,\RR), \by\in\RR^3,
\end{align}
and we recover the multiplication law \eqref{poincpar}
\begin{align}
(1+\Theta_{0} x^a J_a) u \cdot (1+\Theta_{0} y^b J_b)
v=(1+\Theta_{0}(x^a J_a+ \Ad(u) y^b J_b)uv\quad \forall u,v\in
SL(2,\RR), \bx,\by\in\RR^3.\nonumber
\end{align}
Hence, by exponentiating the Lie algebra
$\mathfrak{sl}(2,R_\Lambda)$ for cosmological constant
$\Lambda=0$, one obtains the parametrisation of the
(2+1)-dimensional Poincar\'e group in Sect.~\ref{defnotsect} whose
 description in terms of a formal supersymmetry parameter
$\Theta_0$ satisfying $\Theta^2_0=0$ was first introduced in the
context of (2+1)-dimensional gravity by Martin \cite{martin}.

For $\Lambda>0$, the exponential \eqref{formalexp} can be
evaluated by introducing the generators
 defined in \eqref{jpmbrack}
\begin{align}
J_a^\pm =\tfrac{1}{2}(1
\pm\tfrac{\Theta_\Lambda}{\sqrt{\Lambda}})J_a,
\end{align}
in terms of which the argument of \eqref{formalexp} takes the form
\begin{align}
(x^a+\Theta_\Lambda y^a)J_a=(x^a+\sqrt{\Lambda}
y^a)J_a^++(x^a-\sqrt{\Lambda} y^a)J_a^-.
\end{align}
Using identity \eqref{nulltads}, we recover the splitting of
$\gothh_{\Lambda>0,L}$ into the direct sum
$\mathfrak{sl}(2,\RR)\oplus\mathfrak{sl}(2,\RR)$
\begin{align}
J_a^+ \cdot J_b^-=J_a^- \cdot J_b^+=0\qquad J_a^\pm \cdot
J_b^\pm=\tfrac{1}{2}(1\pm\tfrac{\Theta_\Lambda}{\sqrt{\Lambda}})(\tfrac{1}{4}\eta_{ab}+\tfrac{1}{2}\epsilon_{ab}^{\:\:c})J_c,
\end{align}
and by applying these identities to \eqref{formalexp} we obtain
\begin{align}
\label{adsexp} \exp_{\Lambda>0}((x^a+\Theta_\Lambda y^a)
J_a)=&\sum_{n=0}^\infty \frac{((x^a+\sqrt{\Lambda}y^a)
J^+_a)^n}{n!}+ \sum_{n=0}^\infty
\frac{((x^a-\sqrt{\Lambda}y^a) J^-_a)^n}{n!}\nonumber\\
=&\tfrac{1}{2}(1+\tfrac{\Theta_\Lambda}{\sqrt{\Lambda}}) e^{(x^a
+\sqrt{\Lambda}y^a)J_a}+
\tfrac{1}{2}(1-\tfrac{\Theta_\Lambda}{\sqrt{\Lambda}}) e^{(x^a
-\sqrt{\Lambda}y^a)J_a}.
\end{align}With the identification
\begin{align}
(u_+,u_-)\cong
\tfrac{1}{2}(1+\tfrac{\Theta_\Lambda}{\sqrt{\Lambda}})u_+ +
\tfrac{1}{2}(1-\tfrac{\Theta_\Lambda}{\sqrt{\Lambda}})u_-\qquad\forall
u_\pm\in SL(2,\RR)
\end{align}
we then recover formula \eqref{exmapform} for the exponential map
$\exp: \mathfrak{sl}(2,\RR)\oplus \mathfrak{sl}(2,\RR)\rightarrow
SL(2,\RR)\times SL(2,\RR)$ and the group multiplication law
\eqref{dirprod}. Hence, for all values of the cosmological
constant, the group $H_{\Lambda,L}$ and the exponential map
$\exp_{\Lambda,L}: \gothh_{\Lambda,L}\rightarrow H_{\Lambda,L}$
can be obtained from the identification of the Lie algebra
$\gothh_{\Lambda,L}$ with $\mathfrak{sl}(2, R_\Lambda)$.

The cosmological constant can therefore be implemented in the
Chern-Simons formulation of (2+1)-dimensional gravity by
interpreting it as a parameter in the multiplication law
\eqref{ringmult} of a commutative ring $R_\Lambda$. By
parametrising the elements of this ring in terms of a formal
parameter $\Theta_\Lambda$ satisfying $\Theta_\Lambda^2=\Lambda$
one then obtains a unified description of the Lie algebras
$\gothh_{\Lambda,L}$ and the associated Lie groups
$H_{\Lambda,L}$, which can be identified, respectively, with the
Lie algebra $\mathfrak{sl}(2,R_\Lambda)$ and the associated matrix
groups obtained by exponentiation.

\subsection{Grafting transformations as deformed Dehn twists}

\label{ccgrdt}

We will now apply these results  to demonstrate that the parameter
$\Theta_\Lambda$ which can be viewed as a formal square root of
$\Lambda$ appears as a deformation parameter relating the Dehn
twist and grafting transformations \eqref{dtform} and
\eqref{conctrafosaibi} for all values of the cosmological constant
$\Lambda$. For this we recall the discussion from
Sect.~\ref{grafthamilt} and Sect.~\ref{grdtsect}, where it is
shown that the grafting transformation
$T^t_{\tilde\kappa,\eta,\Lambda,L}:
\text{Isom}(\mathbb{X}_{\Lambda,S})^{2g}\rightarrow
\text{Isom}(\mathbb{X}_{\Lambda,S})^{2g}$ associated to a closed,
simple curve $\eta$ on $S_g$ and the corresponding Dehn twist
$T^t_{\tilde t,\eta,\Lambda,L}:
\text{Isom}(\mathbb{X}_{\Lambda,S})^{2g}\rightarrow
\text{Isom}(\mathbb{X}_{\Lambda,S})^{2g}$ are generated,
respectively,
 by
Wilson loop observables $\tilde\kappa_\eta$ and $\tilde t_\eta$
constructed from bilinear forms $\kappa$ and $\langle\,,\,\rangle$
on $\gothh_{\Lambda,S}$. The fact that the bilinear forms $\kappa$
and $\langle\,,\rangle$ on $\gothh_{\Lambda,L}$ appear as
projections on, respectively, the real and the
$\Theta_\Lambda$-component   of Killing form $g_K$ on
$\mathfrak{sl}(2,R_\Lambda)$ allows one to interpret the Wilson
loop observables $\tilde \kappa_\eta$, $\tilde t_\eta$ as
projections on real and $\Theta_\Lambda$ component of a Wilson
loop observable $\widetilde{g_K}_\eta
\in\cif(\text{Isom}(\mathbb{X}_{\Lambda,L})^{2g})$ which takes
values in the Ring $R_\Lambda$
\begin{align}
\label{ringloopobs}
&\widetilde{g_K}_\eta(A_1,\ldots,B_g)=\tfrac{1}{2}\tilde\kappa_\eta(A_1,\ldots,B_g)+\tfrac{1}{2}\Theta_\Lambda
\tilde t_\eta(A_1,\ldots,B_g).\nonumber
\end{align}
Moreover, we found in Sect.~\ref{grdtsect} that the grafting and
Dehn twist transformations are of a similar form. Both act on the
holonomies $\ai,\bi$ by right-multiplication with functions
$G^t_{\tilde t,\Lambda,L}, G^t_{\tilde\kappa,\Lambda,L}:
\text{Isom}(\mathbb{X}_{\Lambda,L})\rightarrow
\text{Isom}(\mathbb{X}_{\Lambda,L})$ of the holonomies of certain
curves conjugated to $\eta$, which are obtained  as exponentials
of linear maps $\bl_{t}, \bl_\kappa: \gothh_{\Lambda,L}\rightarrow
\gothh_{\Lambda,L}$. For all values of the cosmological constant,
the linear map
 for the Dehn twist is the identity
 $\bl_t=\text{id}_{\gothh_{\Lambda,L}}$, and the associated map $G^t_{\tilde t,\Lambda,L}$
 takes the form
\begin{align}
G^t_{\tilde t,\Lambda,L}(u)=u^{t}\qquad\forall
u\in\text{Isom}(\mathbb{X}_{\Lambda,L}).
\end{align}
The linear map $\bl_\kappa$ and the associated one parameter group
of diffeomorphisms  $G^t_{\tilde\kappa, \Lambda,L}$ for the
grafting transformation are given by, respectively, \eqref{blmaps}
and \eqref{biling}. Unlike the corresponding maps for the Dehn
twists, they show an explicit dependence on the cosmological
constant $\Lambda$. The identification of the Lie algebra
$\gothh_{\Lambda,L}$ with $\mathfrak{sl}(2,R_\Lambda)$  allows us
to relate these maps for the different values of the cosmological
constant and to establish a link with the corresponding maps for
Dehn twists. For this we note that for all values of the
cosmological constant the linear map
 $\bl_\kappa$ on $\gothh_{\Lambda,L}$ given by \eqref{blmaps}  can be identified with a linear map
on the Lie algebra $\mathfrak{sl}(2, R_\Lambda)$, which acts by
multiplication with $\Theta_\Lambda$
\begin{align}
\bl_{\kappa}(\bx+\Theta_\Lambda\by)=\Theta_\Lambda
(\bx+\Theta_\Lambda\by)\qquad\forall \bx+\Theta_\Lambda \by\in
\mathfrak{sl}(2,R_\Lambda).
\end{align}
The discussion in the previous subsection then allows us to
express the associated one-parameter group of transformations
$G^t_{\tilde\kappa,\Lambda,L}:
\text{Isom}(\mathbb{X}_{\Lambda,L})\rightarrow\text{Isom}(\mathbb{X}_{\Lambda,L})
$ via the exponential map \eqref{formalexp}
\begin{align}
\label{formalmult} G^t_{\tilde\kappa,
\Lambda,L}(\exp_\Lambda(\bx+\Theta_\Lambda
\by))=\exp_\Lambda(t\Theta_\Lambda(\bx+\Theta_\Lambda\by))\qquad\forall\bx+\Theta_\Lambda\by\in\mathfrak{sl}(2,R_\Lambda).
\end{align}
Evaluating this expression by setting
$\Theta_\Lambda=i\sqrt{|\Lambda|}$ for $\Lambda<0$ and by using
expressions \eqref{minkring} and \eqref{adsexp} for $\Lambda=0$
and $\Lambda>0$, we recover expression \eqref{biling}. Hence, the
one-parameter group of transformations
$G^t_{\tilde\kappa,\Lambda,L},G^t_{\tilde t,\Lambda,L}:
\text{Isom}(\mathbb{X}_{\Lambda,L})\rightarrow
\text{Isom}(\mathbb{X}_{\Lambda,L})$ are formally related by the
identity
\begin{align}
G^t_{\tilde\kappa,\Lambda,L}=G^{t\Theta_\Lambda}_{\tilde
t,\Lambda,L}.
\end{align}
After inserting this identity in the expressions \eqref{dtform},
\eqref{conctrafosaibi} for the Dehn twist and the grafting
transformation, we find that formally, the transformation of the
holonomies $\ai,\bi$ under grafting  along a closed, simple curve
$\eta$ on $S_g$ with parameter $t$ can be expressed as a Dehn
twist with parameter $\Theta_\Lambda t$
\begin{align}
T^t_{\tilde\kappa, \eta,\Lambda,L}=T^{t\Theta_\Lambda}_{\tilde
t,\eta,\Lambda,L}.
\end{align}
By interpreting the Lie algebra $\gothh_{\Lambda,L}$ of the gauge
group in Chern-Simons formulation as a (2+1)-dimensional Lorentz
algebra over the commutative ring $R_\Lambda$, we therefore
establish a common pattern which relates the grafting and Dehn
twist transformations for all values of the cosmological constant.
The dependence on the cosmological constant is encoded in the
formal parameter $\Theta_\Lambda$ satisfying
$\Theta_\Lambda^2=\Lambda$ which plays the role of a deformation
parameter. In this formalism, the two Wilson loop observables
$\tilde\kappa_\eta$, $\tilde t_\eta$ associated to a closed,
simple curve $\eta$ on $S_g$  arise canonically  as the projection
on the $\Theta_\Lambda$-component and on the real component of the
$R_\Lambda$-valued Wilson loop observable \eqref{ringloopobs}
constructed from the Killing form on $\mathfrak{sl}(2,R_\Lambda)$.
Via the Poisson bracket, these two canonical observables generate
the two basic geometry changing transformations on the phase
space. The former acts as the Hamiltonian for the Dehn twist
transformations \eqref{dtform}, while the latter is the
Hamiltonian for the grafting transformation
\eqref{conctrafosaibi}. Viewed as transformations over the ring
$R_\Lambda$, these two phase space transformations exhibit a
similar structure and can be transformed into each other by
substituting $t\mapsto t\Theta_\Lambda$.

\section{Conclusions}
\label{outlook}

In this paper we clarified the relation between the geometrical
and the Chern-Simons description of (2+1)-dimensional spacetimes
of topology $\RR\times S_g$ for Lorentzian signature and general
cosmological constant and for the Euclidean case with negative
cosmological constant. We showed how the fact that such spacetimes
are obtained as quotients of the
 model spacetimes $\mathbb{X}_{\Lambda,S}$ corresponds to the
trivialisation of the gauge field in the Chern-Simons formalism.
This allowed us to relate the variables encoding the physical
degrees of freedom in the two approaches, the group homomorphism
$h_{\Lambda,\Gamma}^G:\Gamma\rightarrow\text{Isom}(\mathbb{X}_{\Lambda,S})$
in the geometric formulation and the holonomies along a set of
generators of the fundamental group $\pi_1(S_g)$ in the
Chern-Simons description.

We demonstrated how the construction of evolving (2+1)-spacetimes
via grafting along closed, simple geodesics $\eta$ gives rise to a
transformation on the phase space of the associated Chern-Simons
theory. After deriving explicit expression for the transformation
of the holonomies, we showed that this transformation is generated
via the Poisson bracket by one of the two canonical Wilson loop
observables associated to $\eta$, while the other observable
generates Dehn twists. We found a close relation between the
action of these transformations on the phase space which is
reflected in a general symmetry relation for the associated Wilson
loop observables.

Finally, we investigated the role of the cosmological constant in
the geometrical and the Chern-Simons formulation of the theory
with Lorentzian signature. We found that the square root of minus
the cosmological constant can be viewed as a deformation parameter
in the parametrisation of the domains
$U^G_{\Lambda,L}\subset\mathbb{X}_{\Lambda,L}$ and in the group
homomorphisms $h^G_{\Lambda,L}: \Gamma\rightarrow
\text{Isom}(\mathbb{X}_{\Lambda,L})$. In the Chern-Simons
formulation, we obtained a unified description for the different
signs of the cosmological constant by  identifying the Lie
algebras of the gauge groups with the (2+1)-dimensional Lorentz
algebra $\mathfrak{sl}(2, R_\Lambda)$ over a commutative ring
$R_\Lambda$. In this framework,  the cosmological constant arises
as a parameter in the ring's multiplication law and can be
implemented via a formal parameter $\Theta_\Lambda$ satisfying
$\Theta_\Lambda^2=\Lambda$. By extending the Killing form on the
(2+1)-dimensional Lorentz algebra to an $\Ad$-invariant, bilinear
form on $\mathfrak{sl}(2,R_\Lambda)$ and considering the
associated Wilson loop observables with values in $R_\Lambda$, we
found that the Wilson loop observables  generating grafting and
Dehn twists arise canonically as the projections on the real and
the $\Theta_\Lambda$ component of this $R_\Lambda$-valued
observable. Moreover, we found that a grafting transformation with
weight $w$ associated to a closed, simple curve $\eta$ on $S_g$
can be viewed as a Dehn twist around $\eta$ with parameter
$\Theta_\Lambda w$.

These results clarify the relation between spacetime geometry and
the description of the phase space in the Chern-Simons formalism
and provide a geometrical interpretation of the Wilson loop
observables. Moreover, we obtained explicit expressions for the
action of grafting and Dehn twists in Fock and Rosly's description
of the phase space \cite{FR}, which is the basis of the
combinatorial quantisation formalism \cite{AGSI, AGSII}  and the
related approaches \cite{BNR} and \cite{we2} for the group
$SL(2,\CC)$ and semidirect product groups $\prgr$ such as the
(2+1)-dimensional Poincar\'e group. It would therefore be
interesting to see how these results can be applied to the
quantised theory and to use them to investigate concrete physics
questions in quantum (2+1)-gravity.

\subsection*{Acknowledgements}

I thank Bernd Schroers  for comments on the draft of this paper.
Research at Perimeter Institute is supported in part by the
Government of Canada through NSERC and by the Province of Ontario
through MEDT.

\end{document}